\documentclass[opre,nonblindrev,copyedit]{informs2}

\usepackage{amsfonts}
\usepackage{amssymb,color}
\usepackage{amsmath,epsfig}
\usepackage{rawfonts,graphicx,amsfonts,amssymb}
\usepackage{array}
\usepackage{mathrsfs}
\usepackage{url}
\usepackage{multirow}
\usepackage{mathrsfs}
\usepackage{graphicx}
\usepackage{amssymb,amsmath}
\usepackage{color}
\usepackage{verbatim}
\usepackage{bbding,lscape}
\usepackage{algorithm}
\usepackage{subfig}

\ifodd 12
\newcommand{\rev}[1]{{\color{blue}#1}} 
\newcommand{\com}[1]{\textbf{\color{red} (COMMENT by Jianwei: #1) }} 

\newcommand{\response}[1]{\textbf{\color{green} (RESPONSE: #1)}}

\else
\newcommand{\rev}[1]{#1}

\newcommand{\com}[1]{}
\newcommand{\response}[1]{}
\fi

\def\argmin{arg\,min}
\def\argmax{arg\,max}
\def\arginf{arg\,inf}

\def\ag{\text{agent}}
\def\ags{\text{agents}}

\def\pp{\text{principal}}

\def\sp{\textsc{p}}     
\def\a{\textsc{a}}       

\def\EX{\mathbb{E}}

\def\dt{ \mathrm{d}}

\def\x{x^{*}}   
\def\Xset{\mathcal{X}}  
\def\xe{\hat{x}}  
\def\y{y}        
\def\yr{\tilde{y}}  
\def\yrn{\yr_{\n}}

\def\N{N}  
\def\Nset{\mathcal{A}}    
\def\Nsetactive{\mathcal{A}^{\textsc{p}}}    

\def\n{n}    
\def\no{\n_{0}}

\def\yn{\y_{\n}}   
\def\yr{\hat{y}}  
\def\yrn{\hat{y}_{\n}}  

\def\P{R}   

\def\q{q}  
\def\Q{Q}  

\def\qp{\q^{\sp}}  
\def\qpn{\qp_{\n}}
\def\QP{\Q^{\sp}}  
\def\BQP{\boldsymbol{\QP}}  

\def\qn{\q_{\n}}  

\def\qa{\q}
\def\qap{\qa^{\sp}}
\def\QA{\Q}
\def\QAP{\QA^{\sp}}
\def\QB{\Q}
\def\QBP{\QB^{\sp}}

\def\qapn{\q^{\sp}_{\n}}
\def\qapno{\q^{\sp}_{\no}}

\def\qbpo{\q^{\sp}_{1}}
\def\qbpN{\q^{\sp}_{\N}}
\def\qbpi{\q^{\sp}_{i}}
\def\qbpj{\q^{\sp}_{j}}

\def\qbpn{\q^{\sp}_{\n}}
\def\xea{\xe}
\def\xeb{\xe}

\def\PA{\P}
\def\PB{\P}
\def\pia{\pi}
\def\pib{\pi}
\def\Sba{\Sb}
\def\Sbb{\Sb}
\def\Saa{\Sa}
\def\Sab{\Sa}
\def\bqa{\bq}
\def\bqb{\bq}

\def\bqap{\bqa^{\sp}}
\def\bqbp{\bqb^{\sp}}
\def\CA{\C}  
\def\CB{\C}  

\def\bq{\boldsymbol{\q}}
\def\bqp{\boldsymbol{\q^{\sp}}}

\def\by{\boldsymbol{\y}}
\def\byr{\boldsymbol{\yr}}

\def\C{C}  
\def\c{c}

\def\th{\theta}   
\def\thn{\th_{\n}}
\def\thl{\underline{\th}}
\def\thu{\bar{\th}}
\def\thr{\hat{\th}}
\def\thrn{\thr_{\n}}

\def\bth{\boldsymbol{\th}}
\def\bthr{\boldsymbol{\thr}}

\def\F{F}   
\def\f{f}    

\def\Rset{\mathbb{R}}  
\def\Rsetp{\Rset_{+}}   

\def\AMset{\mathcal{M}}

\def\est{\text{observation}}

\def\ests{\text{observations}}

\def\pre{\text{prediction}}

\def\Pf{\phi}   

\def\Ths{\theta}          

\def\BThs{\boldsymbol{\Ths}}

\def\ae{\textsc{ae}}       

\def\U{U} 
\def\UP{\U^{\sp}}
\def\UA{\U^{\a}}  
\def\UAE{\U^{\ae}}  

\def\hp{\B^{\sp}}  
\def\ha{\B^{\a}}  

\def\var{\sigma^2}   
\def\vars{\sigma^2_0}   

\def\B{B}  

\def\Sa{K}
\def\Sb{S}

\def\l{\ell}   
\def\lp{\l^{\sp}} 
\def\la{\l^{\a}} 

\newcommand{\gaussian}[2]{{\cal N}(#1,#2)}
\newcommand{\costHomogeneous}{\th^{\dagger}}
\newcommand{\effortHomogeneous}{\q^{\dag}}

\usepackage{endnotes}
\let\footnote=\endnote

\usepackage{natbib}
 \bibpunct[, ]{(}{)}{,}{a}{}{,}%
 \def\bibfont{\small}%
\TheoremsNumberedThrough     
\ECRepeatTheorems

\EquationsNumberedThrough    

\begin{document}

\RUNTITLE{Parametric Prediction from Parametric Agents}

\TITLE{Parametric Prediction from Parametric Agents}

\ARTICLEAUTHORS{%
	\AUTHOR{Yuan Luo}
	\AFF{Department of Information Engineering, The Chinese University of Hong Kong, \\\EMAIL{yluo@ie.cuhk.edu.hk,}}
	\AUTHOR{Nihar B. Shah}
	\AFF{Department of Electrical Engineering and Computer Sciences, UC Berkeley, \\
		\EMAIL{nihar@eecs.berkeley.edu}}
	\AUTHOR{Jianwei Huang}
	\AFF{Department of Information Engineering, The Chinese University of Hong Kong, \\\EMAIL{jwhuang@ie.cuhk.edu.hk}} 
	\AUTHOR{Jean Walrand}
	\AFF{Department of Electrical Engineering and Computer Sciences, UC Berkeley, \\
		\EMAIL{walrand@berkeley.edu}}
}

\ABSTRACT{%
	We consider a problem of prediction based on opinions elicited from heterogeneous rational agents with private information.
	Making an accurate prediction with a minimal cost requires a \emph{joint} design of the incentive mechanism and the prediction algorithm. Such a problem lies at the nexus of statistical learning theory and game theory, and arises in many domains such as consumer surveys and mobile crowdsourcing.
	In order to elicit heterogeneous agents' private information and incentivize agents with different capabilities to act in the principal's best interest,
	we design an optimal joint incentive mechanism and prediction algorithm called COPE (COst and Prediction Elicitation), the analysis of which offers several valuable engineering insights.
	First, when the costs incurred by the agents are linear in the exerted effort,
	COPE corresponds to a ``crowd contending''  mechanism, where the principal only employs the agent with the highest capability.
	Second, when the costs are quadratic, COPE corresponds to a ``crowd-sourcing'' mechanism that employs multiple agents with different capabilities at the same time.
	Numerical simulations
	show that COPE improves the {\pp}'s profit and the network profit significantly (larger than $30\%$ in our simulations), comparing to those mechanisms that assume all agents have equal capabilities.
}%


\maketitle



\section{Introduction}\label{sec:intro}

\subsection{Background and Motivations}
Prediction markets, which aggregate information elicited from people with private beliefs, have served as a reliable tool for estimating the outcome of specific future events (see \cite{berg2003prediction}). For example, these markets have been used to predict the winners of election (see \cite{wolfers2011forecasting}), future demand for a product (see \cite{hayes1998measuring}), and stock prices and returns (see \cite{gottschlich2014decision}). In these markets, values of traded information depend on future outcomes, and the accuracy of prediction can be verified based on the realized outcomes.


With the emergence of several commercial crowdsourcing platforms such as Amazon Mechanical Turk and Microworkers, collecting information from people to make prediction has become much cheaper, easier and faster. However, the information collected from people (``the agents'') can be highly unreliable due to the agents' insufficient expertise and the lack of appropriate incentives. More specifically, in a crowdsourcing platform, the agents are heterogeneous as they come from different countries and have different skills, which leads to significant variations of the work quality (see \cite{karger2014budget}).
Furthermore, agents may exert different levels of effort to finish the allocated task based on different levels of payments, and different agents may react differently to the same level of payment (see \cite{liu2014crowdsourcing}). The chosen level of effort affects their performance dramatically.
Due to these issues, an appropriate incentive mechanism that exploits the agents' heterogeneity whilst incentivizes appropriate effort levels is crucial to a successful prediction system.

Besides eliciting high quality of information, the prediction performance also depends on the prediction behaviour of the surveyor (``the principal''). Without an appropriate prediction rule, the principal may not be able to effectively utilize the collected information and may obtain an inaccurate prediction results. This motivates us to study the incentive mechanism design together with the prediction rule optimization.

The resultant problem of joint design poses a significant challenge due to the following reasons.

First, due to the incorporation of the prediction problem,
the objective of incentive mechanism changes from eliciting agents' information truthfully to minimizing the prediction error. As the prediction error is  a result of the agents' information and actions, the designed mechanisms not only needs to motivate agents to report their truthful estimation information, but also needs to make sure that agents take appropriate actions. Hence, we cannot directly implement existing strictly proper scoring rules, \emph{e.g.}, the quadratic scoring rule (see \cite{brier1950verification,selten1998axiomatic}), that only promotes truthfulness among agents to address the joint problem.

Second, the designed mechanism needs to solve both moral hazard and adverse selection problems simultaneously. Moral hazard results from the inability of the principal to observe an agent's actions (\emph{i.e.}, effort exerted by the agent), while adverse selection corresponds to the inability of knowing an agent's private information (\emph{i.e.}, the expertise of an agent).
This is different from most incentive mechanisms designed in the existing works, which separate the mechanism design from the prediction problem and address either ``hidden actions'' (\emph{i.e.}, moral hazard)  (see \emph{e.g.}, \cite{fang2007putting,ioannidis2013linear}) or ``hidden information'' (\emph{i.e.}, adverse selection) (see \emph{e.g.}, \cite{frongillo2014elicitation,abernethy2015low}).

Nevertheless, we formulate and optimally solve a ``parametric'' form of this joint design problem.
More specifically, the principal desires to predict a parameter of a known distribution.
Each agent is modeled in a parametric fashion, with her expertise governed by a single parameter that is the agent's private information.
We assume that agents are heterogeneous as they have different levels of expertise.
While each agent aims to maximize her own expected payoff (\emph{i.e.}, the revenue minus the effort cost),
the principal optimizes a joint utility that trades off the prediction error and the monetary costs.
For ease of exposition, we refer to the principal as ``she'' and each agent as ``he''.

\subsection{Results and Contributions}
We focus on the interactions among a principal and multiple agents, and design an appropriate incentive mechanism to facilitate the parametric prediction process.
Specifically, we design a mechanism, which we call ``COPE'' (standing for ``COst and Prediction Elicitation''), that jointly optimizes the principal's payoff in terms of the payments made to the agents and the prediction error incurred.
COPE provides a systematic way for the principal to incentivize all participating agents to report their estimations truthfully and exert appropriate amounts of effort based on their respective capabilities.

We summarize our key results and the main contributions as follows.


\begin{itemize}
	\item
	\emph{Theoretic significance}:
	We relax several critical assumptions that are common in papers in the related literature, \emph{i.e.}, the costs incurred by the agents are all known to the principal, the agents do not incur costs for efforts, and the agents are all homogeneous. Hence, the proposed model pushes this line of theoretical focused research into more realistic settings.

	\item
	\emph{Optimal incentive mechanism design}:
	We study a generic incentive mechanism design problem situated in a prediction process. To study the optimal prediction solution, we design the COPE mechanism, which ensures that all participating agents report their estimations truthfully and exert appropriate amounts of effort based on their respective capabilities, meanwhile maximizes the {\pp}'s expected payoff.
	
	\item
	\emph{Observations and insights}:
	Our results show that, with Gaussian estimation noise,
	when the agent's marginal cost is independent of his amount of exerted effort, the principal should conduct a \emph{crowd-tender} mechanism, by soliciting service only from the agent with the lowest reported cost. On the other hand, when the marginal costs depends on the exerted effort,  the optimal mechanism is in the form of \emph{crowd-sourcing}, where the principal recruits multiple agents to complete the task.
	
	\item \emph{Numerical results}: Simulation in Section \ref{sec:simulation} show that COPE improves both the {\pp}'s profit and the network profit, comparing to those mechanisms that assume all agents have equal capabilities and incentivize agents exert the same amount of effort.	
	Moreover, the performance gap between COPE and the centralized benchmark solution is less than $3\%$ under the quadratic cost function and $10\%$ under the linear cost function.	
	
\end{itemize}

The rest of the paper is organized as follows.
After reviewing the literature in Section \ref{sec:related_work}, we describe the system model in Section \ref{sec:model_general}, and design the incentive mechanism in Section \ref{sec:normal_case}.
In Section \ref{sec:simulation}, we provide the simulations results. We conclude in Section \ref{sec:conclusion}.


\section{Related Work}\label{sec:related_work}
Mechanism design for truthful elicitation of agents' opinions is an extensively studied problem, most recently investigated are in the context of crowdsourcing (see, e.g., \cite{cavallo2012, miller2005eliciting, prelec2004bayesian, shah2015approval,shah2015double}).
In contrast to our work, this line of literature does not consider the prediction aspect, and only focuses on the elicitation problem alone.
Mechanism design for truthful elicitation of agents' opinions is also studied in the context of prediction markets (see, \emph{e.g.}, \cite{wolfers2004prediction,conitzer2009prediction}). These results, however, study the scenario where the agents take the responsibility of aggregating information. Our paper concerns a different setting and objective, in which the principal is in charge of information gathering and making the final prediction.
Hence, the mechanism we design should not only elicit the agents' information but also incentivize agents to exert appropriate effort.


The scenario becomes quite different when prediction must be done by taking incentives into account, and calls for the design of new procedures catering to both aspects. The recent studies~(see, \emph{e.g.}, \cite{fang2007putting, ioannidis2013linear, frongillo2014elicitation,cai2014optimum, abernethy2015low})
address problems in this space. However, the models assumed in these works are different, and generally more restrictive than the models considered in our paper in many respects. In particular,
\cite{fang2007putting} propose a betting mechanism for eliciting the observation and the quality of each agent, under the assumption that the agents are homogeneous with the same cost type. In contrast, we consider the more general and realistic setting where agents can have different types.
\cite{ioannidis2013linear} formulate the noise addition as a non-cooperative game and prove the existence and  uniqueness of the Nash equilibrium.
\cite{frongillo2014elicitation} study how a principal can make predictions by eliciting the agents' confidences, again without considering the costs that may be incurred by the agents.
\cite{abernethy2015low} consider a model where the agents cannot fabricate their observation, but may lie about their costs, and design a mechanism to ensure that agents truthfully report their costs. In contrast, we assume a more general scenario where the agents can be strategic in choosing and reporting their respective observations.
\cite{cai2014optimum} propose a monetary mechanism to collect data and to perform an estimation of a function at one random point. However, they assume that the agent always reports truthfully once he makes an observation. In contrast, our work considers strategic agents and proposes an optimal mechanism to ensure truth-telling by the agents.

\begin{figure*}
	\centering
	\includegraphics[width=\textwidth]{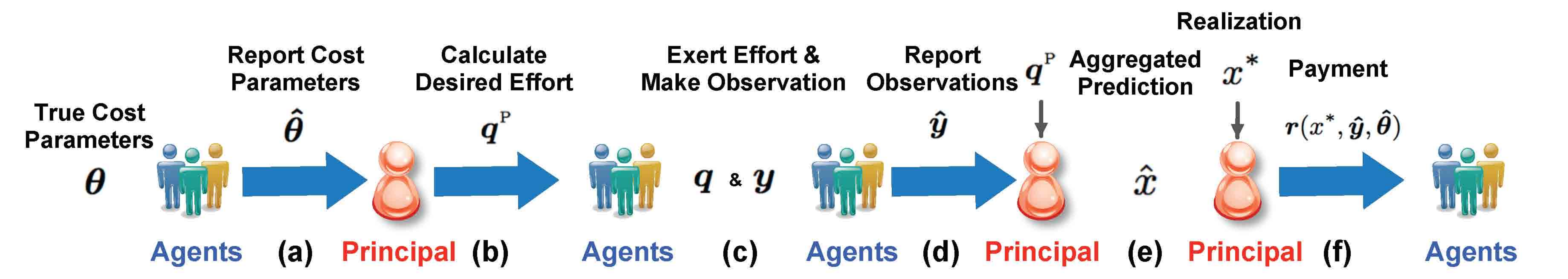}
	\caption{Sequence of interactions between the principal and the agents.}
	\label{fig:event_sequence}
	\vspace{-4mm}
\end{figure*}


\section{Problem Setting}\label{sec:model_general}
We begin with a formal description of the problem formulation.
Through this description, we will set up notations to capture the behavior of the agents, the objective of the principal, the prediction problem, and the mechanism-design problem.

\subsection{System Model}\label{sec:model}
We consider a setting where
the {\pp} wishes to make a \textit{parametric prediction}, that is, to form an informed estimate about a parameter $\x \in \Xset \subseteq \Rset$. Predicting the winner of an election and predicting box office results
for movies are two motivating examples. We assume that $\x$ has a prior distribution that is publicly known, for instance, from the results of an earlier election.
We assume that the {\pp} will come to know the precise value of $\x$ sometime in the future, for instance, upon completion of the election.
In order to make a prediction, the {\pp} queries a set $\Nset = \{ 1, \ldots, \N\}$ of $\N$ agents to report their observations.

Figure \ref{fig:event_sequence} pictorially depicts the interaction between agents and the {\pp}, including the agents' reporting strategy and the {\pp}'s prediction and payment decision. Before we explain each individual components of the figure, we first introduce some notations used to characterize the agents' strategies and types in Figure \ref{fig:event_sequence}.

\textbf{Effort Level and Cost Type}:
When queried by the {\pp}, an agent can put in some effort to form an ``observation" whose value is known only to that agent. We assume that the observation of any agent is noisy, and the distribution of the {\est} $\yn$ comes from a parameterised family of distributions $\Pf( \x, \qn )$, where $\qn$ represents the effort exerted by agent $\n$ to make {\est}. The higher value of $\qn$, the more effort {\ag} $\n$ exerts, and thus the better quality of {\ag} $\n$'s {\est}. An example that we focus on subsequently in the paper is additive Gaussian noise (see, \emph{e.g.}, \cite{fang2007putting,cai2014optimum}), \emph{i.e.}, $\yn \sim \gaussian{ \x }{ \frac{1}{\qn} }$,
where $\yn$ follows from Gaussian distribution with mean $\x$ and variance $1/\qn$.
Conditioned on $\x$, the observations of the agents are assumed to be mutually independent.
As a shorthand, we let $\by = (\yn, \forall \n \in \Nset)$ be the observation vector and $\bq = (\qn, \forall \n \in \Nset)$ be the effort vector.
We further assume that  agents do not collude with each other, as each agent submits his observation to the crowdsourcing platform anonymous and does not know the identities of others.



It is costly for each agent to exert a high level effort when making an observation, and this cost not only depends on the effort chosen by the agent, but also is affected by the agent's cost-type parameter $\thn$.
The cost types of different agents are allowed to be different, capturing the heterogeneity of the agents.
A smaller value of $\thn$ implies a higher capability of {\ag} $\n$.
Specifically, we consider a publicly known cost function $\C: \Rsetp \times \Rsetp \rightarrow \Rsetp$, and assume that the cost incurred by an agent $\n \in \Nset$ with the cost type $\thn$ when exerting an effort $\qn$ is $\C(\qn, \thn)$.
We will study two types of cost functions, \emph{i.e.}, the linear and quadratic cost functions in Section \ref{sec:special_case}, and generalize the results to general cost functions in Section \ref{sec:generalization}.

The cost types $\{\thn\}_{\n \in \Nset}$ are assumed to be randomly, independently and identically distributed on support $[\thl, \thu]$ for some $0 \leq \thl < \thu < \infty$. This distribution is assumed to be public knowledge. In this paper we focus on the case where the distribution is uniform on the interval $[ \thl, \thu ]$.
Uniform assumption has been frequently used in the past incentive mechanism design (\emph{e.g.,} \cite{fang2007putting,sheng2013profit,koutsopoulos2013optimal}), and our analysis also holds for the general class of log-concave distributions such as exponential distribution and normal distribution.

Next, we will discuss each individual components of Figure \ref{fig:event_sequence} sequentially.




\textbf{Reporting Observations and Making Payments}:
The {\pp} employs monetary incentives to ensure that agents make their observations and report them to the principal.
In order to incentivize agents to participate the prediction task, the payment to an agent must, at the least, cover the cost incurred by that agent in putting effort to make the observation. 
However, since each agent's cost parameter is known only to that agent, the principal needs to ask each agent to report his own cost type (Figure~\ref{fig:event_sequence}a).
The agents are strategic, and any agent $\n\in\Nset$  may report a cost type $\thrn$ that is different from his true cost type $\thn$.
Let $\bthr = (\thrn, \n \in \Nset)$ be the reporting cost type vector.

As we will see subsequently, incentivizing different agents to put different levels of effort depending on their respective cost types allows for a significantly better prediction performance. The principal must incentivize these different effort levels, and the choice of these effort levels is based on the agents' reported cost types $\bthr$ (Figure~\ref{fig:event_sequence}b).
Let function $\QP: [\thl, \thu]^{\N} \rightarrow \Rsetp$ denote the effort that the principal requires an agent to exert.
The function $\QP$ depends on the cost types reported by the agents:  $\QP(\thrn, \bthr_{-\n})$ represents the effort required from agent $\n \in \Nset$, where $\bthr_{-\n} = [ \thr_1, \ldots, \thr_{\n-1}, \thr_{\n+1}, \ldots, \thr_{\N} ]^{T}$ is the reported cost parameters of all {\ags} except {\ag} $\n$. Here (and elsewhere in the paper), we use the superscript ``P'' to represent the {\pp}.
We let $\BQP(\bthr) = ( \QP(\thrn, \thr_{-\n}), \forall \n \in \Nset )$ be the effort vector required by the {\pp}.


Each {\ag} $\n \in \Nset$ is strategic and may exert an effort $\qn \neq \QP(\thrn, \bthr_{-\n})$ to suit his own interests. When choosing the effort to exert, the agent may also exploit the fact that the principal cannot directly observe the actual effort exerted. Upon exerting the chosen effort $\qn$, the agent obtains an observation $\yn$ (Figure~\ref{fig:event_sequence}c). The principal seeks to know the value of the observation  $\yn$, but the agent may report a strategically chosen value $\yrn \neq \yn$ to the principal (Figure~\ref{fig:event_sequence}d) that suits the agent's own interests. We adopt the shorthand $\byr = ( \yrn, \n \in \Nset)$ as the reporting vector.
Based on the information obtained, the principal must make a prediction for the value of $\x$ (Figure~\ref{fig:event_sequence}e).


The {\pp} makes payment to each agent once she observes the true value of $\x$. Specifically, we define the payment function as
$\P: \Rset \times \Rset \times [\thl, \thu]^{\N} \times  \rightarrow \Rsetp$; the payment to agent $\n$ is $\P (  \x, \yrn, \thrn, \bthr_{-\n})$, which depends on the value of $\x$, the agent $\n$'s
reported {\est} $\yrn$, and all agents' reported cost parameters $\bthr$.

As indicated above, the model considered is a one-shot model, \emph{i.e.}, once the agents simultaneously report their observations, the principal determines the payments based on the agents' reported cost parameters and observations only, with no further adjustments on reported values or the payments.

\textbf{Payoff of Agent}:
Given the payment function announced by the {\pp},
each $\ag$ $\n$'s
payoff $\UA: \Rset \times [\thl, \thu] \times \Rsetp \times \Rset \times [\thl, \thu]^{\N} \rightarrow \Rset$ is defined as the difference between the payment received from the {\pp} and the cost incurred in making the observation, and is given as
\begin{equation}\label{eq:ag_payoff_contract}
\begin{aligned}
\UA( \x, \thrn, \qn, \yrn, \thn,  \bthr_{-\n}) = & \P (  \x, \yrn, \thrn, \bthr_{-\n} ) - \C\big(  \qn, \thn \big),
\end{aligned}
\end{equation}
Here the superscript ``A'' indicates a term associated to the agents.
Equation \eqref{eq:ag_payoff_contract} shows
that {\ag} $\n \in \Nset$'s payoff also depends on other {\ags}' reported cost parameters $\bthr_{-\n}$.
When each {\ag} $\n \in \Nset$ chooses his strategy, \emph{i.e.}, his cost reported value $\thrn$, exerted effort $\qn$, and the reported observation $\yrn$, to maximize his expected payoff.
The expected payoff of the agent $\n$ is calculated as
\begin{equation}\label{eq:ag_payoff_contract_expect}
\begin{aligned}
\textstyle
\EX [ \UA(  \x, \thrn, \qn, \yrn, \thn,  \bthr_{-\n}) ]
& =  \EX [ \P (  \x, \yrn, \thrn, \bthr_{-\n} )]
- C(  \qn, \thn ),
\end{aligned}
\end{equation}
where the expectation is taken with respect to the distributions of $\x$ and all agents' cost parameters $\bth_{-\n}$. Recall that each {\ag} $\n$ only knows his own cost parameter $\thn$, and only has distributional information about other agents' cost parameters.


\textbf{Payoff of the Principal}:
Let function $\xe: \Rset \times \Rsetp \rightarrow \Rset$ characterize the prediction made by the principal based on the agents' reported observations $\byr$ and the effort assumed to be exerted by the agents $\BQP(\bthr)$.
Due to the inability of the principal to observe the agents' exerted effort, the principal makes prediction based on her own knowledge (\emph{i.e.,} the agents reported observation $\byr$ and the effort $\BQP(\bthr)$ required from agents). Later in Section \ref{sec:normal_case}, we will show that the agents' true effort level are the same as that required by the principal by employing our proposed mechanism.
Let $\lp: \Rset \times \Rset \rightarrow \Rsetp$ be the loss function that characterizes the penalty term for mistakes in the {\pre}.
For instance, one could consider the squared loss
$\lp(\x,\xe(\byr, \bqp)) = ( \x-\xe(\byr, \bqp ) )^2$
as the penalty for the principal, where $\qapn = \QAP(\thrn, \bthr_{-\n})$ and $\bqp = (\qapn, \forall \n \in \Nset )$. We use $\qap$ as the shorthand notation for $\QAP(\thrn, \bthr_{-\n})$, in order to show that $\qpn$, for any $\n \in \Nset$, is a decision variable for the principal. In Section \ref{sec:assumption_general_cases}, we will show how the principal determines the desired effort from the agents by taking the first order derivative of her Bayes risk with respect to $\qpn$.

We measure the utility gained by the principal through the prediction in terms of the \emph{Bayes risk} incurred under this loss function.
The reason that we use Bayes risk  is that it yields a real number (not a function of $\xe$ or $\byr$) for each prediction, and the principal's posterior expected prediction loss is equivalent to the Bayes risk (see \cite{robert2007bayesian,berger2013statistical}).
If all {\ags} report their true {\ests} (\emph{i.e.}, $\byr = \by$) and
exert efforts as desired by the principal (\emph{i.e.},  $\bq = \bqp = \BQP(\bthr) $), then the \emph{Bayes risk} $\hp: \Rset \times \Rset  \rightarrow \Rsetp$ is (see \cite{robert2007bayesian, berger2013statistical})
\vspace{-2mm}
\begin{equation}\label{eq:utility_pp_new}
\hp (  \bqp ) =  \inf_{\xe} \EX [  \lp (  \x,\xe(  \by, \bqp )  ) ],
\vspace{-3mm}
\end{equation}
where the expectation is taken with respect to $\x$ and $\by$. 
By assuming that the principal's utility with perfect prediction is zero, the Bayes risk chacterzies the penality for the principal's mistakes in the prediction, and the prinicial's expected utility is just $-  \hp (  \bqp )$. Correspondingly,
the payoff of the principal $\UP: \Rset \times \Rset^{\N}\times [\thl, \thu]^{\N}  \rightarrow \Rset$ is then defined as the difference between her utility obtained from prediction and the monetary payments to all {\ags}:
\begin{equation}\label{eq:payoff_pp}
\vspace{-3mm}
\UP(\x, \bqp, \by, \bthr) = -  \hp (  \bqp ) -  \sum_{\n\in\Nset}  \P (  \x, \yn, \thrn, \bthr_{-\n} ).
\end{equation}
Here, we assumed without loss of generality that the monetary payment and the prediction loss is normalized to be on the same scale. 

\begin{table*}[t]
	\centering
	\caption{Notations}
	\label{table:notaion}
	\footnotesize
	\begin{tabular}{|c|p{10cm}|}
		\hline
		\centering $\x$
		& A parameter needs to be estimate
		\\
		\hline
		\centering $\xe$
		& Prediction based by the {\pp}
		\\
		\hline
		\centering $\qn$
		& Effort exerted by agent $\n \in \Nset$
		\\
		\hline
		\centering $\qpn$
		& Effort desired by the principal from agent $\n \in \Nset$, calculated by the function $\QP:[\thl, \thu]^{\N} \rightarrow \Rsetp$
		\\
		\hline
		\centering $\yn$
		& True observation of agent $\n \in \Nset$
		\\
		\hline
		\centering $\yrn$
		& Reported observation of agent $\n \in \Nset$
		\\
		\hline
		\centering $\thn$
		& True cost type of agent $\n \in \Nset$
		\\
		\hline
		\centering $\thrn$
		& Reported cost type of agent $\n \in \Nset$
		\\
		\hline
		\centering  $\C(\qn, \thn)$
		& Cost incurred by an agent $\n \in \Nset$ with the cost type $\thn$ when exerting an effort $\qn$
		\\
		\hline
		\centering $\QP(\thrn, \bthr_{-\n})$
		& Effort desired by the principal from agent $\n \in \Nset$ given all agents reported cost type
		\\
		\hline
		\centering $\hp (  \bqp )$
		& Bayes risk of the principal's prediction
		\\
		\hline
	\end{tabular}\label{tab:notations}
	\vspace{-4mm}
\end{table*}

For clarity, we list the key notations in Table \ref{tab:notations}.

\subsection{Design Objective}\label{sec:requirement}
Before explaining the design objective,  we begin by defining two standard game-theoretic terms that are required for subsequent discussions.

\begin{definition}\label{def:BIC}
	(BIC: Bayesian Incentive Compatibility)
	A mechanism satisfies the Bayesian incentive compatibility (BIC) if for every {\ag} $\n \in \Nset$, his expected payoff
	satisfies (see \cite{fudenberg1991game,myerson1979incentive})
	\vspace{-4mm}
	\begin{align}\label{eq:IC_requirement}
	&\!\!\!\! \EX \big[ \UA( \x, \thn, \QP(\thn, \bth_{-\n}), \yn, \thn,  \bth_{-\n} ) \big] \! \geq \! \EX \big[\UA( \x, \thrn,\qn , \yrn, \thn,  \bth_{-\n} )\big] ,~~ \nonumber\\
	&\qquad{} \qquad{} \qquad{} \qquad{}\qquad{}\qquad{}\qquad{}\qquad{}\qquad{}\forall (\thrn, \qn  ,\yrn)\! \neq\! (\thn, \QP(\thn, \bth_{-\n}),\yn),
	\end{align}
	\vspace{-4mm}
	where the expectation is taken with respect to $\x$ and all other {\ags} cost parameters $\bth_{-\n}$.
\end{definition}

BIC means that for any agent $\n$, reporting the true cost parameter, exerting the effort requested by the {\pp}, and  reporting true observation will maximize his expected payoff, given common knowledge about the distribution on agents cost parameters and when other agents are truthfully report their cost parameters.

\begin{definition}\label{def:BIR}
	(BIR: Bayesian Individual Rationality)
	~{}
	A mechanism satisfies the Bayesian incentive rationality (BIR), if the expected payoff of every agent $\n \in \Nset$ is non-negative, given that he reports truthfully, exerts effort as the principal desires, and assumes that all other agents report their cost parameters truthfully, that is (see \cite{fudenberg1991game,myerson1979incentive}),
	\begin{equation}
	\label{eq:IR_requirement}
	\EX \big[ \UA( \x, \thn, \QP(\thn, \bth_{-\n}), \yn, \thn,  \bth_{-\n}  ) \big] \geq 0, \quad \forall \n \in \Nset,
	\end{equation}
	where the expectation is taken with respect to $\x$ and all other {\ags}' cost parameters $\bth_{-\n}$.
\end{definition}

Assuming (without loss of generality) that the payoff of an agent not participating in this process equals zero, BIR means that an agent will participate only if his expected payoff is at least as much as that of a non-participating agent.

Based on the revelation principal (see \cite{myerson1979incentive}), the problem of finding a mechanism that maximizes the {\pp}'s expected payoff can be restricted to the set of mechanisms where agents are willing to reveal their private information to the {\pp}. Moreover, the {\pp} cannot force agents to accept the task.
Hence, the problem that we want to solve is formalized as follows. The goal is to design a mechanism, say $\AMset$, that maximizes the principal's expected while ensuring truthful reports from the agents:
\begin{equation}\label{eq:optimal_mechanism}
\begin{aligned}
\sup_{ \AMset  } ~& \EX \big [ \UP(\x, \by, \bth)\big]\\
\mbox{subject~to:~~} & \mathrm{BIC~and~BIR~in~(\ref{eq:IC_requirement})~and~(\ref{eq:IR_requirement}),}
\end{aligned}
\end{equation}
where the expectation is taken with respect to $\x$, $\by$ and $\bth$, and the BIR condition makes sure that the agents are willing to participate in the game.
In words, the goal is to design a mechanism such that:
(i) the {\pp}'s  payoff is maximized in expectation;
(ii) the {\pp} can elicit truthful information from all agents;
and (iii) the {\pp} can incentivise suitable effort  from the agents based on their respective cost parameters.

%


\section{The COPE Mechanism}\label{sec:normal_case}
In this section, we present our mechanism ``COPE'' (COst and Prediction Elicitation) that provides an optimal solution to the problem~\eqref{eq:optimal_mechanism} of parametric prediction from parametric agents. We will first consider two specific settings in order to illustrate the key ideas behind COPE, and to obtain some concrete engineering insights. We will then proceed to present COPE in full generality.

\subsection{Two Example Settings}\label{sec:special_case}
We consider the following specific setting in this section. We consider the Gaussian case, where we assume the prior $\x \sim \gaussian{\mu_0}{\sigma_0^2}$, and the observation of every agent $n$ follows the distribution $y_n \sim \gaussian{\x}{\frac{1}{q_n}}$, independent of all other events. The values of $\mu_0$ and $\sigma_0$ are assumed to be public knowledge.
We assume $\thn \sim \mbox{Uniform}[\thl, \thu]$, independent for every $\n \in \Nset$. We consider the squared $\ell_2$-loss to measure the prediction error, namely, $\lp(\x, \xe) = (\x - \xe)^2$. In what follows, we consider two cost functions: (i) the linear cost function $\CA(\q, \th)= \q\th$, and (ii) the quadratic cost function $\CB(\q, \th) = \frac{1}{2} \th\q^{2}$. 



\subsubsection{Linear Cost Function $\CA(\q, \th) = \q \th$}\label{sec:cost_exampe1}
We first consider
the linear cost function $\CA(\q, \th) = \q \th$ and discuss the corresponding COPE mechanism. 
Algorithm~\ref{algo_interaction_general} presents the higher-level structure of the mechanism which corresponds to the steps in Figure 1. The optimality  of the mechanism crucially relies on the careful construction of specific functions referred to in the algorithm, and these constructions are described below.

\begin{algorithm}[h]
	\small
	{Step 1: The {\pp} announces a payment function $\PA$. }\\
	~{Step 2: Every {\ag} $\n\in \Nset$ independently reports a cost type $\thrn \in [\thl, \thu]$ to the principal.}\\
	~{Step 3: The {\pp} sends each {\ag} $\n\in \Nset$ a contract, which specifies the effort level
		$\QP(\thrn, \bthr_{-\n})$ along with values of functions $\pi(\thrn, \bthr_{-\n})$, $\Sa(\thrn, \bthr_{-\n})$, and $\Sb(\thrn, \bthr_{-\n})$ that comprise the function $\PA$.}\\
	~{Step 4: Each agent $\n \in \Nset$ exerts effort $\qn$ and makes an observation $\yn$.}\\
	~{Step 5: Each {\ag} $\n \in \Nset$ reports an estimate $\yrn$.}\\
	~{Step 6: The {\pp} makes {\pre} $\xe$.}\\
	~{Step 7: The true value $\x$ is realized.}\\
	~{Step 8: The {\pp} makes the payment $\PA(\x, \yrn, \thrn, \bthr_{-\n} ) $ to every {\ag} $\n \in \Nset$.}
	\caption{COPE}
	\label{algo_interaction_general}
\end{algorithm}
\vspace{-3mm}

Recall that the function $\QAP: [\thl, \thu] \times [\thl, \thu]^{\N-1}  \rightarrow \Rsetp$ specifies the effort that the {\pp} requires an {\ag} to exert, based on the cost parameters reported by all agents.  In Theorem \ref{them:optimal_cost_I} subsequently, we show that when the cost function is linear, the principal requires \emph{only one} agent to exert effort.
If there are multiple agents achieving the same minimum value of $\thrn$, the {\pp} would randomly choose one agent to exert effort.
This property is reflected in the following choice of function $\QAP$:
\begin{equation}\label{eq:effort_cost_I}
\textstyle
\QAP( \thrn, \bthr_{-\n} ) = \left\{
\begin{array}{l l}
\max\{ (2\thrn-\thl)^{-\frac{1}{2}} - \sigma_0^{-2} , 0\}& \quad \text{if}~\n = \arg\min_{m \in \Nset} \thr_{m}\\
0 & \quad \text{otherwise}.
\end{array} \right.
\end{equation}
The function $\QAP$ is designed to strike an optimal balance between the prediction error and the monetary expenditure, accommodating the fact that the agents are heterogeneous.
We define $\no = \argmin_{m \in \Nset} \thr_{m}$, that is, $\no$ is the agent with the lowest reported cost parameter.

We now characterize the function $\PA$ that governs the payment made by the principal to the agents. The payments to all agents other than agent $\no$ are zero, since these agents are not involved in the observation and prediction procedure. The payment made to agent $\no$ is
\begin{equation}
\label{eq:payment_costI}
\PA (  \x, \yr_{\no}, \thr_{\no}, \bthr_{-\no}) =
\pia(\thr_{\no}, \bthr_{-\no})   - ( \x - \yr_{\no} )^2 \cdot \Saa(  \thr_{\no}, \bthr_{-\no} )\\
+ \Sba(\thr_{\no}, \bthr_{-\no}),
\end{equation}
where
\begin{align}
\pia(\thr_{\no}, \bthr_{-\no}) &= \pia(\thr_{\no}) =   \thr_{\no}(2\thr_{\no}-\thl)^{-\frac{1}{2}} - \thu \sigma_0^{-2} + 2 [  (2\thu - \thl )^{\frac{1}{2}} -  (2\thrn - \thl)^{\frac{1}{2}} ]
,
\nonumber \\
\Saa(\thr_{\no}, \bthr_{-\no}) &=\Saa(\thr_{\no}) = \thr_{\no} (2\thr_{\no}-\thl)^{-1}, \text{~and~}\Sba(\thr_{\no}, \bthr_{-\no}) = \Sba(\thr_{\no})= \thr_{\no} (2\thr_{\no}-\thl)^{-\frac{1}{2}}. 
\end{align}

Let us explain the main ideas behind the above choices. The detailed proof can be found in Appendix.
First, The term $( \x - \yr_{\no} )^2$ in~\eqref{eq:payment_costI} ensures that the agent reports his observation truthfully. This is because no other terms in~\eqref{eq:payment_costI} depend on $\yr_{\no}$, and the value of calculated by $\Saa(\thr_{\no})$ is always positive.
Hence, when the agent chooses the reporting observation strategy to maximize his expected payoff, he focuses on minimizing the term $\EX_{\x}[ ( \x - \yr_{\no} )^2 ]$ whose value is minimum only when the agent reports his truthful observation, \emph{i.e.}, $\yr_{\no} = \y_{\no}$.
Second, we can verify that the expected payoff of the agent is maximized only when the agent truthfully reports his cost parameter, given the agent reports his true observation.
Third, the choices of functions $\Saa$ and $\Sba$ ensure that the agent exerts an effort as desired by the principal. As the term $( \x - \yr_{\no} )^2$ makes the agent reports his true observation, we can verify that the expected payoff of the agent is maximized only when the agent chooses $\q_{\no} = \QP(\th_{\no}, \bth_{-\no})$.
Finally, the function $\pia$ is designed to ensure that the {\pp}'s expected payoff defined in~\eqref{eq:payoff_pp} is maximized while ensuring BIC and BIR condition is satisfied. As we shown that the term $( \x - \yr_{\no} )^2$ and the choices of functions $\Saa$ and $\Sba$ guarantee the truthful behavior of the agent, \emph{i.e.}, $\yr_{\no} = \yrn$ and $\q_{\no} = \QP(\th_{\no}, \bth_{-\no})$,
the expected value of $\PA$ simply equals to the value of the function $\pia$. In such case, we can focus on deriving the function of $\pia$ to maximize the expected payoff the {\pp}.

Then we characterize the prediction decision made by the {\pp}.
After collecting all agents reported observations, the {\pp} makes the prediction $\xe$ based on the following equation:
\begin{align}\label{eq:prediction_pp_linear}
\xe(\yr_{\no}, \qapno ) = \frac{  {\mu_0} \cdot  { 1/{\var_0}} +  { \qapno} \cdot {g(\yr_{\no})} }{ { 1/{\var_0}} +   \qapno },
\end{align}
where $\qapno = \QAP(\thr_{\no}, \bthr_{-\no})$, and function $g: \Rset \rightarrow \Rset$ is defined as
$g(\yr_{\no}) = \yr_{\no} + \frac{ ( \yr_{\no} - \mu_0 )}{ ( \qapno \cdot \vars )  }$.

Here, the predictor $\xea$ employed by the principal is the standard Bayes estimator  operating on the agents' responses and this predictor does not affect the payoff of the agents.

We prove that the proposed COPE mechanism is optimal.

\begin{theorem}\label{them:optimal_cost_I}
	Under the linear cost function $\CA(\q,\th) = \q \th$, COPE satisfies the BIC and BIR condition defined in \eqref{eq:IC_requirement} and \eqref{eq:IR_requirement} and maximizes the {\pp}'s expected payoff defined in \eqref{eq:optimal_mechanism}.
	
\end{theorem}

We provide the detailed proof in Appendix \ref{them:optimal_cost_I_proof}.
An important consequence of the theorem is that the optimal mechanism in the case of linear costs awards the task to the single agent with the lowest bid. This corresponds to  what we call a ``crowd-tender'' system where all agents submit their cost parameters, and the lowest bidder is awarded the task. We will discuss the intuition  behind such a result in more details in Section \ref{sec:insights}.

\subsubsection{Quadratic Cost Function $\CB(\q, \th) = \frac{1}{2}{\th \q^2}$}\label{sec:example_costII}
We now consider a quadratic cost function $\CB(\q, \th) = \frac{1}{2}{\th \q^2}$ and present COPE for this setting.
The higher level structure of COPE is again given by Algorithm~\ref{algo_interaction_general}, and the specific functions referred to in the algorithm is provided below.

Under COPE, the function $\QBP: [\thl, \thu] \times [\thl, \thu]^{\N-1}  \rightarrow \Rsetp$ that governs the effort that the {\pp} requires an {\ag} to exert is given as
\begin{align}\label{eq:effort_cost_II}
\QBP (\thrn, \bthr_{-\n}) = (2\thrn - \thl)^{-1} (W(\bthr) )^{-2},
\end{align}
where $W$ is the solution of the equation $[ W(\bth) ]^3 - \frac{1}{\vars} [ W(\bth) ]^2 = \sum_{m \in \Nset} \frac{ 1 }{ 2\th_m - \thl}$.
An explicit (although cumbersome) solution of $W$ is provided in \eqref{eq:aggregate_effort_type_II_proof_xx} of Appendix~\ref{them:optimal_cost_II_proof}.

As in the case of linear costs, the function $\QBP$ is designed to optimally harness the heterogenity of the agents in order to minimize the prediction error with a small enough payment. Note that in contrast to the linear case~\eqref{eq:effort_cost_I}, here the {\pp} requires every agent to exert a positive effort.

We define the function $\PB$ that governs the payment to any agent $\n$ as
\begin{equation}
\label{eq:reward_fun_special_II}
\begin{aligned}
&\PB(\x, \yrn, \thrn, \bthr_{-\n})  = \pib(\thrn,\bthr_{-\n})   -  ( \x - \yrn )^2 \cdot \Sab(  \thrn, \bthr_{-\n} )  + \Sbb(\thrn, \bthr_{-\n}),
\end{aligned}
\end{equation}
where 
\vspace{-.3cm}
\begin{align}
\pib(\thrn,\bthr_{-\n}) =  & \frac{1}{2} \big( \thrn \cdot \big[ \QBP (\thrn, \bthr_{-\n}) \big]^2 + \int_{\thrn}^{\thu}   \big[ \QBP (z, \bthr_{-\n}) \big]^2  \dt{z}  \big), \nonumber\\
\Sab(\thrn, \bthr_{-\n}) = &\big[\QBP(\thrn, \bthr_{-\n}) + 1/{\var_0} \big]^2  \thrn \cdot \QBP(\thrn, \bthr_{-\n}), \nonumber\\
\Sbb(\thrn, \bthr_{-\n}) = &\big[\QBP(\thrn, \bthr_{-\n}) + 1/{\var_0} \big]  \thrn \cdot \QBP(\thrn, \bthr_{-\n}).
\end{align}
These functions have a form similar to those in the case of linear costs~\eqref{eq:payment_costI}, except that these functions depend on the reported cost parameters of all $\N$ agents, whereas the corresponding functions in the linear cost setting depended only on the reported cost parameter of one agent. The remaining higher level intuition behind this construction is similar to that behind the linear-cost case described in the previous section.

The principal uses the Bayes estimate as her predictor:
\begin{equation}\label{eq:prediction_principal_quadratic}
\xeb(\byr, \bqbp ) = \frac{ (1-N)\mu_0/\vars + \sum_{\n \in \Nset}{ ( 1/\vars + \qbpn )} \cdot {\yrn}  }{ { 1/{\var_0}} +  \sum_{\n \in \Nset} { \qbpn } },
\end{equation}
where $\qapn = \QAP(\thrn, \bthr_{-\n})$.


The following theorem establishes the optimality guarantee of COPE under quadratic costs.
\begin{theorem}\label{them:optimal_cost_II}
	Under the quadratic cost function $\CB(\q,\th) = \frac{1}{2}\th \q^2$, COPE satisfies the BIC and BIR condition defined in \eqref{eq:IC_requirement} and \eqref{eq:IR_requirement} and maximizes the {\pp}'s expected payoff defined in \eqref{eq:optimal_mechanism}.
	
\end{theorem}

The detailed proof is provided in Appendix \ref{them:optimal_cost_II_proof}.
As COPE under the quadratic cost function requires all agents to exert certain effort and to report their observation, this corresponds to a ``crowd-sourcing'' system.

\subsubsection{Engineering Takeaways}\label{sec:insights}
Our results show that interestingly, it is optimal for the {\pp} to call for a \emph{crowd-tender} when the cost function is linear, while it is optimal to design a \emph{crowd-sourcing} mechanism when the cost function is quadratic.
Informally, the cost function acts as a regularizer on the choice of effort levels $\qp$, and the dichotomy of these two cost functions is related to the sparsity inducing properties of the $\ell_1$-regularizer, and the lack thereof of the (squared) $\ell_2$-regularizer.



\subsection{General Setting}\label{sec:generalization}
In this section, we will present COPE under more general forms of the cost function, the noise distribution, the prior distribution, and the prediction loss function. Under these general conditions, the structure of the mechanism remains identical to Algorithm~\ref{algo_interaction_general}. We will show that COPE is optimal and feasible under certain regularity conditions.



\subsubsection{Assumptions}\label{sec:assumption_general_cases}
~

\textbf{Cost Function}
We first define the general cost function of the agent $\n\in\Nset$. Specifically, for {\ag} $\n \in \Nset$, his cost function $\C: \Rsetp \times \Rsetp \rightarrow \Rsetp$ is $\C(  \qn, \thn  ) = \int_0^{\qn} \c( z, \thn ) \dt{z}$,
where $\c( z, \thn )$ is the the marginal cost function.
We assume that the marginal cost function $\c: \Rsetp \times \Rsetp \rightarrow \Rsetp$ satisfies:
\begin{equation}\label{cost_function_prop}
\begin{aligned}
\textstyle
&\frac{  \partial{ \c(  \q, \th  ) } }{  \partial{  \q  }   } > 0,
\frac{  \partial{ \c(  \q, \th  ) } }{  \partial{  \th  }   } > 0,
\frac{  \partial^2{ \c(  \q, \th  ) } }{  \partial{  \th^2  }   } \geqslant 0,
\frac{  \partial^2{ \c(  \q, \th  ) } }{  \partial{  \q }\partial{\th}   } \geqslant 0.
\end{aligned}
\end{equation}
where the first inequality shows a nondecreasing marginal cost in agent's effort level,
the second and third inequalities show that the marginal cost is monotonically increasing and convex in the cost parameter $\thn$, {the last inequality implies that the marginal cost with respect to cost parameter $\thn$ is increasing in effort $\q$}.
These assumptions are widely used to model the cost function (see, \emph{e.g.}, \cite{che1993design,chen2008sourcing} and references therein).


The cost types $\{\thn\}_{\n \in \Nset}$ are assumed to be randomly, independently and identically distributed with a cumulative distribution function $\F: [ \thl, \thu ] \rightarrow \Rsetp$ and probability density function $\f: [\thl,\thu] \rightarrow \Rsetp$.
The functions $\F$ and $\f$ are public knowledge to all agents.
We also assume that the c.d.f. function $\F$ is continuous, differentiable, and log concave in $[\thl, \thu ]$.
This is a regularity condition often assumed in auction contexts (see, \emph{e.g.}, \cite{myerson1981optimal}). This assumption is satisfied by a wide range of distributions, such as the uniform, gamma, and beta distributions. See \cite{rosling2002inventory} for an extensive discussion on log concave probability distributions.

\textbf{Observation and Loss Function}
We assume that the distribution of the {\est} $\yn$ of any agent $\n\in\Nset$ comes from a parameterized family of distributions $\Pf( \x, \qn )$, where $\qn$ represents the effort exerted by agent $\n$ to make {\est}.
A typical parameterized distribution, for example, is the Gaussian distribution with mean $\x$ and variance $1/\qn$.

Let $\xe: \Rset^{\N} \times \Rsetp^{\N} \rightarrow \Rset^{\N}$ be the prediction function that characterizes the prediction made by the {\pp}, and $\lp: \Rset \times \Rset \rightarrow \Rsetp$ be the loss function that characterizes the penalty term for mistakes in the principal's prediction.
A typical loss function, for example, is the squared loss $\lp(\x,\xe(  \by, \bqp )) = [ \x-\xe(  \by, \bqp ) ]^2$, where $\qapn = \QAP(\thrn, \bthr_{-\n})$ and $\bqp = (\qapn, \forall \n \in \Nset )$.
We measure the utility gained by the principal through the prediction in terms of the \emph{Bayes risk}. Here, the Bayes risk is calculated under the loss function $\lp$.
Specifically, if all {\ags} report their true {\ests} (\emph{i.e.}, $\byr = \by$) and
exert efforts as desired by the principal (\emph{i.e.},  $\bq = \bqp = \BQP(\bthr) $), then the \emph{Bayes risk} $\hp: \Rset \times \Rset  \rightarrow \Rsetp$ is
\begin{equation}\label{eq:utility_pp_new_general}
\vspace{-4mm}
\hp (  \bqp ) =  \inf_{\xe} \EX [  \lp (  \x,\xe(  \by, \bqp )  ) ],
\end{equation}
where the expectation is taken with respect to $\x$ and $\by$.  We assume that such a Bayes estimator minimizing the Bayes risk exists, and that the Bayes risk is finite.\footnote{{We want to emphasize that whether the Bayes risk exists is a open problem. \cite{scharf1991statistical,figueiredo2004lecture} consider some special case such as Gaussian distribution. Characterizing the general condition for the existence and uniqueness of Bayes risk will be interesting future work.}}

We will also assume the existence of a function $\la: \Rset \times \Rset \rightarrow \Rsetp$ using which the principal may measure the accuracy of an agent's report. Specifically, we assume that if $\yn$ is generated according to agent $\n$'s observation distribution, then we assume that $\la$ satisfies
\vspace{-4mm}
\begin{align}\label{Bayes_risk_assumtion_general}
\yn \in \arg \inf_{\yn^{\dag}} \EX [  \la (  \x, \yn^{\dag}(\yn)  ) ],
\end{align}
where function $\yn^{\dag}: \Rset \rightarrow \Rset$ characterizes the reporting strategy of the agent $\n \in \Nset$ given his true observation is $\yn$, the infimum is over all measurable functions of the observation $\yn$, and the assumption says that the identity function is a minimizer of the expected value of $\la$ when its first argument is $\x$.
For instance, we considered $\la(\x,\yrn) = (\x-\yrn)^2$ earlier in the two motivating examples involving the Gaussian distribution.

We let $\ha: \Rset \times \Rset  \rightarrow \Rsetp$ denote the associated Bayes risk:
\begin{equation}\label{eq:utility_ag_new_general}
\vspace{-4mm}
\ha (  \qn ) =  \inf_{\yn} \EX [  \la (  \x,\yn) ],
\end{equation}
where the expectation is taken with respect to $\x$ and $\yn$, and the distribution of $\yn$ depends on the agent's exerted effort $\qn$.
The Bayes risk of the principal, \emph{i.e.}, $\hp (  \bqp )$, characterizes the principle's expected payoff loss due to the difference between the true value $\x$ and her own prediction $\xe$; while the Bayes risk of the agent, \emph{i.e.}, $\ha (\qn)$, characterizes the agent $\n \in \Nset$'s expected payoff loss due to the difference between the true value $\xe$ and his reporting prediction $\yrn$.

We assume that the Bayes risk of the principal and the agent satisfy the following monotonicity conditions:
\begin{equation}\label{bayes_risk_general}
\begin{aligned}
\textstyle
&\frac{  \partial{ \hp(  \bqp  ) } }{  \partial{  \qpn  }   } \leqslant 0,
\frac{  \dt{ \ha(  \qn  ) } }{  \dt{  \qn  }   }  \leqslant 0,
\frac{  \partial^2{ \hp(  \bqp ) } }{  \partial{  {\qpn}^2 } } \geqslant 0,
\frac{  \dt^2{ \ha(  \qn  ) } }{  \dt{  {\qn}^2  }   } \geqslant 0,
\frac{  \partial^2{ \hp(  \bqp ) } }{  \partial{  {\qpn} }\partial{\qp_m} } \geqslant 0,~~ \forall m \neq \n.
\end{aligned}
\end{equation}

In Section \ref{sec:special_case}, we can verify that under the Gaussian distribution, the Bayes risk of the principal is $\hp(\bqp) = \frac{1}{ { 1/{\var_0}} +  \sum_{\n \in \Nset} { \qbpn }  }$ and the Bayes risk of the agent is $\ha(\qn ) = \frac{1}{{ 1/{\var_0}} +  \qn }$, both satisfy \eqref{bayes_risk_general}.

We assume that the principal has designed a mechanism that can control the agents' exerted effort and elicit agents to report their observation truthfully. Hence, from the principal's point of view, $\qpn$, for any $\n \in \Nset$ is a decision variable for the principal, and we can take the derivative of the principal's Bayes risk with respect to $\qpn$ in \eqref{bayes_risk_general}. However, in reality, the agents would strategically choose their exerted effort. Hence, we need to design a mechanism that involves a carefully designed function $\QP$, so that the agents would put the effort as the principal expected and truthfully report their observations.

Given these preliminaries, we now present our mechanism for this setting.

\subsubsection{Mechanism}\label{sec:mechanism_general}
Our proposed mechanism COPE for the general setting also follows Algorithm~\ref{algo_interaction_general}, with the specific functions detailed below.

The function $\QBP: [\thl, \thu] \times [\thl, \thu]^{\N-1}  \rightarrow \Rsetp$ that governs the effort that the {\pp} requires an {\ag} to exert is given as the solution of the equation
\begin{align}\label{eq:optimal_contrac_equivalent_general}
\textstyle \max_{ \BQP} ~& \EX_{\BThs} \bigg[  -    \hp \big(  \BQP(\bthr) \big) -  \sum_{\n \in \Nset} \C\big(  \QBP (\thrn, \bthr_{-\n}) , \thn \big) - \sum_{\n \in \Nset} \frac{  \partial{  \C\big(  \QBP (\thrn, \bthr_{-\n}), \thn  \big)}  }{  \partial{\thn}   } \cdot \frac{  F(\thn)  }{  f(\thn) }   \bigg] \nonumber\\
\mathrm{s.t.~} & \QBP (\thrn, \bthr_{-\n}) \mathrm{~is~nonincreasing~in~}\thrn, \forall \n \in \Nset.~
\end{align}
We define the function $\PB(\cdot)$ that governs the payment to any agent $\n$ as
\begin{equation}
\label{eq:reward_fun_general}
\begin{aligned}
\textstyle&\PB(\x, \yrn, \thrn, \bthr_{-\n})  = \pib(\thrn,\bthr_{-\n})   -  \la(\x, \yrn) \cdot \Sab(  \thrn, \bthr_{-\n} )  + \Sbb(\thrn, \bthr_{-\n}),
\end{aligned}
\end{equation}
where 
\vspace{-.3cm}
\begin{align}
\textstyle \pib(\thrn,\bthr_{-\n}) =  &  \C\big( \QP(\thn, \bth_{-\n}) , \thn \big) + \int_{\thn}^{\thu} \frac{ \partial{  \C(\QP(z,\bth_{-\n}), \eta )     }   }{ \partial{\eta}  } \dt{z}, \label{eq:reward_fun_general_pi_proof}\\
\textstyle \Sab(\thrn, \bthr_{-\n}) = & \left.-\frac{ \c(  \QP(\thrn, \bthr_{-\n}), \thrn )   }{  \dt{ \ha(  \qn  ) }/\dt{\qn}}\right|_{\qn=\QP(\thrn,\bthr_{-\n})}, \label{eq:reward_fun_general_Sa_proof} \\
\textstyle\Sbb(\thrn, \bthr_{-\n}) = & \left. -\frac{ \c(  \QP(\thrn, \bthr_{-\n}), \thrn ) \cdot \ha( \qn )   }{  \dt{ \ha(  \qn  ) }/\dt{\qn}}\right|_{\qn=\QP(\thrn,\bthr_{-\n})}, \label{eq:reward_fun_general_Sb_proof}
\end{align}

The principal uses the Bayes estimate as her predictor: $\xeb(\byr, \bqbp ) = \arg \inf_{\xe} \EX \bigg[  \lp\big(  \x,\xe(  \byr, \bqp)  \big) \bigg]$.
The detailed form of the {\pp}'s estimation depends on the distribution of each agent's observation $\Pf( \x, \qn )$ defined in Section \ref{sec:model} and the loss function $\lp$. With the Gaussian distribution and quadratic loss function adopted in Section \ref{sec:normal_case}, the {\pp}'s predictor is calculated as $\xeb(\byr, \bqbp ) = \frac{ (1- |\Nsetactive|)\mu_0/\vars + \sum_{\n \in \Nsetactive}{ ( 1/\vars + \qbpn )} \cdot {\yrn}  }{ { 1/{\var_0}} +  \sum_{\n \in \Nsetactive} { \qbpn } } $, where $\qapn = \QAP(\thrn, \bthr_{-\n})$ is the decision variable of the principal, $\Nsetactive$ is the set of agents recruited by the {\pp} to report their estimation, and $|\Nsetactive|$ is the number of agents in the set $\Nsetactive$.


\subsubsection{Guarantees}

The following theorem establishes the optimality guarantees of COPE.
\begin{theorem}\label{thm:optimal_general}
	COPE satisfies the BIC and BIR condition defined in \eqref{eq:IC_requirement} and \eqref{eq:IR_requirement} and maximizes the {\pp}'s expected payoff defined in \eqref{eq:optimal_mechanism} if
	\begin{equation}\label{eq:bid_assumption_general}
	\begin{aligned}
	\frac{  \partial{  \c\big( \QP (\thn, \bth_{-\n}), \thn \big)  }  }{  \partial{\thn}  } \leq 0,
	\end{aligned}
	\end{equation}
	where function $\c$ characterizes the agent's magical cost and is defined in \eqref{cost_function_prop}.
\end{theorem}

Condition (\ref{eq:bid_assumption_general}) implies that the marginal cost of the agent should decrease with the agent's cost type, so that COPE can induce the truthful behavior of the agents. We note that condition \eqref{eq:bid_assumption_general} is satisfied under the Gaussian case when $\th$ follows from uniform distribution, as discussed in
Appendix \ref{them:optimal_cost_I_proof} and \ref{them:optimal_cost_II_proof}.


\section{Simulations}\label{sec:simulation}

We conduct numerical studies to evaluate the performance of COPE. In particular, we investigate the amount of gain that can be achieved by (optimally) exploiting the heterogeneity of the agents.
We first consider an \emph{integrated} system, where the {\pp} and all agents act as an integrated decision maker to maximize their aggregate profit (called \emph{network profit}). We denote the network profit achieved under the integrated system as the centralized benchmark.\footnote{\rev{To the best of our knowledge, this is the first paper that studies how to incentivize all agents to report their truthful estimations and exert appropriate amounts of effort based on their respective capabilities during a prediction process. Hence, we have not found an algorithm in the existing literature as a fair benchmark to compare  with the COPE performance.}}
Then we describe the details of the homogeneous benchmark mechanism under both the linear and quadratic cost function. Finally we compare the performance of COPE to the homogeneous benchmark in terms of principal's expected payoff, expected prediction error and total payment made by the principal to the agents.

\subsection{Centralized Mechanism}\label{sec:simulation_centralized_solution}
In the integrated system, the integrated player (the {\pp} and all agents) knows the precise value of $\bth$ (\emph{i.e.,} all agents' cost types). Moreover, all participated agents would exert the effort that maximizes the network profit. Specifically, the expected network profit is defined as the difference between her utility obtained from prediction and the cost of all {\ags}:
\begin{align}\label{eq:network_profit_benchmark}
\U^{\textsc{np}}(\bq, \bth) = -  \hp (  \bq ) -  \sum_{\n\in\Nset} \C \big(  \qn, \thn \big),
\end{align}
where the utility gained through the prediction is measured in terms of the {Bayes risk} given in \eqref{eq:utility_pp_new_general}.
Let function $\q^{o}: \Rsetp \times [\thl,\thu] \rightarrow \Rsetp$ be the solution that maximizes the network profit defined in \eqref{eq:network_profit_benchmark}. By focusing on the case $\x \sim \gaussian{\mu_0}{\sigma_0^2}$, we have the optimal solution under linear cost function as:
\begin{equation}\label{eq:effort_cost_benchmark_linear}
\textstyle
\qn^{o} = \left\{
\begin{array}{l l}
\max\{ 1/\sqrt{\thn} - 1/\vars,0\} ,& \quad \text{if}~\n = \arg \min_{m \in \Nset} \th_{m},\\
0, & \quad \text{otherwise}.
\end{array} \right.
\end{equation}
The optimal solution under quadratic cost function is
\begin{align}\label{eq:effort_cost_benchamrk_quadratic}
\qn^{o} = &\frac{ 1 }{ \thn } \cdot  \frac{1}{\big[W^{o}(\bth) \big]^2},
\end{align}
where the function $W^{o}: [\thl, \thu]^{\N} \rightarrow \Rsetp$ is the solution of the below equation:
\begin{align}\label{eq:aggregate_effort_benchmark_quadratic}
\big[ W^{o}(\bth) \big]^3 - \frac{1}{\vars} \cdot \big[ W^{o}(\bth) \big]^2 - \sum_{m \in \Nset} \frac{ 1 }{ \th_m } =0.
\end{align}

In the integrated system, the integrated player makes the prediction as
\begin{align}
\xe = \frac{ \mu_0/\vars + \sum_{\n\in \Nset}{ \yn^{o} \cdot \qn^{o}   }  }{ 1/\vars + \sum_{\n \in \Nset}\qn^{o}  },
\end{align}
where $\yn^{o}$ is agent $\n \in \Nset$'s true observation.

\subsection{Homogenous Mechanism}\label{sec:simulation_homogenous_appendix}
The homogenous mechanism assumes all agents to be identical (although in practice they are not), and hence does not elicit the cost parameters of individual agents. In the absence of this knowledge, the principal operates under the belief that every agent's cost parameter equals $\costHomogeneous \in [\thl, \thu]$. The principal thus chooses payment function $R_{\mbox{\tiny hom}} := \alpha(\costHomogeneous) - \beta(\costHomogeneous) \cdot (\x - \yrn)^2$, where the function $\alpha:[\thl, \thu] \rightarrow \Rsetp$ and the function $\beta:[\thl,\thu] \rightarrow \Rsetp$ are chosen to incentivize every agent $\n$ to exert optimal effort and report observations truthfully in a manner that maximizes the principal's payoff.

We focus on the case where $\x \sim \gaussian{\mu_0}{\sigma_0^2}$. Let function $\effortHomogeneous: \Rsetp \times [\thl,\thu] \rightarrow \Rsetp$ be the effort that the principal requires every agent to exert, based on the principal's belief that every agent's cost parameter equals to $\costHomogeneous$. Then the principal makes the prediction as
\begin{align}\label{eq:prediction_homogeneous_mechanism_pp}
\xe = \frac{ \mu_0/\vars + \q^{\dag} \sum_{\n\in \Nset}{ g( \yrn )   }  }{ 1/\vars + \N \cdot \q^{\dag}  },
\end{align}
where $\yrn$ is the agent $\n$'s reported observation, and the function $g:\Rset \rightarrow \Rset$ is defined as
\begin{align}
g( \yrn ) = \yrn + \frac{ \yrn - \mu_0   }{  \q^{\dag} \cdot \vars  }.
\end{align}

\textbf{Linear Cost Function:}
Under the linear cost function, the choices of functions $\effortHomogeneous$, $\alpha$, and $\beta$ are
\begin{equation*}
\effortHomogeneous( N, \costHomogeneous  ) = \frac{1}{N}\big(  \frac{1}{\sqrt{\costHomogeneous}}  - \frac{1}{\vars}  \big),
\end{equation*}
\begin{equation*}
\alpha(\costHomogeneous) = {  \big(1/\vars + \effortHomogeneous  \big)   } \cdot \costHomogeneous \effortHomogeneous  + \costHomogeneous \effortHomogeneous,~~\beta(\costHomogeneous) = {  \big(1/\vars + \effortHomogeneous  \big)^2   } \cdot \costHomogeneous
\end{equation*}


The principal chooses the function $\alpha(\cdot)$ to make sure that the agent $\n$ with the cost type $\costHomogeneous$ is willing to participate the prediction task, and chooses the function $\beta(\cdot)$ to make sure that the agent $\n$ exerts the effort $\qn = \effortHomogeneous( N, \costHomogeneous  )$ as the principal desires.

Recall that the actual cost parameter of the agent $\n\in\Nset$ is $\thn$. Hence, the agent $\n$ will exert effort $\qn = \sqrt{  {\beta(\costHomogeneous)}/{\thn} }  - {1}/{\vars}$ and report $\yrn = \frac{  \mu_0/\vars + \yn \cdot \qn  }{  1/\vars + \qn }$ to maximize his expected payoff. Besides, if the expected payoff of the agent $\n$ is negative, he will not participate this prediction task.

Also recall that the principal knows the prior information of $\x \sim \gaussian{\mu_0}{\vars}$. Hence, the principal can always achieve a payoff of $-1/\vars$ by not making any payments, and simply choosing the prior mean has her prediction. Hence, the principal does not pay anything and simply sets $\xe = \mu_0$  if her expected payoff is smaller than $-1/\vars$.


\textbf{Quadratic Cost Function:}
Under the quadratic cost function, $\effortHomogeneous$ is the solution of the following equation:
\begin{equation*}
\frac{   1  }{   \big( 1/\vars + \N \cdot \q^{\dag} \big)^2  } -   \th^{\dag} \cdot \q^{\dag} = 0
\end{equation*}
The functions $\alpha(\cdot)$ and $\beta(\cdot)$ are :
\begin{align}
\alpha(\costHomogeneous) = {  \big(1/\vars + \effortHomogeneous  \big)   } \cdot \costHomogeneous \effortHomogeneous  + \frac{1}{2} \costHomogeneous \big[ \effortHomogeneous \big]^2,~~\beta(\costHomogeneous) = {  \big(1/\vars + \effortHomogeneous  \big)^2   } \cdot \costHomogeneous \effortHomogeneous.\nonumber
\end{align}


Recall that the actually cost parameter of the agent $\n\in\Nset$ is $\thn$. Hence, the agent $\n$ will exert effort $\qn$ to maximize his own expected payoff, where $\qn$ is the solution of
\begin{align}
\frac{ \beta  }{  \big(1/\vars + \qn  \big)^2   } - \thn \qn = 0 \nonumber.
\end{align}
Besides, if the expected payoff of the agent $\n$ is negative, he will not participate this prediction task. Also, the principal does not pay anything and simply sets $\xe = \mu_0$  if her expected payoff is smaller than $-1/\vars$.

\subsection{Numerical Results}
In the simulations, we draw $\x \sim \gaussian{0}{1}$, and set $\thl=0$ and $\thu=1$. We vary the number of agents from $\N = 3$ to $\N = 19$. {Each point in the plots is an average across $50000$ trials.} Without loss of generality, we have normalized the principal's payoff (see~\eqref{eq:payoff_pp}) so that it equals zero in the ideal (unachievable) case of zero prediction error and a zero payment. Note that the principal can always achieve  a payoff of $-1$ by not making any payments, and simply choosing the prior mean has her prediction.

\begin{figure}[ht]
	\centering
	\subfloat[Linear cost function]{%
		\includegraphics[width=3in]{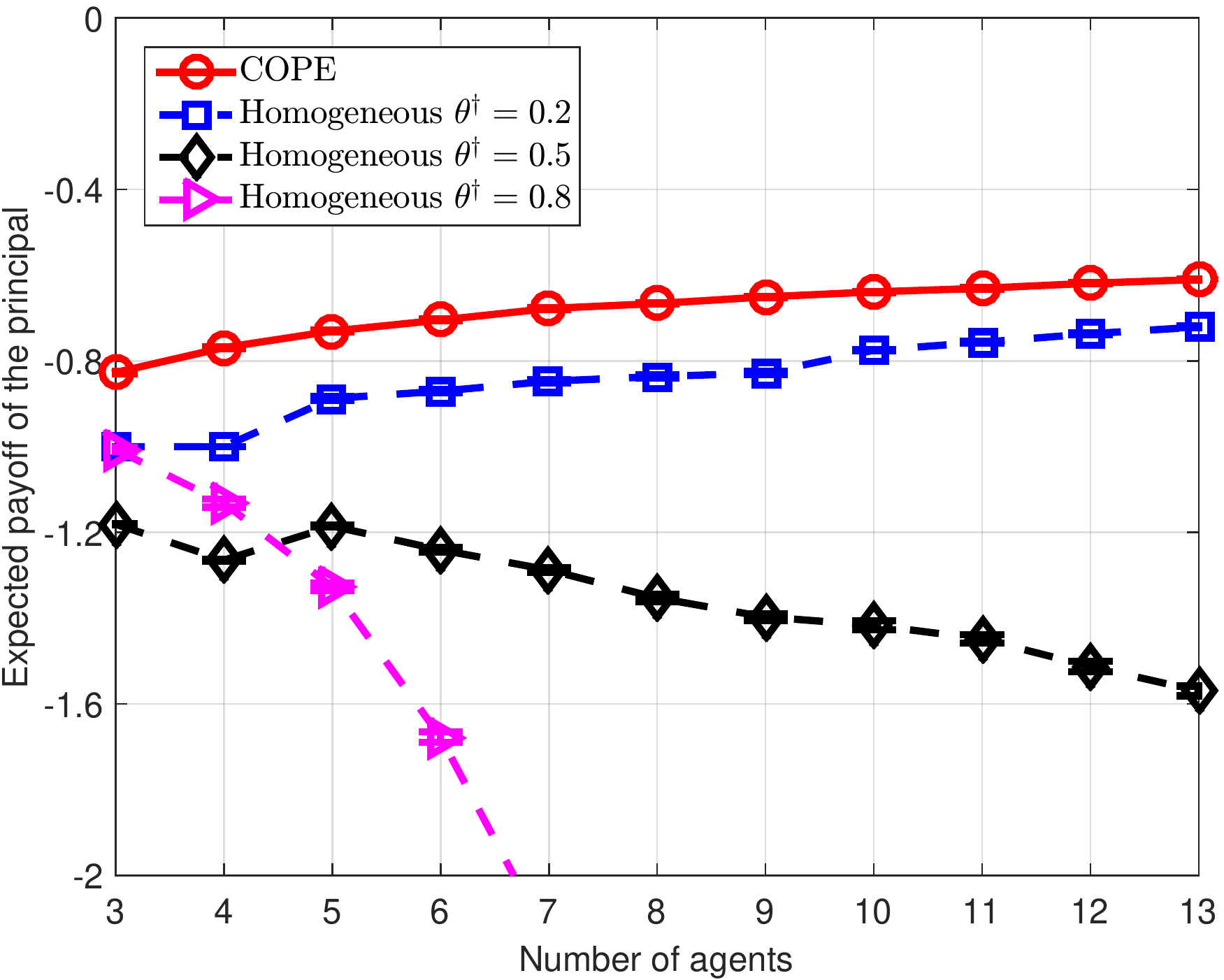}\label{fig:payoff_pp_linear}
	}%
	\subfloat[Quadratic cost function]{%
		\includegraphics[width=3in]{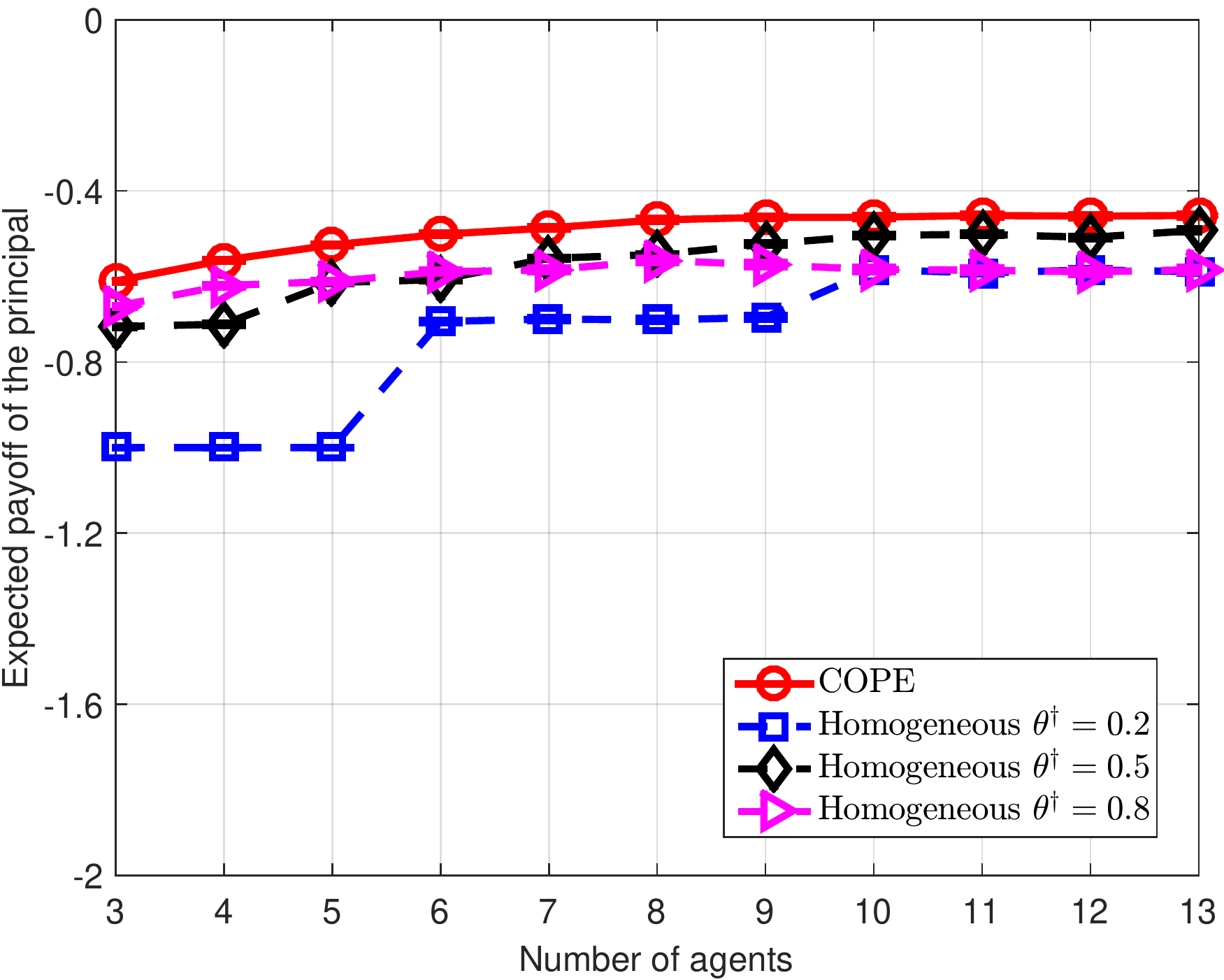}\label{fig:payoff_pp_quadatic}
	}%
	
	\medskip
	
	\caption{The principal's expected payoff under COPE and the homogeneous mechanism.}\label{fig:payoff_pp}
	
\end{figure}

Figure~\ref{fig:payoff_pp} depicts
the {\pp}'s expected payoff achieved under COPE and under the homogeneous mechanism for different values of $\costHomogeneous$ when the cost function is linear and quadratic, respectively.
We use the red line with circle markers to denote COPE, the blue dash line with square markers to denote homogeneous mechanism with $\costHomogeneous = 0.2$, the dark dash line with diamond markers to denote homogeneous mechanism with $\costHomogeneous = 0.5$, and the magenta line with right-pointing triangle markers to denote homogeneous mechanism with $\costHomogeneous = 0.8$.

Figure~\ref{fig:payoff_pp} shows
that COPE can improve the {\pp}'s payoff  by exploring the heterogeneity of users. \rev{The improvement is at least $10\%$ under the linear cost function and $5\%$ under the quadratic cost function.}
	
By comparing Figure~\ref{fig:payoff_pp_linear} to Figure~\ref{fig:payoff_pp_quadatic}, we can see that the value of the belief of principal (i.e., $\costHomogeneous$) under the homogeneous mechanism will affect the final result. Under the linear cost function, a lower value of $\costHomogeneous$ results in a better performance in terms of the principal's payoff. However, under the quadratic cost function, a higher value of $\costHomogeneous$ results in a better performance. The reasons are as follows.

Under the homogeneous mechanism, the value of $\costHomogeneous$ will determine the number and types of agents joining the task.
	Having a higher value of $\costHomogeneous$ would incentivize more agents to participate. This is because, for an agent $\n \in \Nset$ whose cost parameter $\thn < \costHomogeneous$, he can put less effort to achieve the same performance as the agent with cost parameter $\costHomogeneous$ can.
	
Under the linear cost function, similar as COPE, finding the most capable one would be optimal for the principal, as the marginal cost is nonnegative even the agent does not put any effort. Hence, having a lower value of $\costHomogeneous$ would eliminate more agents, and have a higher chance to find the agent with $\thn \leq \costHomogeneous$.

	On the contrary, under the quadratic cost function, it would be optimal to recruit as many agents as possible to improve the prediction accuracy. The benefit brought by the accuracy improvement would be higher than the additional payment to agents. Hence, having a higher value of $\costHomogeneous$ would help the principal recruit more agents.

	\begin{figure}[htp]
	\centering
	\subfloat[Linear cost function]{%
		\includegraphics[width=3.1in]{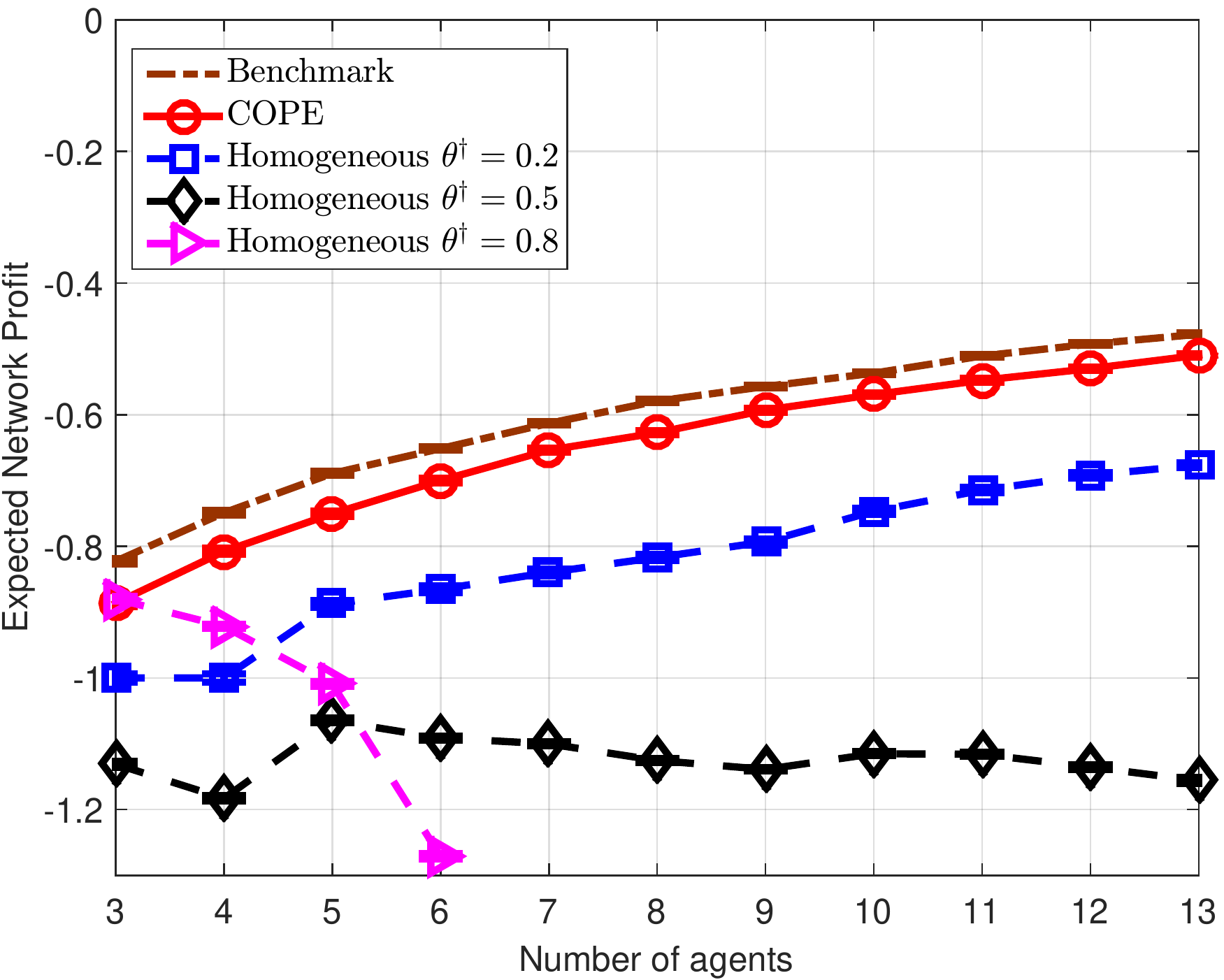}\label{fig:error_np_linear}
	}%
	\subfloat[Quadratic cost function]{%
			\includegraphics[width=3.1in]{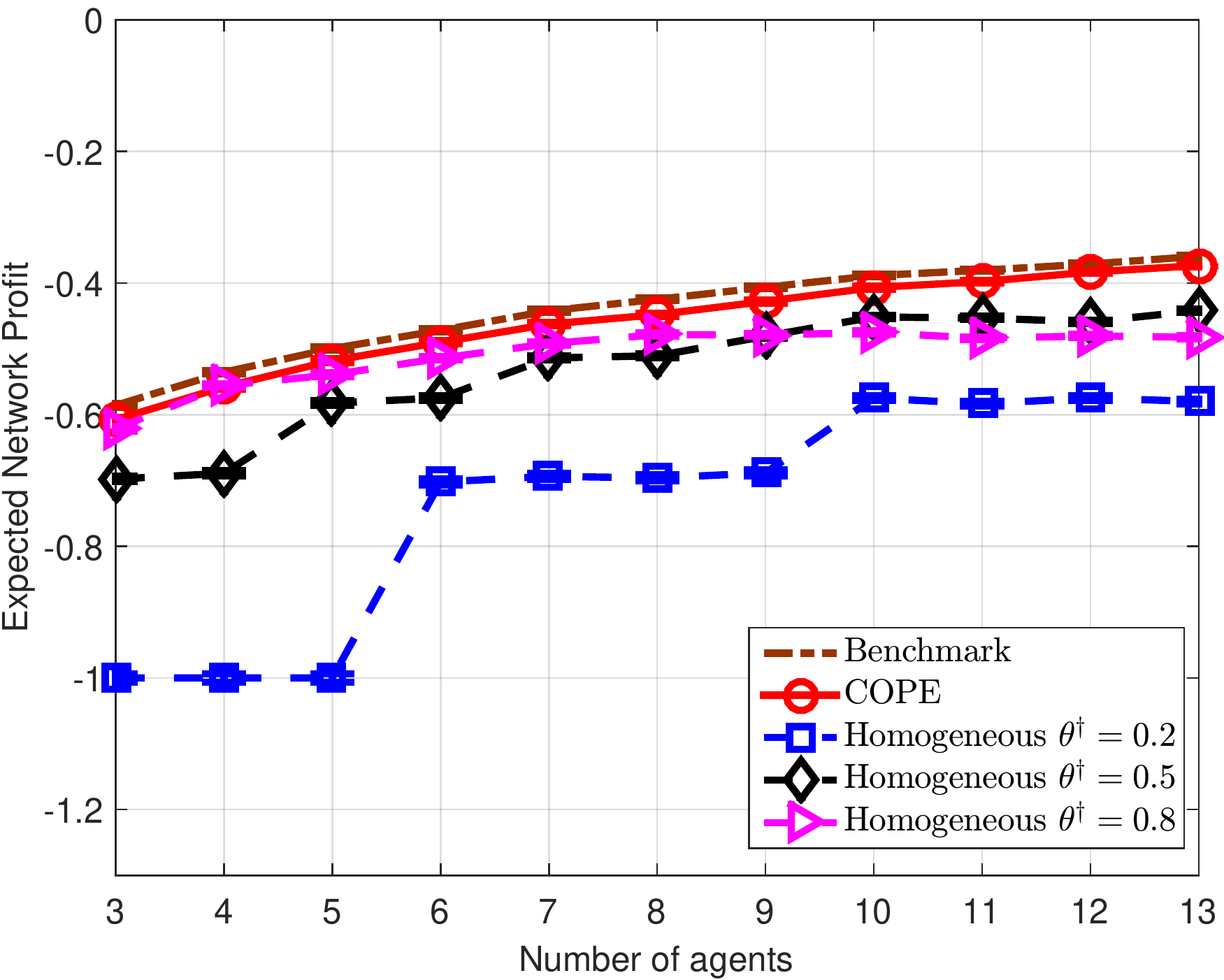}\label{fig:error_np_quadatic}
	}%
	
	
	\medskip
	
	\caption{The network profit under COPE and the homogeneousmechanism.}\label{fig:np}
	
\end{figure}

Figure~\ref{fig:np} depicts
the network profit achieved under COPE and under the homogeneous mechanism for different values of $\costHomogeneous$ when the cost function is linear and quadratic, respectively.
We use the dash brown line to denote the benchmark solution where the {\pp} and all agents acted as an integrated player.

From Figure \ref{fig:np},
we have the following observations.
\begin{itemize}
	\item \rev{COPE can achieve a network profit close to the integrated benchmark solution, \emph{e.g.,} the gap is less than $10\%$ under the linear cost function and $3\%$ under the quadratic cost function.}
	
	\item Network profit achieved under COPE increases with the number of agents. This is because the increasing number of agents allows the {\pp} to have a higher chance to incentivize agents with high capability to improve the prediction accuracy.  This is true under both costs, even though only one agent will be recruited under the linear cost.
	
	\item COPE leads to a much higher network profit, compared to the homogeneous mechanism. The reason is that COPE explores the heterogeneity of agents by eliciting their cost type information.
\end{itemize}

\rev{
	Recall that the principal makes the prediction based on \eqref{eq:prediction_homogeneous_mechanism_pp} in the homogeneous mechanism. As the actually effort exerted by the agents are different from that desired by the principal (\emph{i.e.,} $\qn \neq \effortHomogeneous$), the prediction made by the principal is inaccurate. On the contrary, COPE elicits the cost types of agents as well as incentivizes each agent to exert the appropriate level of effort, which results in a better performance than the homogeneous mechanism. Due to the joint effect of the number of agents and the value of $\costHomogeneous$, the performance of homogeneous mechanism is close to that of COPE (\emph{e.g.,} $\N <4$ under the linear cost function and $\N < 6$ under the quadratic cost function) in terms of expected network profit. However, it is difficult to determine the proper choice of $\costHomogeneous$ in terms of the number of agents. Finding the optimal value of $\costHomogeneous$ given the number of agents will be an interesting future work.
}



\section{Conclusions}\label{sec:conclusion}
We study the parametric prediction market under information asymmetry. To elicit the truthful information of participating agents and exploit the heterogeneity  in the agents, we propose mechanism COPE, which ensures agents to exert effort desired by the {\pp} and report their true observation. Our analysis indicate that, under the Gaussian estimation noise scenario, when the costs incurred by the agents are linear in the amount of exerted effort, the principal should require service from only one agent with the lowest reported cost. On the other hand, when the costs are quadratic in the exerted effort,  it is optimal for the principal to recruit multiple agents to complete the task.
We also present the general form of COPE that is optimal for a wide variety of settings (\emph{e.g.,} general cost function and the noise distribution).

In this work, we have focused on the parametric setting, where the {\pp} recruits agents to estimate a parameter (\emph{e.g.,} the winner of a election) that the realized value can be observed in the future. As in some cases, such as rating the quality of a book, the true outcome cannot be easily observed or verified. In the future, we will consider how to incentivize agents' behaviour when collecting subjective data.
Moreover, in order to give theoretical insights into the problem of estimation from strategic agents,
we use one parameter, \emph{i.e.,} the cost type $\th$ to characterize the heterogeneity of the agents in terms of their cost. Relaxing the parametric assumption (\emph{e.g.,} characterizing agents' types with random functions) and designing mechanisms with theoretical guarantees is extremely challenging. In practice, heuristics (e.g., \cite{brousseau2002economics,chiappori1997empirical}) are employed to circumvent the parametric assumption when using these mechanisms. In the future, we would consider how to relax such parametric assumption.



\begin{APPENDIX}{Full Proofs}



	\section{Proof of Theorem \ref{them:optimal_cost_I}: Linear Costs}\label{them:optimal_cost_I_proof}

	The proof will proceed in four steps. The first three steps show that our mechanism incentivizes the agents to be truthful, and the fourth step proves optimality of our mechanism. First, we show that irrespective of what an agent reports as his cost parameter, and irrespective of the effort he exerts, the agent is always incentivized to report his true observation. We follow this up and show that irrespective of the effort that an agent exerts, he is always incentivized to report his cost parameter correctly. The third step completes the proof of truthfulness, showing that under truthful reporting of the cost parameter and the observation, in our mechanism, an agent is always incentivized to exert precisely the effort as desired by the principal. Finally, we show that among all mechanisms that ensure truthful reports, our mechanism maximizes the principal's expected utility.
	
	We assume that the random variables $\{ \thn \}_{\n\in\Nset}$ are independently and identically distributed on support $[\thl,\thu]$, with a cumulative distribution function $F:[\thl, \thu] \rightarrow \Rsetp$ and a probability density function $f:[\thl, \thu] \rightarrow \Rsetp$. We further assume that
	the c.d.f. function $F$ is continues, differentiable, and log concave in $[\thl, \thu]$. This assumption is satisfied by a wide range of distributions, such as uniform, gamma, and beta distributions.

	\textbf{Step 1. Truthful reporting of observation under COPE}
	
	We will analyze the strategies of the agent who is recruited by the principal and the agents who are not recruited by the principal, respectively.
	
	We first study the observation reporting strategy of the agent $\no$ who is recruited and rewarded by the principal, where $\no = \argmin_{m \in \Nset} \thr_{m}$.
	
	We will show that the agent $\no$ will choose
	\begin{align}\label{eq:agent_report}
	\yr_{\no} = \frac{  {\mu_0} \cdot  { 1/{\var_0}} +  {\y_{\no}} \cdot { \q_{\no}} }{ { 1/{\var_0}} +   \q_{\no} }
	\end{align}
	to maximize his expected payoff given his exerting effort $\q_{\no}$ and own observation $\y_{\no}$.
	
	As shown in \eqref{eq:payment_costI}, $\pia(\thr_{\no})$, $\Saa(\thr_{\no})$ and $\Sba(\thr_{\no})$ are independent of $\yr_{\no}$. Moreover, the value of calculated by $\Saa(\thr_{\no})$ is always positive. Hence, when the agent $\no$ makes reporting observation strategy to maximize his expected payoff, i.e.,
	\begin{align}
	&\yr_{\no} \in \argmax \EX \big[ \pia(\thr_{\no})   - \Saa(  \thr_{\no} ) \cdot ( \x - \yr_{\no} )^2 + \Sba(\thr_{\no}) \big] - \C\big(  \q_{\no}, \th_{\no} \big), \nonumber
	\end{align}
	where the expectation is taken with respect to $\x$ and cost parameters $\bth_{-\no} = [\Ths_1, \ldots, \Ths_{\no-1}, \Ths_{\no+1}, \ldots, \Ths_{\N}]^{T}$ except agent $\no$, it is equivalent for the agent $\no$ to choose reporting strategy such that
	\begin{align}
	\yr_{\no} \in \argmin \EX_{\x}[ ( \x - \yr_{\no} )^2 ].
	\end{align}
	
	Based on theory of Bayesian estimation (see \cite{myerson1979incentive, lehmann1998theory}),
	only when $\yr_{\no} = \frac{  {\mu_0} \cdot  { 1/{\var_0}} +  {\y_{\no}} \cdot { \q_{\no}} }{ { 1/{\var_0}} +   \q_{\no} }$, the value of $\EX_{\x}[ ( \x - \yr_{\no} )^2 ]$ is minimized and the expected value is 
	\begin{align}
	\EX_{\x}[ ( \x - \yr_{\no} )^2 ] = \frac{1}{1/{\var_0} +  { \q_{\no}}}. \nonumber
	\end{align}
	
	We then study the observation reporting strategy of other agents who are not recruited and rewarded by the principal. For agent $\n \in \Nset, \n \neq \no$, he will put zero effort as he does not receive any reward from the principal. In this case, only when reporting his observation $\yrn = \mu_0$ can minimize $\EX_{\x}[ ( \x - \yrn )^2 ]$. The expected value of $\EX_{\x}[ ( \x - \yrn )^2 ]$ is
	\begin{align}
	\EX_{\x}[ ( \x - \yrn )^2 ] = \frac{1}{1/{\var_0} }, ~~\n \in \Nset, \n \neq \no. \nonumber
	\end{align}

	\textbf{Step 2. Truthful reporting of the cost parameter under COPE}
	
	We first show that the agent $\no$ will truthfully reveals his cost type.
	We first rewrite functions $\pia$, $\Saa$, and $\Sba$ as follows.
	\begin{align}
	\pia(\thr_{\no}, \bth_{-\no}) =   & \thr_{\no} \cdot { \QAP( \thr_{\no}, \bth_{-\no}) }  + \int_{\thr_{\no}}^{\thu} { \QAP( z, \bth_{-\no}) } \dt{z} ,\nonumber\\
	\Saa(\thr_{\no},\bth_{-\no}) = &\big[  \QAP( \thr_{\no}, \bth_{-\no}) + 1/\vars  \big]^2 \cdot \thr_{\no}, \nonumber\\
	\Sba(\thr_{\no}, \bth_{-\no}) = &\big[ \QAP( \thr_{\no}, \bth_{-\no}) + 1/\vars  \big] \cdot \thr_{\no}. \nonumber
	\end{align}

	The expected payoff of the agent who has a cost type $\th_{\no}$ but reports $\thr_{\no}$ is:
	\begin{equation}\label{eq:agent_expect_bidding}
	\begin{aligned}
	&\EX_{\{\x, \y_{\no},\BThs_{-\no}\}} \big[  \UA( \x, \thr_{\no}, \q_{\no},\y_{\no}, \th_{\no},  \bth_{-\no})  \big] \\
	&=\EX_{\BThs_{-\no} } \big[ \pia(\thr_{\no}, \bth_{-\no}) - \Saa(  \thr_{\no} , \bth_{-\no}) \cdot \frac{1}{1/{\var_0} +  { \q_{\no}}}  + \Sba(\thr_{\no}, \bth_{-\no}) - \q_{\no} \th_{\no} \big].
	\end{aligned}
	\end{equation}
	
	For notation convenience, we define the function $\UAE: \Rset \times [\thl, \thu] \times \Rsetp \times [\thl, \thu]^{\N} \rightarrow \Rsetp$ as
	\begin{align}\label{eq:agent_expected_payoff_new_typeI_proof}
	\UAE( \thr_{\no}, \q_{\no}, \th_{\no}, \bth_{-\no}  ) &=
	\big[ \pia(\thr_{\no}, \bth_{-\no}) - \Saa(  \thr_{\no} , \bth_{-\no})  \frac{1}{1/{\var_0} +  { \q_{\no}}}  + \Sba(\thr_{\no}, \bth_{-\no}) - \q_{\no} \th_{\no} \big],
	\end{align}
	where $\bth_{-\no}$ are the random variables of all agents' cost type except that of agent $\no$. By comparing \eqref{eq:agent_expect_bidding} to \eqref{eq:agent_expected_payoff_new_typeI_proof}, the expected payoff of the agent $\n$ is
	\begin{equation*}
	\EX_{\{\x, \y_{\no},\BThs_{-\no}\}} \big[  \UA( \x, \thr_{\no}, \q_{\no},\y_{\no}, \th_{\no},  \bth_{-\no})  \big] = \EX_{\BThs_{-\no}} \big[  \UAE( \thr_{\no}, \q_{\no}, \th_{\no}, \bth_{-\no}  ) \big].
	\end{equation*}
	
	By the mean value theorem, we have:
	\begin{equation}\label{eq:mean_value}
	\begin{aligned}
	&\EX\big[  \UAE( \th_{\no}, \q_{\no}, \th_{\no},\bth_{-\no} )  \big]- \EX \big[  \UAE( \thr_{\no}, \q_{\no},\th_{\no}, \bth_{-\no} )  \big] = \EX \bigg[  \frac{ \partial{ \UAE( \eta, \q_{\no},\th_{\no}, \bth_{-\no}  )} }{\partial{\eta}}  \bigg] \cdot (  \th_{\no} - \thr_{\no} ),
	\end{aligned}
	\end{equation}
	where the expectation is taken with respect to $\bth_{-\no}$, and $\eta$ lies between $\th_{\no}$ and $\thr_{\no}$.
	
	We further have:
	\begin{equation}\label{eq:compare_mean_value}
	\begin{aligned}
	&\EX_{\BThs_{-\no}} \bigg[  \frac{ \partial{ \UAE( \eta, \q_{\no},\th_{\no}, \bth_{-\no}  )} }{\partial{\eta}}  \bigg]  \\
	&=\EX_{\BThs_{-\no}} \bigg[    \frac{ \partial{ } }{\partial{\eta}} \bigg(   \eta{ \QAP( \eta, \bth_{-\no}) }  + \int_{\eta}^{\thu} { \QAP( z, \bth_{-\no}) } \dt{z}  - \frac{  \big [ { \QAP( \eta, \bth_{-\no}) }  + 1/{\vars}  \big]^2 }{   1/{\vars} + \q_{\no}  } \eta \\
	& \qquad{} \qquad{} \quad{} +    \big[ { \QAP( \eta, \bth_{-\no}) }  + 1/{\vars} \big] \eta -  \q_{\no} \th_{\no} \  \bigg)     \bigg]\\
	&= \EX_{\BThs_{-\no}} \bigg[    2 \eta \frac{ \partial{ \QAP( \eta, \bth_{-\no}) }  }{  \partial{\eta} }  -  \frac{ \big[ { \QAP( \eta, \bth_{-\no}) }  + 1/{\vars}  \big]^2}{   1/{\vars} + \q_{\no}  } \\
	&\qquad{}\qquad{} \quad{} - 2\frac{ [ { \QAP( \eta, \bth_{-\no}) }  + 1/{\vars} ] }{   1/{\vars} + \q_{\no}  }\cdot  \frac{ \partial{  \QAP( \eta, \bth_{-\no}) }  }{  \partial{\eta} } \cdot \eta + \big[ { \QAP( \eta, \bth_{-\no}) }  + 1/{\vars} \big]    \bigg]\\
	&= \EX_{\BThs_{-\no}} \bigg[  \bigg( 1 - \frac{ \QAP (\eta, \bth_{-\no})  + 1/{\var_0}}{ \q_{\no} + 1/{\var_0}}  \bigg)\cdot \bigg(  2 \eta \cdot \frac{ \partial{ \QAP (\eta, \bth_{-\no}) } }{ \partial{\eta} } +  \QAP (\eta, \bth_{-\no}) + 1/{\var_0}  \bigg) \bigg].
	\end{aligned}
	\end{equation}
	
	If we have
	\begin{align}\label{sufficient_condition_linear}
	- \frac{  \partial{ \QAP (\eta, \bth_{-\no}) }/\big( \QAP (\eta, \bth_{-\no}) + 1/\vars\big) }{ \partial{\th_{\no}}/\th_{\no}  }
	\geq  \frac{  1 }{ 2 },
	\end{align}
	then we have
	\begin{equation*}\label{eq:bid_assumption}
	\begin{aligned}
	2 \eta \cdot \frac{ \partial{ \QAP (\eta, \bth_{-\no}) } }{ \partial{\eta} } +  \QAP (\eta, \bth_{-\no}) + 1/{\var_0}   \leq 0.
	\end{aligned}
	\end{equation*}
	
	\begin{lemma}\label{lemma:sufificient_condition_linear_unifor_proof}
		If $\thn \sim \mbox{Uniform}[\thl, \thu]$ which is independent for every $\n \in \Nset$, then \eqref{sufficient_condition_linear} is satisfied.
	\end{lemma}
	
	\emph{Proof}:
	First consider the case $\N = 1$. Since there is only one agent, hence the principal can only select that agent. So $\QAP$ is simply $\Q$ of that agent:
	\begin{align*}
	\Q(\th) =  {1}/{\sqrt{\th +  {\F(\th)}/{\f(\th)}}} - 1/{\sigma^2_0},
	\end{align*}
	hence
	\begin{align*}
	-\frac{\partial \Q(\th)}{\partial \th} \frac{\th}{\Q(\th) + 1/{\sigma^2_0}}&= \frac{1 + \frac{\partial}{\partial \th} \left(\frac {\F(\th)}{\f(\th)} \right) }{2[{\th +  {\F(\th)}/{\f(\th)}}]} \frac{1}{{\sqrt{\th +  {\F(\th)}/{\f(\th)}}}}  \frac{\th}{1/{\sqrt{\th +  {\F(\th)}/{\f(\th)}}}}\\
	& = \frac{1}{2} \frac{1 + \frac{\partial}{\partial \th} \left(\frac {\F(\th)}{\f(\th)} \right) }{  1 + \frac{1}{ \th} \left(\frac {\F(\th)}{\f(\th)} \right)  }
	\geq \frac{1}{2},
	\end{align*}
	where the final inequality holds for uniform distribution.
	
	We now consider $\N > 1$. Observe that the calculation above will be violated only when the cost parameter of some other agent is infinitesimally close to $\th_{\no}$ (since in that case, $\frac{  \partial{ \QAP(\th_{\no},\bth_{-\no}) } }{ \partial{\th_{\no}}}$ is different from that calculated above). However given our assumptions that  the distribution of $\th$ \rev{follows some known distribution such as uniform and normal distributions}, and given that the number of agents $\N$ is finite, $\th_{\no}$ will be well separated from the cost types of all other agents with probability $1$. $\Box$
	
	\rev{As the agent is selfish, he will exert effort $\q_{\no}$ to maximize his expected payoff. Hence, the agent's exerted effort can be obtained by taking the first order derivative of (\ref{eq:agent_expect_bidding}) with respect to ${\q_{\no}}$ and setting it to zero, which is}
	\begin{equation}\label{eq:h1_func_bid_linear}
	\begin{aligned}
	( 1/\vars + \qapno )^2 \cdot \thr_{\no} = ( 1/\vars + \q_{\no} )^2 \cdot \th_{\no},
	\end{aligned}
	\end{equation}
	where $\qapno$ is the shorthand notation for $\QAP(\thr_{\no}, \bth_{-\no})$.
	
	Based on \eqref{eq:h1_func_bid_linear}, we have (i) if $\thr_{\no} > \th_{\no} $, $\qapno < \q_{\no} $, (ii) if $\thr_{\no}  < \th_{\no} $, $\qapno> \q_{\no} $, and (iii) if $\thr_{\no}  = \th_{\no} $, $\qapno = \q_{\no} $.

	Hence, If $\thr_{\no} > \th_{\no}$, the equation \eqref{eq:compare_mean_value} is negative and \eqref{eq:mean_value} is positive.
	This inequality also holds for $\thr_{\no} < \th_{\no}$, by a similar argument. Therefore, agent $\no$ will truthfully report his own cost parameter.
	
	We then show that an agent $\n \in \Nset, \n \neq \no$ will truthfully reveal his cost type.
	Recall that the principal does not recruit and reward the agent $\n \in \Nset, \n \neq \no$. Hence, the payment to the agent $\n \in \Nset, \n \neq \no$ is zero.
	The we have
	\begin{align}
	&\EX_{\BThs_{-\n}} \big[  \UAE( \thn, \qn, \thn,\bth_{-\n} )  \big]  - \EX_{\BThs_{-\n}} \big[  \UAE( \thrn, \qn,\thn, \bth_{-\n} )  \big] =0, \forall \n \in \Nset , \n \neq \no, \nonumber
	\end{align}
	which shows that there is no difference between truthfully reporting cost type or not in terms of expected payoff for the agent $\n$. Without loss of generality, we assume that in this case, the agent will truthfully report their cost types.

	\textbf{Step 3.  Incentivize agent to exert precisely the effort as desired by the principal under COPE}
	
	As we have proved in Step 2 that the agent $\no$ would truthfully report his cost type ($\thrn = \thn$), then we will show that the agent $\no$ exerts an effort level such that $\q_{\no} = \qapno$ would maximize his expected payoff, which is given as
	\begin{equation}\label{eq:ag_payoff_contract_expect_updated}
	\begin{aligned}
	&\EX\big[ \UAE( \th_{\no}, \qapno, \th_{\no},  \bth_{-\no}) \big]  =  \pia(\th_{\no},\bth_{-\no})   - \Saa(  \th_{\no}, \bth_{-\no} )  \frac{1}{1/{\var_0} +  { \q_{\no}}}  + \Sba(\th_{\no}, \bth_{-\no}) - \q_{\no} \th_{\no},
	\end{aligned}
	\end{equation}
	where the expectation is taken with respect to $\BThs_{-\no}$.
	
	It can be verified that \eqref{eq:ag_payoff_contract_expect_updated} is concave in $\q_{\no}$.
	Hence, by taking the first order derivative of \eqref{eq:ag_payoff_contract_expect_updated} with respect to $\q_{\no}$, we have
	\begin{align}\label{eq:agent_FOC}
	&\frac{\partial}{\partial{\q_{\no}}}\EX \big[ \UAE( \th_{\no}, \qapno, \th_{\no}, \bth_{-\no}) \big]  =  \bigg[  \frac{1/{\var_0} +  { \qapno}}{1/{\var_0} + \q_{\no}} \bigg]^2 \cdot \th_{\no} - \th_{\no}.
	\end{align}
	
	We can verify that the value of \eqref{eq:agent_FOC} equals to zero only when $\q_{\no} = \qapno$. Hence, agent $\no$ will exert the effort as the principal desires to maximize his expected payoff. Then \eqref{eq:agent_report} is rewritten as
	\begin{align}\label{eq:agent_report2}
	\yr_{\no} = \frac{  {\mu_0} \cdot  { 1/{\var_0}} +  {\y_{\no}} \cdot { \qapno} }{ { 1/{\var_0}} +   \qapno }.
	\end{align}
	
	Because the principal knows the value of $\mu_0$, $\var_0$, and $\qapno$, she can infer the agent $\no$'s truth observation $\y_{\no}$ from \eqref{eq:agent_report2}.
	
	\textbf{Step 4. Maximize the principal's expected utility under COPE}
	
	Then we look at the expected payoff of the principal. The following lemma describes that COPE is the optimal mechanism that maximizes the principal's expected utility.
	\begin{lemma}\label{lemma:optimal_predictor_proof}
		The optimal predictor $\xe =  \frac{  {\mu_0} \cdot  { 1/{\var_0}} +  \sum_{\n\in \Nset}{\yn} \cdot { \qapn} }{ { 1/{\var_0}} +   \sum_{\n \in \Nset}\qapn }$ defined in COPE maximizes the principal's expected utility, and the Bayes risk of the principal's prediction is $\hp \big(  \bqp \big) = \frac{ 1 }{ 1/{\var_0} +  \sum_{\n\in\Nset}{ \qapn}}$.
	\end{lemma}
	
	\emph{Proof}:
	Recall that $\qapn = \QAP(\thn, \bth_{-\n})$. 
	Given all agents'  observation $\by$ and agents' exert effort $\bqap$, the principal's updated belief on the realization of $\x$ can be expressed as
	\begin{align}
	&\x |( \by, \bqap ) \sim N\bigg(  \frac{  {\mu_0} \cdot  { 1/{\var_0}} +  \sum_{\n\in \Nset}{\yn} \cdot { \qapn} }{ { 1/{\var_0}} +   \sum_{\n \in \Nset}\qapn }, \frac{  1  }{ { 1/{\var_0}} +   \sum_{\n \in \Nset}\qapn }  \bigg).\nonumber
	\end{align}
	
	To maximize the expected utility for the prediction, the principal solves
	\begin{align}
	&\max_{\xe}\EX \big[ v -  (\x - \xe)^2 | ( \by, \bqap ) \big] \nonumber\\
	&= \max_{\xe} \bigg(  v - \bigg\{  \EX\big[  {\x}^2 |  ( \by, \bqap )  \big]  - 2 \xe \EX\big[  {\x} |  ( \by, \bqap ) \big] + \xe^2  \bigg\}  \bigg)\nonumber\\
	&= \max_{\xe} \bigg(  v - \bigg[ \xe -   \frac{  {\mu_0} \cdot  { 1/{\var_0}} +  \sum_{\n\in \Nset}{\yn} \cdot { \qapn} }{ { 1/{\var_0}} +   \sum_{\n \in \Nset}\qapn } \bigg]^2 -  \frac{  1  }{ { 1/{\var_0}} +   \sum_{\n \in \Nset}\qapn } \bigg)\nonumber\\
	&\leq v -  \frac{  1  }{ { 1/{\var_0}} +   \sum_{\n \in \Nset}\qapn }\nonumber
	\end{align}
	
	The equality holds only when
	\begin{align}
	\xe =  \frac{  {\mu_0} \cdot  { 1/{\var_0}} +  \sum_{\n\in \Nset}{\yn} \cdot { \qapn} }{ { 1/{\var_0}} +   \sum_{\n \in \Nset}\qapn }.
	\end{align}
	
	Hence, the optimal predictor that maximizes the {\pp}'s expected utility is
	\begin{equation}\label{eq:pp_predict_opt_proff}
	\xea\big(\by, \bqap \big) = \frac{  {\mu_0} \cdot  { 1/{\var_0}} +  \sum_{\n\in \Nset}{\yn} \cdot { \qapn} }{ { 1/{\var_0}} +   \sum_{\n \in \Nset}\qapn }，
	\end{equation}
	and the Bayes risk is
	%
	%
	\begin{equation*}\label{eq:MSE_estimate}
	\hp \big(  \bqp \big) = \inf_{\xea} \EX[ ( \x - \xea )^2 ]  = \frac{ 1 }{ 1/{\var_0} +  \sum_{\n\in\Nset}{ \qapn}},
	\end{equation*}
	where the expectation is taken with respect to $\x$ and $\by$.
	
	Recall that under the linear cost function, the principal only recruits agent $\no$ to exert effort, in such case, $\qapn = 0$, $\forall \n \in \Nset$, $\n \neq \no$. Also recall that the principal can infer the true observation of the agent $\no$ through the function $g: \Rset \rightarrow \Rset$, and such an observation is defined as $\y_{\no} = g(\yr_{\no}) = \yr_{\no} + { ( \yr_{\no} - \mu_0 )}/{ ( \qapno \vars )  }$.
	Then putting back to \eqref{eq:pp_predict_opt_proff} we can get the conclusion. $\Box$
	

	We then show that the desired effort level $\QAP(\thn, \bth_{-\n})$ defined in \eqref{eq:effort_cost_I} and the function $\pia(\thn, \bth_{-\n})$ defined in \eqref{eq:payment_costI} can maximize the {\pp}'s expected payoff and satisfy BIC and BIR conditions.
	

	Notice that as the agent $\no$ exerts effort such that $\q_{\no} = \qapno$ and reports $\yr_{\no} = \frac{  {\mu_0} \cdot  { 1/{\var_0}} +  {\y_{\no}} \cdot { \qapno} }{ { 1/{\var_0}} +   \qapno }$, the expected payment function is reduced to
	\begin{align}\label{eq:payment_agent_typeI_proof_xx}
	&\EX_{\{\x, \y_{\no},\BThs_{-\no}\}} \big[  \PA (  \x, \y_{\no}, \q_{\no}, \th_{\no}, \bth_{-\no} )   \big] \nonumber \\
	&=\EX_{\BThs_{-\no} } \big[ \pia(\th_{\no},\bth_{-\no}) - \Saa(  \th_{\no} ,\bth_{-\no}) \cdot \frac{1}{1/{\var_0} +  { \q_{\no}}}  + \Sba(\th_{\no},\bth_{-\no})  \big] = \EX_{\BThs_{-\no} } \big[ \pia(\th_{\no},\bth_{-\no})  \big].
	\end{align}
	For other agent $\n \in \Nset, \n \neq \no$, as the principal does not require him to do the observation, we first assume that the expected payment to him is as follows,
	\begin{align}\label{eq:payment_agent_typeI_proof_xxx}
	\EX_{\{\x, \y_{\no},\BThs_{-\no}\}}[ \PA (  \x, \yn, \thn, \bth_{-\n} ) ] =
	\EX_{\BThs_{-\no} } \big[ \pia(\thn, \bth_{-\n}) \big], \forall \n \in \Nset, \n \neq \no.
	\end{align}
	Later we will show that $\pia(\thn, \bth_{-\n}) = 0, \forall \n \neq \no$.
	
	The expected payoff of agent $\n \in \Nset$ is
	\begin{align}
	\EX_{\BThs_{-\n}} \big[ \UAE\big( \thr_{\no}, \qapno, \th_{\no}, \bth_{-\no}  \big) \big] = &\EX_{\BThs_{-\n}} \big[  \pia(\thrn, \bth_{-\n}) - \thn  { \QAP(\thrn, \bth_{-\n})}  \big],\nonumber
	\end{align}
	where $\qapno$ is the shorthand notation of $\QAP(\thrn, \bth_{-\n})$.
	For notation convenience, we adopt $\UAE\big( \pia(\thrn, \bth_{-\n}), \QAP(\thrn, \bth_{-\n}), \thn  \big)$ in the later proof of Theorem \ref{them:optimal_cost_I}, where the function $\UAE$ is rewritten as
	\begin{align}\label{eq:payoff_agent_simple}
	&\EX_{\BThs_{-\n}} \big[ \UAE\big( \pia(\thrn, \bth_{-\n}), \QAP(\thrn, \bth_{-\n}), \thn  \big) \big]  = \EX_{\BThs_{-\n}} \big[  \pia(\thrn, \bth_{-\n}) - \thn  { \QAP(\thrn, \bth_{-\n})}  \big].
	\end{align}
	Correspondingly, BIC and BIR conditions, i.e., \eqref{eq:IC_requirement} and \eqref{eq:IR_requirement}, can be rewritten as
	\begin{align}
	&\EX_{\BThs_{-\n}} \big[ \UAE\big( \pia(\thn, \bth_{-\n}), \QAP(\thn, \bth_{-\n}), \thn  \big) \big] \geq \EX_{\BThs_{-\n}} \big[ \UAE\big( \pia(\thrn, \bth_{-\n}), \QAP(\thrn, \bth_{-\n}), \thn  \big) \big], ~\forall \thrn \neq \thn, \label{eq:IC_requirement_simple}\\
	&\EX_{\BThs_{-\n}} \big[ \UAE\big( \pia(\thn, \bth_{-\n}), \QAP(\thn, \bth_{-\n}), \thn  \big) \big] \geq 0,\ \forall \thn \in [\thl, \thu].\label{eq:IR_requirement_simpe}
	\end{align}
	
	
	Base on Lemma \ref{lemma:optimal_predictor_proof}, \eqref{eq:payment_agent_typeI_proof_xx}, and \eqref{eq:payment_agent_typeI_proof_xxx}, the expected payoff of the principal is
	\begin{align}
	\EX_{\x, \by, \BThs} [ \UP(\x, \bqap, \by, \bthr) ] &=   -  \hp \big(  \bqp \big) -  \EX_{\x, \by, \BThs} \big[  \sum_{\n\in\Nset}  \P (  \x, \yn, \thn, \bth_{-\n} ) \big] \nonumber\\
	& =  -   \frac{ 1 }{ 1/{\var_0} +  \sum_{\n\in\Nset}{ \qapn}} - \EX_{\BThs} \big[ \sum_{\n\in\Nset} \pia(\thn, \bth_{-\n}) \big].\nonumber
	\end{align}
	
	Recall that $\qapn = \QAP(\thn, \bth_{-\n})$, the principal's optimal problem defined in \eqref{eq:optimal_mechanism} can be rewritten as
	\begin{equation}\label{eq:optimal_mechanism_proof}
	\begin{aligned}
	\sup_{\{ \QAP(\bth), \pia(\bth) \}, \forall \thn\in\BThs, \forall \n \in \Nset } ~&\EX [ \UP(\x, \bqap, \by, \bthr) ] ,\\
	\mathrm{subject~to:~~} & \mathrm{BIC~and~BIR~in~(\ref{eq:IC_requirement_simple})~and~(\ref{eq:IR_requirement_simpe}).}
	\end{aligned}
	\end{equation}

	In the following lemmas, we characterize an equivalent formulation for the feasible region defined by BIC and BIR. Using these lemmas, we show that $\QAP(\thn, \bth_{-\n})$ defined in \eqref{eq:effort_cost_I} and $\pia(\thn, \bth_{-\n})$ defined in \eqref{eq:payment_costI}  are the optimal solution that solves the principal's problem in \eqref{eq:optimal_mechanism_proof}.
	\begin{lemma}\label{lemma:feasibility}
		The solution of \eqref{eq:optimal_mechanism_proof} is feasible if and only if it satisfies the following conditions for all $\thn\in[\thl, \thu]$, $\forall \n \in \Nset$:
		\begin{itemize}
			\item the expected payoff of agent $\n$ is
			\begin{align}\label{eq:agent_profit_contract_feasi}
			&\EX_{\BThs_{-\n}} \bigg[ \UAE\big( \pia(\thn, \bth_{-\n}), \QAP(\thn, \bth_{-\n}), \thn  \big) \bigg] =  \EX_{\BThs_{-\n}} \bigg[  \int_{\thn}^{\thu} \QAP(x, \bth_{-\n}) \dt{x}  \bigg];
			\end{align}
			\item $\QAP(\thn, \bth_{-\n} )$ is non-increasing in $\thn$.
		\end{itemize}
	\end{lemma}
	
	\emph{Proof}:
	The proof of Lemma \ref{lemma:feasibility} is as follows.
	We first show that BIC and BIR imply the condition in \eqref{eq:agent_profit_contract_feasi}.
	
	Notice that the first derivative of (\ref{eq:payoff_agent_simple}) is:
	\begin{equation}\label{eq:illustr_tmp}
	\begin{aligned}
	& \frac{ \partial{  \EX_{\BThs_{-\n}} \big[ \UAE\big( \pia(\thrn, \bth_{-\n}), \QAP(\thrn, \bth_{-\n}), \thn  \big) \big]   } }{\partial{\thn}}=  \EX_{\BThs_{-\n}} \big[ - \QAP(\thrn, \bth_{-\n}) \big] \leq 0.
	\end{aligned}
	\end{equation}
	
	Then, for any $\thn^1 > \thn^2$, we have
	\begin{equation}\label{eq:illstr_tmp2}
	\begin{aligned}
	\EX_{\BThs_{-\n}}  \big[  \UAE( \pia(\thn^1, \bth_{-\n}), \QAP(\thn^1, \bth_{-\n}), \thn^1)  \big] &\leq
	\EX_{\BThs_{-\n}} \big[  \UAE( \pia(\thn^1, \bth_{-\n}), \QAP(\thn^1, \bth_{-\n}), \thn^2) \big]\\
	& \leq \EX_{\BThs_{-\n}} \big[  \UAE( \pia(\thn^2, \bth_{-\n}), \QAP(\thn^2, \bth_{-\n}), \thn^2)  \big];
	\end{aligned}
	\end{equation}
	where the first inequality is because \eqref{eq:illustr_tmp} and the second is from the BIC condition defined in \eqref{eq:IC_requirement_simple}.
	{In other words, for the agent $\n\in\Nset$ whose cost parameter $\thl \leq \thn \leq \thu$, we have
		\begin{align}\label{eq:illustrate_IR_sufficient_appendix_typeI}
		\EX_{\BThs_{-\n}}  \big[  \UAE( \pia(\thu, \bth_{-\n}), \QAP(\thu, \bth_{-\n}), \thu)  \big]
		&\leq
		\EX_{\BThs_{-\n}} \big[  \UAE( \pia(\thn, \bth_{-\n}), \QAP(\thn, \bth_{-\n}), \thn) \big]\nonumber\\
		& \leq \EX_{\BThs_{-\n}} \big[  \UAE( \pia(\thl, \bth_{-\n}), \QAP(\thl, \bth_{-\n}), \thl)  \big].
		\end{align}
		
		Recall that the BIR condition is
		\begin{align}\label{eq:illustrate_IR_sufficient_appendix_typeI_xx}
		\EX_{\BThs_{-\n}} \big[ \UAE\big( \pia(\thn, \bth_{-\n}), \QAP(\thn, \bth_{-\n}), \thn  \big) \big] \geq 0,\ \forall \thn \in [\thl, \thu],
		\end{align}
		which implies that, for the agent $\n\in\Nset$ with any value $\thn \in [\thl, \thu]$, his expected payoff should at least be zero.
		Then the expected payoff of the agent $\n$ with cost parameter $\thu$ must be binding at zero. Otherwise, the principal can reduce the $\pia(\thu, \bth_{-\n})$ by a small value of $\delta >0$, which does not violate the constraint of \eqref{eq:illustrate_IR_sufficient_appendix_typeI_xx} but raises the principal's expected payoff. Hence, we have
		\begin{equation}\label{eq:IR_proof_xxx}
		\begin{aligned}
		\EX_{\BThs_{-\n}} \big[  \UAE( \pia(\thu, \bth_{-\n}), \QAP(\thu, \bth_{-\n}), \thu)  \big] &= 0.
		\end{aligned}
		\end{equation}
	}
	
	Let $\UAE( \thn, \bth_{-\n} )  =  \UAE\big( \pia(\thn, \bth_{-\n}), \QAP(\thn, \bth_{-\n}), \thn \big)$. From the BIC condition, we have
	\begin{align}
	&\EX_{\BThs_{-\n}} \big[  \UAE( \thn, \bth_{-\n} )  \big] = \max_{\thrn}  \EX_{\BThs_{-\n}} \big[   \UAE\big( \pia(\thrn, \bth_{-\n}), \QAP(\thrn, \bth_{-\n}), \thn\big)  \big].\nonumber
	\end{align}
	By using the envelope theorem, we have:
	\begin{equation}\label{eq:evelop_theorem_typeI_appendix}
	\begin{aligned}
	\frac{\partial{\EX_{\BThs_{-\n}} \big[  \UAE( \thn, \bth_{-\n} )  \big] } }{ \partial{\thn}}
	&= \left.\frac{\partial{\EX_{\BThs_{-\n}} \big[ \UAE( \pia(\thrn, \bth_{-\n}), \QAP(\thrn, \bth_{-\n}), \thn)  \big]}}{\partial{\thn}}\right|_{\thrn = \thn} =  \EX_{\BThs_{-\n}} \big[ - \QAP(\thn, \bth_{-\n}) \big],
	\end{aligned}
	\end{equation}
	where $\thn$ is a parameter.
	By integrating both sides {from the value of $\thn$ to $\thu$} and using \eqref{eq:IR_proof_xxx} and the assumption that the random variable $\thn$ of the agent $\n$ is independent for every $\n \in \Nset$ , we get
	\begin{equation}
	\begin{aligned}
	&\EX_{\BThs_{-\n}} \big[ \UAE\big( \pia(\thn, \bth_{-\n}), \QAP(\thn, \bth_{-\n}), \thn  \big) \big] =  \EX_{\BThs_{-\n}} \bigg[  \int_{\thn}^{\thu} \QAP(x, \bth_{-\n}) \dt{x}  \bigg].
	\end{aligned}
	\end{equation}
	
	{We prove that $\QAP(\thn, \bth_{\n})$ is nonincreasing in $\thn$ by contradiction. Let $p_{\n}$ be the shorthand notation for $\pi( \thn, \bth_{-\n} )$. Suppose for any $\thn^1 > \thn^2$, we have $\QAP(\thn^1,\bth_{-\n}) > \QAP(\thn^2,\bth_{-\n})$.  }
	Because
	\begin{equation}\label{eq:second_order_derivative_qth_typeI_simple}
	\begin{aligned}
	\frac{\partial^2{  \UAE\big( p_{\n}, \qapn, \thn\big)   }}{\partial{\qapn}\partial{\thn}} =  -  1& < 0 ,
	\end{aligned}
	\end{equation}
	\begin{equation}\label{eq:second_order_derivative_qq_typeI_simple}
	\begin{aligned}
	\frac{\partial^2{ \UAE\big( p_{\n}, \qapn, \thn\big)   }}{\partial{\qapn}^2}  = 0,~~~~
	\end{aligned}
	\end{equation}
	%
	we have
	\begin{align}
	0 &=\left.\frac{\partial{ \UAE\big( p_{\n}, \qapn, \thn^1\big)  }}{\partial{\qapn}}\right|_{\qapn=\QAP(\thn^1,\bth_{-\n})} \nonumber\\
	&= \left.\frac{\partial{ \UAE\big( p_{\n}, \qapn,  \thn^1\big)  }}{\partial{\qapn}}\right|_{\qapn=\QAP(\thn^2,\bth_{-\n})} \nonumber\\
	&< \left.\frac{\partial{\UAE\big( p_{\n}, \qapn,  \thn^2\big)  }}{\partial{\qapn}}\right|_{\qapn=\QAP(\thn^2,\bth_{-\n})},
	\end{align}
	{where the first equality is due to BIC when the agent $\n$'s cost parameter $\thn$ has the value of $\thn^1$, the second equality is due to \eqref{eq:second_order_derivative_qq_typeI_simple}, and the inequality is due to \eqref{eq:second_order_derivative_qth_typeI_simple}.
		
		However, based on the BIC condition, if the agent $\n$'s cost parameter $\thn$ has the value of $\thn^2$, then we should have
		\begin{align}
		\left.\frac{\partial{ \UAE\big( p_{\n}, \qapn, \thn^2\big)  }}{\partial{\qapn}}\right|_{\qapn=\QAP(\thn^2,\bth_{-\n})}  = 0, \nonumber
		\end{align}
		which holds true for all scalar values of $p_{\n}$.
		Hence, for any $\thn^1 > \thn^2$, $\QAP(\thn^1,\bth_{-\n}) \leq \QAP(\thn^2,\bth_{-\n})$.
	}
	

	Then we need to prove that (\ref{eq:agent_profit_contract_feasi}) implies BIC and BIR defined in (\ref{eq:IC_requirement_simple}) and (\ref{eq:IR_requirement_simpe}).
	
	BIR is verified by putting $\thn$ back to \eqref{eq:agent_profit_contract_feasi}. Besides, by putting $\thn = \thu$ back to \eqref{eq:agent_profit_contract_feasi}, we have
	\begin{align}
	&\EX_{\BThs_{-\n}} \bigg[ \UAE\big( \pia(\thu, \bth_{-\n}), \QAP(\thu, \bth_{-\n}), \thu  \big) \bigg] =  0. \nonumber
	\end{align}

	Then we prove that (\ref{eq:agent_profit_contract_feasi}) implies BIC.
	Notice that we have:
	\begin{equation*}
	\begin{aligned}
	&\EX_{\BThs_{-\n}} \bigg[  \UAE\big( \pia(\thrn, \bth_{-\n}), \QAP(\thrn, \bth_{-\n}), \thn\big)   \bigg]  \\
	&\overset{1}{=} \EX_{\BThs_{-\n}} \bigg[ - \int_{\thn}^{\thu}  \frac{\partial{ \UAE\big( \pia(\thrn, \bth_{-\n}), \QAP(\thrn, \bth_{-\n}), z \big)  }}{\partial{z}}\dt{z}    \bigg]   \\
	&\overset{2}{=}  \EX_{\BThs_{-\n}} \bigg[ \UAE\big( \pia(\thrn, \bth_{-\n}), \QAP(\thrn, \bth_{-\n}), \thrn\big)  -  \int_{\thn}^{\thrn}  \frac{\partial{ \UAE\big( \pia(\thrn, \bth_{-\n}), \QAP(\thrn, \bth_{-\n}), z \big)  }}{\partial{z}}\dt{z}   \bigg] \\
	&\overset{3}{=}  \EX_{\BThs_{-\n}} \bigg[  \int_{\thrn}^{\thu} \QAP(\eta, \bth_{-\n}) \dt{\eta}  -  \int_{\thn}^{\thrn}  \frac{\partial{ \UAE\big( \pia(\thrn, \bth_{-\n}), \QAP(\thrn, \bth_{-\n}), z \big)  }}{\partial{z}}\dt{z}   \bigg] \\
	&\overset{4}{=}  \EX_{\BThs_{-\n}} \bigg[  -\int_{\thu}^{\thn} \QAP(\eta, \bth_{-\n}) \dt{\eta} -  \int_{\thn}^{\thrn} \QAP(\eta, \bth_{-\n}) \dt{\eta} +  \int_{\thn}^{\thrn}  \QAP(\thrn, \bth_{-\n}) \dt{z}   \bigg] \\
	&\overset{5}{=}  \EX_{\BThs_{-\n}} \bigg[ \UAE\big( \pia(\thn, \bth_{-\n}), \QAP(\thn, \bth_{-\n}), \thn \big) + \int_{\thn}^{\thr}  \big( \QAP(\thrn, \bth_{-\n}) -  \QAP(\eta, \bth_{-\n}) \bigg) \dt{\eta} \big], \\
	\end{aligned}
	\end{equation*}
	where the third equality and the fifth equality are obtained by \eqref{eq:agent_profit_contract_feasi}.
	
	If $\thrn > \thn$, then the above equation is non-positive (because $\QAP(\eta, \bth_{-\n})$ is non-increasing in $\eta$), hence
	\begin{align}
	&\EX_{\BThs_{-\n}} \big[  \UAE( \pia(\thrn, \bth_{-\n}), \QAP(\thrn, \bth_{-\n}), \thn)   \big] < \EX_{\BThs_{-\n}} \big[ \UAE( \pia(\thn, \bth_{-\n}), \QAP(\thn, \bth_{-\n}), \thn)\big].\nonumber
	\end{align}
	This inequality also holds for $\thrn < \thn$ by a similar argument. Therefore, the two condition imply BIC. $\Halmos$
	
	
	Then based on Lemma \ref{lemma:feasibility}, we have the following Lemma.
	\begin{lemma}\label{lemma:optimility}
		The optimisation problem in (\ref{eq:optimal_mechanism_proof}) has the following equivalent formulation:
		\begin{align}\label{eq:optimal_contrac_equivalent}
		\max_{ \{ \QAP(\bth)\}, \forall \thn\in\BThs } & \EX_{\BThs} \bigg[ -   \frac{ 1 }{ 1/{\var_0} +  \sum_{\n\in\Nset}{ \QAP(\thn, \bth_{-\n})}}  - \sum_{\n \in \Nset} \QAP(\thn, \bth_{-\n}) \cdot \thn
		- \sum_{\n \in \Nset}  \QAP(\thn, \bth_{-\n}) \cdot \frac{  F(\thn)  }{  f(\thn) }   \bigg],\nonumber\\
		\mathrm{s.t.~} & \QAP(\thn, \bth_{-\n}) \mathrm{~is~nonincreasing~in~}\thn,~~
		\end{align}
		where the expectation is taken with respect to $\bth$.
	\end{lemma}
	
	\emph{Proof}:
	The proof of Lemma \ref{lemma:optimility} is as follows.
	The expected payoff of the principal can be written as:
	\begin{equation}\label{eq:optimiliayt_linear_proof_xx}
	\begin{aligned}
	&\EX_{\BThs} \bigg[    -   \frac{ 1 }{ 1/{\var_0} +  \sum_{\n\in\Nset}{ \QAP(\thn, \bth_{-\n})}}  - \sum_{\n \in \Nset} \QAP(\thn, \bth_{-\n}) \cdot \thn  -\sum_{\n \in \Nset} \UA\big( \pia(\thn, \bth_{-\n}), \QAP(\thn, \bth_{-\n}), \thn\big)   \bigg] \\
	&=\EX_{\BThs} \bigg[    -   \frac{ 1 }{ 1/{\var_0} +  \sum_{\n\in\Nset}{ \QAP(\thn, \bth_{-\n})}} - \sum_{\n \in \Nset} \QAP(\thn, \bth_{-\n}) \cdot \thn  - \sum_{\n \in \Nset}   \int_{\thn}^{\thu} \QAP(x, \bth_{-\n}) \dt{x}  \bigg], \\
	\end{aligned}
	\end{equation}
	where the expectation is taken with respect to $\bth$.
	Notice that
	\begin{equation*}
	\begin{aligned}
	\EX_{\thn} \big[   \int_{\thn}^{\thu} \QAP(x, \bth_{-\n}) \dt{x} \big] &=
	\int_{\thl}^{\thu}  \int_{z}^{\thu} \QAP(x, \bth_{-\n}) \dt{x} \cdot f(z) \dt{z}
	= \int_{\thl}^{\thu} F(z) \QAP(z, \bth_{-\n}) \dt{z} \\
	&= \int_{\thl}^{\thu} \frac{F(z)}{f(z)} \QAP(z, \bth_{-\n}) f(z)\dt{z}  = \EX_{\thn} \big[   \frac{F(\thn)}{f(\thn)} \QAP(\thn, \bth_{-\n}) \big],
	\end{aligned}
	\end{equation*}
	where the first equation is obtained by using integration by parts. Then by applying the above equation to \eqref{eq:optimiliayt_linear_proof_xx} and the fact that $\{\thn\}_{\n \in \Nset}$ are assumed to be random, independently and identically distributed on support $[\thl, \thu]$, we can get the conclusion. $\Halmos$
	
	
	Based on Lemma \ref{lemma:optimility}, the principal's problem reduces to choosing the desired effort $\qapn= \QAP(\thn, \bth_{-\n})$ for each agent $\n \in \Nset$.
	We first consider the problem in \eqref{eq:optimal_contrac_equivalent} without the constraint. If the optimal solution to this unconstrained problem is increasing, then it is also an optimal solution to the constrained problem.
	
	
	\begin{lemma}
		$\QAP(\thn, \bth_{-\n})$ defined in \eqref{eq:effort_cost_I} and $\pia(\thn, \bth_{-\n})$ defined in \eqref{eq:payment_costI}  are the optimal solution that solves the principal's problem in \eqref{eq:optimal_mechanism_proof}
	\end{lemma}
	
	\emph{Proof}:
	We first prove that for the agent $\forall \n \in \Nset$,
	\begin{equation*}\label{eq:effort_cost_I_proof}
	\textstyle
	\QAP( \thn, \bth_{-\n} ) = \left\{
	\begin{array}{l l}
	\max\{  1/\sqrt{\gamma(\thn)} - 1/\vars,0\},  & \quad \text{if}~\no = \argmin_{m \in \Nset} \th_{m},\\
	0, & \quad \text{otherwise},
	\end{array} \right.
	\end{equation*}
	is the optimal solution of \eqref{eq:optimal_contrac_equivalent} by contradiction.

	As $\qapn= \QAP(\thn, \bth_{-\n})$, the problem in \eqref{eq:optimal_contrac_equivalent} is equivalent to
	\begin{align}\label{eq:reduece_probelm_typeI_proof_xxxx}
	\min_{ \bqap \geq 0 }~ & \frac{ 1 }{ 1/{\var_0} +  \sum_{\n\in\Nset}{ \qapn}} + \sum_{\n \in \Nset} \qapn \cdot \gamma(\thn),\nonumber\\
	\mathrm{s.t.~} & \qapn \mathrm{~is~nonincreasing~in~}\thn,~~~~
	\end{align}
	where $\gamma(\thn) = \thn +  {\F(\thn)}/{\f(\thn)}$.
	
	Without loss of generality, let $\gamma(\th_1) \leq \gamma(\th_2) \ldots \leq \gamma(\th_{\N})$. If the principal's desired efforts from all agents are positive, then the solution is
	\begin{align}\label{eq:solut_suppose_typeI}
	{\qp_1} = {\q^{\sp,*}_{1}}, {\qp_2} = {\q^{\sp,*}_{2}}, \ldots, {\qp_{\N}} = {\q^{\sp,*}_{\N}}.
	\end{align}
	
	Suppose that there is another solution such that
	\begin{align}
	\left\{
	\begin{array}{ll}
	{\q^{\sp,\dag}_{1}} = {\q^{\sp}_{1}} + {\q^{\sp}_{j}}, &~\\
	{\q^{\sp,\dag}_{i}} =  {\q^{\sp}_{i}}, &  \forall i \neq 1, j, i\in\Nset,\\
	{\q^{\sp,\dag}_{j}}  = 0. &~
	\end{array}\right.
	\end{align}
	Then we can verify that
	\begin{align}
	\sum_{\n \in \Nset} {\q^{\sp,\dag}_{\n}}  = \sum_{\n \in \Nset} {\q^{\sp}_{\n}}~\text{and}~\sum_{\n \in \Nset}\big( {\q^{\sp,\dag}_{\n}} \cdot \gamma(\th_{\n}) \big) \leq \sum_{\n \in \Nset} \big( \qp_{\n} \cdot \gamma(\th_\n) \big).\nonumber
	\end{align}
	Hence, \eqref{eq:solut_suppose_typeI} is not an optimal solution. Then we let $\qapn = 0, \forall \n >1$, the problem in \eqref{eq:reduece_probelm_typeI_proof_xxxx} becomes
	\begin{align}
	\min_{ {\q^{\sp}_{1}} }~ & \frac{ 1 }{ 1/{\var_0} +  {\q^{\sp}_{1}} } + {\q^{\sp}_{1}} \cdot \gamma(\th_1),\nonumber\\
	\mathrm{s.t.~} & {\q^{\sp}_{1}} \geq 0.~~~~
	\end{align}
	
	By solving the above problem we can get that ${\q^{\sp}_{1}} = \max\{ 1/\sqrt{\gamma(\th_1)} - 1/\vars,0\}$. As we define $\no = \argmin_{m \in \Nset} \th_{m}$ and the assumption that $F$ is log-concave in $[\thl, \thu]$, we have $\qp_{\no} = \max\{ 1/\sqrt{\gamma(\th_{\no})} - 1/\vars , 0\}$.


	According to \eqref{eq:agent_profit_contract_feasi}, we have
	\begin{align}\label{eq:payment_typeI_proof}
	&\EX_{\BThs_{-\n}} \big[ \pia(\thn,\bth_{-\n}) - \QAP(\thn, \bth_{-\n}) \cdot \thn  \big] =  \EX_{\BThs_{-\n}} \bigg[  \int_{\thn}^{\thu} \QAP(x, \bth_{-\n}) \dt{x}  \bigg]. \nonumber
	\end{align}
	
	Then the optimal payment function given the agent $\no$ and $1/\sqrt{\gamma(\th_{\no})} - 1/\vars \geq 0$ is
	\begin{align}
	\pia(\th_{\no}) &=   \th_{\no}/\sqrt{\gamma(\th_{\no})} - \th_{\no}/\vars + \int_{\th_{\no}}^{\thu} \bigg(  \frac{1}{\sqrt{\gamma(z)}}  - \frac{1}{\vars} \bigg)\dt{z}= \th_{\no}/\sqrt{\gamma(\th_{\no})} - \thu/\vars + \int_{\th_{\no}}^{\thu} \frac{1}{\sqrt{\gamma(z)}}  \dt{z}, \nonumber
	\end{align}
	\vspace{-1cm}
	\begin{equation*}
	~~
	\end{equation*}
	and the payment will be zero if $1/\sqrt{\gamma(\th_{\no})} - 1/\vars < 0$.
	
	For other agents, the payments will be zero as they are not involved in the observation and prediction. $\Halmos$
	
	As in Theorem \ref{them:optimal_cost_I} , we assume that $\thn \sim \mbox{Uniform}[\thl, \thu]$, which  is independent for every $\n \in \Nset$. Putting the expression of $F$ and $f$ back to the above equations, we can have the conclusion.


	\section{Proof of Theorem \ref{them:optimal_cost_II}: Quadratic Costs}\label{them:optimal_cost_II_proof}

	The proof is similar to that in Section \ref{them:optimal_cost_I_proof}. The difference is as follows.
	
	First, the function $\pib: [\thl, \thu]^{\N} \rightarrow \Rsetp$, $\Sab, \Sbb: [\thl, \thu]^{\N} \times \Rsetp \rightarrow \Rsetp$ are defined as
	\begin{align}
	\pib(\thrn,\bth_{-\n}) =  & \frac{1}{2} \cdot \bigg[ \thrn \cdot \big[ \QBP (\thrn, \bth_{-\n}) \big]^2 + \int_{\thrn}^{\thu} \bigg(  \big[ \QBP (z, \bth_{-\n}) \big]^2 \bigg) \dt{z} \bigg], \label{eq:reward_fun_special_II_pi_proof}\\
	\Sab(\thrn, \bth_{-\n}) = &\big[\QBP(\thrn, \bth_{-\n}) + 1/{\var_0} \big]^2  \thrn \cdot \QBP(\thrn, \bth_{-\n}),  \label{eq:reward_fun_special_II_Sa_proof}\\
	\Sbb(\thrn, \bth_{-\n}) = &\big[\QBP(\thrn, \bth_{-\n}) + 1/{\var_0} \big]  \thrn \cdot \QBP(\thrn, \bth_{-\n}).\label{eq:reward_fun_special_II_Sb_proof}
	\end{align}
	
	\textbf{Step 1. Truthful reporting of observation under COPE}
	
	We will analyze the strategies of the agent $\n$, $\forall \n \in \Nset$.
	We will show that the agent $\n$ will choose
	\begin{align}\label{eq:agent_report_typeII}
	\yrn= \frac{  {\mu_0} \cdot  { 1/{\var_0}} +  {\yn} \cdot { \qn} }{ { 1/{\var_0}} +   \qn }
	\end{align}
	to maximize his expected payoff given his exerting effort $\qn$ and own observation $\yn$.
	
	As $\pib(\thrn,\bth_{-\n})$, $\Sab(\thrn, \bth_{-\n}) $ and $\Sbb(\thrn, \bth_{-\n})$ are independent of $\yrn$ and the value of calculated by $\Sab(\thrn, \bth_{-\n})$ is always positive. Hence, when the agent $\n$ makes reporting observation strategy to maximize his expected payoff, i.e.,
	\begin{align}
	&\yrn \in \argmax \EX \big[ \pib(\thrn,\bth_{-\n})   - \Sab(\thrn, \bth_{-\n})  ( \x - \yrn )^2 + \Sbb(\thrn, \bth_{-\n}) \big] - \C\big(  \qn, \thn \big), \nonumber
	\end{align}
	where the expectation is taken with respect to $\x$ and cost parameters $\BThs_{-\n} = [\Ths_1, \ldots, \Ths_{\n-1}, \Ths_{\n+1}, \ldots, \Ths_{\N}]^{T}$ except agent $\n$, it is equivalent for the agent $\n$ to choose reporting strategy such that
	\begin{align}
	\yrn \in \argmin \EX_{\x}[ ( \x - \yr_n )^2 ].
	\end{align}
	
	The value of $\EX_{\x}[ ( \x - \yrn )^2 ]$ is minimized when $\yrn= \frac{  {\mu_0} \cdot  { 1/{\var_0}} +  {\yn} \cdot { \qn} }{ { 1/{\var_0}} +   \qn }$. The expected value in this case is
	\begin{align*}
	\EX_{\x}[ ( \x - \yrn)^2 ] = \frac{1}{1/{\var_0} +  { \qn}}.
	\end{align*}

	\textbf{Step 2. Truthful reporting of cost parameter under COPE}
	
	We will show that the agent $\n$, $\forall \n \in \Nset$ will truthfully reveals his cost type.
	The expected payoff of the agent who has a cost type is $\thn$ but reports $\thrn$ is:
	\begin{equation}\label{eq:agent_expect_bidding_typeII}
	\begin{aligned}
	&\EX_{\{\x, \yn,\BThs_{-\n}\}} \big[  \UA( \x, \thrn, \qn,\yn, \thn,  \bth_{-\n})  \big] \\
	&=\EX_{\BThs_{-\n} } \big[ \pib(\thrn, \bth_{-\n}) - \Sab(  \thrn, \bth_{-\n}) \cdot \frac{1}{1/{\var_0} +  { \qn}}  + \Sbb(\thrn, \bth_{-\n}) - \frac{1}{2} \thn \qn^2 \big].
	\end{aligned}
	\end{equation}
	
	For notation convenience, we define
	\begin{align}
	&\UA( \thrn, \qn, \thn, \bth_{-\n}  ) =
	\big[ \pib(\thrn, \bth_{-\n}) -   \frac{\Sab(  \thrn , \bth_{-\n})}{1/{\var_0} +  { \qn}}  + \Sbb(\thrn, \bth_{-\n}) - \frac{1}{2} \thn \qn^2 \big]\nonumber
	\end{align}
	
	By the mean value theorem, we have:
	\begin{equation}\label{eq:mean_value_typeII}
	\begin{aligned}
	&\EX \big[  \UA( \thn, \qn, \thn,\bth_{-\n} )  \bigg]- \EX \bigg[  \UA( \thrn, \qn,\thn, \bth_{-\n} )  \big] = \EX_{\BThs_{-\n}} \big[  \frac{ \partial{ \UA( \eta, \qn,\thn, \bth_{-\n}  )} }{\partial{\eta}}  \big]  (  \thn - \thrn ),
	\end{aligned}
	\end{equation}
	where the expectation is taken with respect to $\bth_{-\n}$, and $\eta$ lies between $\thn$ and $\thrn$.

	We further have
	\begin{align}\label{eq:compare_mean_value_typeII}
	&\EX_{\BThs_{-\n}} \bigg[  \frac{ \partial{ \UA( \eta, \qn, \thn, \bth_{-\n}  )} }{\partial{\eta}}  \bigg] \nonumber\\
	&=\EX_{\BThs_{-\no}} \Bigg[   \frac{ \partial{ } }{\partial{\eta}} \bigg(   \frac{1}{2} \cdot \eta \big[ { \QAP( \eta, \bth_{-\n}) } \big]^2 + \int_{\eta}^{\thu} \big[ { \QAP( z, \bth_{-\n}) } \big]^2 \dt{z} \nonumber\\
	& \qquad{} \qquad{}  +    \big[ { \QAP( \eta, \bth_{-\no}) }  + 1/{\vars} \big] \eta \big[ { \QAP( \eta, \bth_{-\n}) } \big] - \frac{  \big [ { \QAP( \eta, \bth_{-\n}) }  + 1/{\vars}  \big]^2 }{   1/{\vars} + \q_{\n}  } \eta  \big[ { \QAP( \eta, \bth_{-\n}) } \big]
	-  \frac{1}{2}\thn \qn^2 \  \bigg)     \Bigg] \nonumber\\
	&= \EX_{\BThs_{-\n}} \Bigg[  \bigg( 1 - \frac{ \QBP (\eta, \bth_{-\n}) + 1/\vars }{ \qn + 1/\vars }  \bigg)   \bigg(  2  \QBP (\eta, \bth_{-\n})  \cdot \eta \cdot \frac{ \partial{ \QBP (\eta, \bth_{-\n}) } }{ \partial{\eta} }   +  \big[ \QBP (\eta, \bth_{-\n}) + 1/\vars \big] \cdot \QBP (\eta, \bth_{-\n}) \nonumber\\
	&\qquad{}\qquad{} +  \big[ \QBP (\eta, \bth_{-\n}) + 1/\vars \big] \cdot \eta \cdot \frac{ \partial{}}{ \partial{\eta} } \big[   \QBP (\eta, \bth_{-\n}) \big]    \bigg)  - \frac{1}{2}\big[ { \QAP( \eta, \bth_{-\n}) } \big]^2 \Bigg].
	\end{align}
	
	\rev{
		We can check that if
		\begin{align}\label{sufficient_condition_quadratic}
		-  \frac{  \partial{  \QBP (\eta, \bth_{-\n}) }/(  \QBP (\eta, \bth_{-\n}) + 1/\vars )}{ \partial{\thn} /\thn  } \geq  \frac{  1 }{ 2 }, \forall \n\in\Nset,
		\end{align}
		we have
		\begin{align}
		&2  \QBP (\eta, \bth_{-\n})  \cdot \eta \cdot \frac{ \partial{ \QBP (\eta, \bth_{-\n}) } }{ \partial{\eta} }  +  \big[ \QBP (\eta, \bth_{-\n}) + 1/\vars \big] \cdot \QBP (\eta, \bth_{-\n}) \leq 0. \nonumber
		\end{align}
		
		Later we will show that $\QBP (\eta, \bth_{-\n})$ is non-increasing in $\eta$, $\forall \n\in \Nset$ (i.e., Lemma \ref{lemma:feasibility_typeII}), hence,
		\begin{align}
		2  \QBP (\eta, \bth_{-\n})   \eta  \frac{ \partial{ \QBP (\eta, \bth_{-\n}) } }{ \partial{\eta} }   &+  \big[ \QBP (\eta, \bth_{-\n}) + 1/\vars \big]  \QBP (\eta, \bth_{-\n}) +  \frac{\eta}{\vars}   \frac{ \partial{}}{ \partial{\eta} } \big[   \QBP (\eta, \bth_{-\n}) \big]  \leq 0.\nonumber
		\end{align}
	}

	\begin{lemma}\label{lemma:sufificient_condition_quadratic_unifor_proof}
		If $\thn \sim \mbox{Uniform}[\thl, \thu]$ which is independent for every $\n \in \Nset$, then \eqref{sufficient_condition_quadratic} is satisfied.
	\end{lemma}

	\emph{Proof}:
	First, for an agent $\n \in \Nset$,
	\begin{align}
	\QBP (\thn, \bth_{-\n}) = &\frac{ 1 }{ \thn + \frac{\F(\thn)}{\f(\thn)}  } \cdot  \frac{1}{\big[W(\bth) \big]^2},
	\end{align}
	where the function $W: [\thl, \thu]^{\N} \rightarrow \Rsetp$ is the solution of the below equation:
	\begin{align}\label{eq:aggregate_effort_typeII_proof}
	\big[ W(\bth) \big]^3 - \frac{1}{\vars} \cdot \big[ W(\bth) \big]^2 - \sum_{m \in \Nset} \frac{ 1 }{ \th_m + \frac{\F(\th_m)}{\f(\th_m)}  } =0.
	\end{align}
	hence
	\begin{align*}
	-\frac{\partial \QBP(\thn, \bth_{-\n})}{\partial \thn} \frac{\thn}{\QBP(\thn, \bth_{-\n}) + 1/{\sigma^2_0}}
	& \geq \frac{1 + \frac{\partial}{\partial \th} \left(\frac {\F(\th)}{\f(\th)} \right) }{  1 + \frac{1}{ \th} \left(\frac {\F(\th)}{\f(\th)} \right)  } \cdot \frac{1}{ 1 + \frac{ [ W(\bth) ]^2 }{{\sigma^2_0}} \big(\thn + \frac{\F(\thn)}{\f(\thn)} \big) } \\
	%
	& \geq \frac{1}{2},
	\end{align*}
	where the final inequality holds for uniform distribution. $\Halmos$

	\rev{
		As the agent is selfish, he will exert effort $\q_{\no}$ to maximize his expected payoff. Hence, the agent's exerted effort can be obtained by taking the first order derivative of (\ref{eq:agent_expect_bidding_typeII}) with respect to ${\qn}$ and setting it to zero, which is}
	\begin{equation}\label{eq:h1_func_bid}
	\begin{aligned}
	( 1/\vars + \qbpn )^2 \cdot \qbpn \cdot \thrn = ( 1/\vars + \qn )^2 \cdot \qn \cdot \thn,
	\end{aligned}
	\end{equation}
	where $\qbpn$ is a shorthand for $\QBP(\thrn, \bth_{-\n})$.
	
	Based on \eqref{eq:h1_func_bid}, we have (i) if $\thrn> \thn $, $\qbpn < \qn$, (ii) if $\thrn  < \thn $, $\qbpn> \qn $, and (iii) if $\thrn  = \thn $, $\qbpn = \qn $.

	Then if $\thrn > \thn$, the equation \eqref{eq:compare_mean_value_typeII} is negative and \eqref{eq:mean_value_typeII} is positive.
	This inequality also holds for $\thrn < \thn$, by a similar argument. Therefore, the agent $\n$ will truthfully report his own cost type.
	
	\textbf{Step 3.  Incentivize agents to exert precisely the efforts as desired by the principal under COPE}
	
	Then we will show that an agent $\n$, $\forall \n \in \Nset$ exerts effort such that $\qn = \qbpn$ would maximize his expected payoff as follows.
	\begin{equation}\label{eq:agent_expect_bidding_typeII_updated}
	\begin{aligned}
	&\EX_{\{\x, \yn,\BThs_{-\n}\}} \big[  \UA( \x, \qn,\yn, \thn,  \bth_{-\n})  \big] \\
	&=\EX_{\BThs_{-\n} } \big[ \pib(\thn, \bth_{-\n}) - \Sab(  \thn, \bth_{-\n}) \cdot \frac{1}{1/{\var_0} +  { \qn}}  + \Sbb(\thn, \bth_{-\n}) - \frac{1}{2} \thn \qn^2 \big],
	\end{aligned}
	\end{equation}
	where the expectation is taken with respect to $\BThs_{-\n}$, $\x$, and $\yn$.
	
	It can be verified that \eqref{eq:agent_expect_bidding_typeII_updated} is concave in $\qn$.
	Hence, by taking the first order derivative of \eqref{eq:agent_expect_bidding_typeII_updated} with respect to $\qn$, we have
	\begin{align}\label{eq:agent_FOC_typeII}
	&\frac{\partial}{\partial{\qn}}\EX \big[ \UA( \x, \qn,\yn, \thn,  \bth_{-\n}) \big]  =  \bigg[  \frac{1/{\var_0} +  { \qbpn}}{1/{\var_0} + \qn} \bigg]^2 \cdot \thn \cdot \qbpn - \th_{\no} \cdot \qn.
	\end{align}
	
	We can verify that the value of \eqref{eq:agent_FOC_typeII} equals to zero only when $\qn = \qbpn$. Hence, agent $\n$ will exert the effort as the principal desires to maximize his expected payoff. Then \eqref{eq:agent_report_typeII} is rewritten as
	\begin{align}\label{eq:agent_report2_typeII}
	\yrn = \frac{  {\mu_0} \cdot  { 1/{\var_0}} +  {\yn} \cdot { \qbpn} }{ { 1/{\var_0}} +   \qbpn }.
	\end{align}
	
	Because the principal knows the value of $\mu_0$, $\var_0$, and $\qbpn$, she can infer the agent $\n$'s truth observation $\yn$ from \eqref{eq:agent_report2_typeII}.

	\textbf{Step 4. Maximize the principal's expected utility under COPE}
	Then we look at the expected payoff of the principal.
	We will show that the desired effort level $\QBP(\thn, \bth_{-\n})$ defined in \eqref{eq:effort_cost_II} and the function $\pib(\thn, \bth_{-\n})$ defined in \eqref{eq:reward_fun_special_II_pi_proof} can maximize the {\pp}'s expected payoff and satisfy BIC and BIR condition.
	
	Notice that when an agent $\n$, $\forall \n \in \Nset$ exerts effort such that $\qn = \qbpn$ and reports $\yrn= \frac{  {\mu_0} \cdot  { 1/{\var_0}} +  {\yn} \cdot { \qbpn} }{ { 1/{\var_0}} +   \qbpn }$, the expected payment function is reduced to
	\begin{align}\label{eq:payment_agent_typeII_proof_xx}
	&\EX_{\{\x, \yn,\BThs_{-\n}\}} \big[  \PB (  \x, \yn, \qn, \thn, \bth_{-\n} )   \big] \nonumber \\
	&=\EX_{\BThs_{-\n} } \big[ \pib(\thn,\bth_{-\n})   - \Sab(\thn, \bth_{-\n})  ( \x - \yrn )^2 + \Sbb(\thn, \bth_{-\n})  \big] = \EX_{\BThs_{-\n} } \big[ \pib(\thn,\bth_{-\n})  \big].
	\end{align}
	
	The expected payoff of the agent $\n$ is rewritten as
	\begin{align}\label{eq:payoff_agent_simple_typeII}
	&\EX_{\BThs_{-\n}} \big[ \UA\big( \pib(\thrn, \bth_{-\n}), \QBP(\thrn, \bth_{-\n}), \thn  \big) \big] = \EX_{\BThs_{-\n}} \big[  \pib(\thrn, \bth_{-\n}) - \frac{1}{2}\thn \cdot \big[ { \QBP(\thrn, \bth_{-\n})} \big]^2 \big],
	\end{align}
	and
	the BIC and BIR conditions, i.e., \eqref{eq:IC_requirement} and \eqref{eq:IR_requirement}, can be rewritten as
	\begin{align}
	&\EX_{\BThs_{-\n}} \big[ \UA\big( \pib(\thn, \bth_{-\n}), \QBP(\thn, \bth_{-\n}), \thn  \big) \big] \geq \EX_{\BThs_{-\n}} \big[ \UA\big( \pib(\thrn, \bth_{-\n}), \QBP(\thrn, \bth_{-\n}), \thn  \big) \big] ,~\forall \thrn \neq \thn, \label{eq:IC_requirement_simple_typeII}\\
	&\EX_{\BThs_{-\n}} \big[ \UA\big( \pib(\thn, \bth_{-\n}), \QBP(\thn, \bth_{-\n}), \thn  \big) \big] \geq 0,\ \forall \thn.\label{eq:IR_requirement_simpe_typeII}
	\end{align}
	
	Base on Lemma \ref{lemma:optimal_predictor_proof} and \eqref{eq:payment_agent_typeII_proof_xx}, the expected payoff of the principal is
	\begin{align}
	\EX_{\x, \by, \BThs} [ \UP(\x, \bqbp, \by, \bthr) ] &=   -  \hp \big(  \bqbp \big) -  \EX_{\x, \by, \BThs} \big[  \sum_{\n\in\Nset}  \P (  \x, \yn, \thn, \bth_{-\n} ) \big] \nonumber\\
	& =  -   \frac{ 1 }{ 1/{\var_0} +  \sum_{\n\in\Nset}{ \qbpn}} - \EX_{\BThs} \big[ \sum_{\n\in\Nset} \pia(\thn, \bth_{-\n}) \big].\nonumber
	\end{align}
	
	Recall that $\qbpn = \QBP(\thn, \bth_{-\n})$, the principal's optimal problem defined in \eqref{eq:optimal_mechanism} can be rewritten as
	\begin{equation}\label{eq:optimal_mechanism_typeII_proof}
	\begin{aligned}
	\sup_{\{ \QBP(\bth), \pib(\bth) \}, \forall \thn\in\BThs, \forall \n \in \Nset } ~&\EX [ \UP(\x, \bqbp, \by, \bthr) ] ,\\
	\mathrm{subject~to:~~} & \mathrm{BIC~and~BIR~in~(\ref{eq:IC_requirement_simple_typeII})~and~(\ref{eq:IR_requirement_simpe_typeII}).}
	\end{aligned}
	\end{equation}
	
	In the following lemmas, we characterize an equivalent formulation for the feasible region defined by BIC and BIR. Using these lemmas, we show that $\QBP(\thn, \bth_{-\n})$ defined in \eqref{eq:effort_cost_II} and $\pib(\thn, \bth_{-\n})$ defined in \eqref{eq:reward_fun_special_II_pi_proof}  are the optimal solution that solves the principal's problem in \eqref{eq:optimal_mechanism_typeII_proof}.
	\begin{lemma}\label{lemma:feasibility_typeII}
		The solution of \eqref{eq:optimal_mechanism_typeII_proof} is feasible if and only if it satisfies the following conditions for all $\thn\in[\thl, \thu]$:
		\begin{itemize}
			\item The expected payoff of the agent $\n$, $\forall \n \in \Nset$ is
			\begin{align}\label{eq:agent_profit_contract_feasi_typeII}
			&\EX_{\BThs_{-\n}} \big[ \UA\big( \pib(\thn, \bth_{-\n}), \QBP(\thn, \bth_{-\n}), \thn  \big) \big] =  \frac{1}{2}\EX_{\BThs_{-\n}} \bigg[  \int_{\thn}^{\thu}  \big[ \QBP(x, \bth_{-\n}) \big]^2  \dt{x}  \bigg],
			\end{align}
			\item $\QBP(\thn, \bth_{-\n} )$ is non-increasing in $\thn$.
		\end{itemize}
	\end{lemma}
	
	\emph{Proof}:
	The proof of Lemma \ref{lemma:feasibility_typeII} is as follows.
	We first show that BIC and BIR imply the condition in \eqref{eq:agent_profit_contract_feasi_typeII}.
	
	Notice that the first derivative of (\ref{eq:payoff_agent_simple_typeII}) is:
	\begin{equation}\label{eq:illustr_tmp_typeII}
	\begin{aligned}
	& \frac{ \partial{  \EX_{\BThs_{-\n}} \big[ \UA\big( \pib(\thrn, \bth_{-\n}), \QBP(\thrn, \bth_{-\n}), \thn  \big) \big]   } }{\partial{\thn}}=  \EX_{\BThs_{-\n}} \big[ - \frac{1}{2} \big[ \QBP(\thrn, \bth_{-\n})\big]^2 \big] \leq 0.
	\end{aligned}
	\end{equation}
	
	Then, for any $\thn^1 > \thn^2$, we have
	\begin{equation}\label{eq:illustr_tmp2_typeII}
	\begin{aligned}
	\EX_{\BThs_{-\n}}  \big[  \UA( \pib(\thn^1, \bth_{-\n}), \QBP(\thn^1, \bth_{-\n}), \thn^1)  \big] &\leq
	\EX_{\BThs_{-\n}} \big[  \UA( \pib(\thn^1, \bth_{-\n}), \QBP(\thn^1, \bth_{-\n}), \thn^2) \big]\\
	& \leq \EX_{\BThs_{-\n}} \big[  \UA( \pib(\thn^2, \bth_{-\n}), \QBP(\thn^2, \bth_{-\n}), \thn^2)  \big],
	\end{aligned}
	\end{equation}
	where the first inequality is due to \eqref{eq:illustr_tmp_typeII} and the second is due to the BIC condition defined in \eqref{eq:IC_requirement_simple_typeII}.
	
	Recall that the BIR condition is
	\begin{align}\label{eq:illustrate_IR_sufficient_appendix_typeII_xx}
	\EX_{\BThs_{-\n}} \big[ \UA\big( \pib(\thn, \bth_{-\n}), \QBP(\thn, \bth_{-\n}), \thn  \big) \big] \geq 0,\ \forall \thn \in [\thl, \thu],
	\end{align}
	which implies that, for an agent $\n\in\Nset$ with any value $\thn \in [\thl, \thu]$, his expected payoff should be nonnegative.
	Then the expected payoff of the agent $\n$ with cost parameter $\thu$ must be binding at zero. Otherwise, the principal can raise the $\pib(\thu, \bth_{-\n})$ by a small value of $\delta >0$, which does not violate the constraint of \eqref{eq:illustrate_IR_sufficient_appendix_typeII_xx} but raises the principal's expected payoff. Hence, we have
	\begin{equation}\label{eq:IR_proof_xxx_typeII}
	\begin{aligned}
	\EX_{\BThs_{-\n}} \big[  \UA( \pib(\thu, \bth_{-\n}), \QBP(\thu, \bth_{-\n}), \thu)  \big] &= 0.
	\end{aligned}
	\end{equation}
	
	Let $\UA( \thn, \bth_{-\n} )  =  \UA\big( \pib(\thn, \bth_{-\n}), \QBP(\thn, \bth_{-\n}), \thn \big)$. From BIC condition, we have
	\begin{align}
	&\EX_{\BThs_{-\n}} \big[  \UA( \thn, \bth_{-\n} )  \big] = \max_{\thrn}  \EX_{\BThs_{-\n}} \bigg[   \UA\bigg( \pib(\thrn, \bth_{-\n}), \QBP(\thrn, \bth_{-\n}), \thn\bigg)  \bigg].\nonumber
	\end{align}
	By using the envelope theorem, we have:
	\begin{equation*}
	\begin{aligned}
	\frac{\partial{\EX_{\BThs_{-\n}} \big[  \UA( \thn, \bth_{-\n} )  \big] } }{ \partial{\thn}} &= \left.\frac{\partial{\EX_{\BThs_{-\n}} \big[ \UA( \pib(\thrn, \bth_{-\n}), \QBP(\thrn, \bth_{-\n}), \thn)  \big]}}{\partial{\thn}}\right|_{\thrn = \thn}  =  \EX_{\BThs_{-\n}} \bigg[ - \frac{1}{2} \big[ \QBP(\thn, \bth_{-\n}) \big]^2 \bigg],
	\end{aligned}
	\end{equation*}
	where $\thn$ is a parameter.
	By integrating both sides {from the value of $\thn$ to $\thu$}
	and using \eqref{eq:IR_proof_xxx_typeII}, we get
	\begin{equation}
	\begin{aligned}
	&\EX_{\BThs_{-\n}} \bigg[ \UA\big( \pib(\thn, \bth_{-\n}), \QBP(\thn, \bth_{-\n}), \thn  \big) \bigg] =  \frac{1}{2}\EX_{\BThs_{-\n}} \bigg[  \int_{\thn}^{\thu} \big[ \QBP(x, \bth_{-\n}) \big]^2 \dt{x}  \bigg]
	\end{aligned}
	\end{equation}
	
	We prove that $\QBP(\thn, \bth_{\n})$ is nonincreasing in $\thn$ by contradiction. Let $p_{\n}$ as the shorthand notation for $\pi( \thn, \bth_{-\n} )$. Suppose for any $\thn^1 > \thn^2$, we have $\QBP(\thn^1,\bth_{-\n}) > \QAP(\thn^2,\bth_{-\n})$.
	Because
	\begin{equation}\label{eq:second_order_derivative_qth_typeII_simple}
	\begin{aligned}
	\frac{\partial^2{  \UA\big( p_{\n}, \qbpn, \thn\big)   }}{\partial{\qbpn}\partial{\thn}} =  -  \qbpn& < 0 , \mathrm{~~and~~}
	\end{aligned}
	\end{equation}
	\begin{equation}\label{eq:second_order_derivative_qq_typeII_simple}
	\begin{aligned}
	\frac{\partial^2{ \UA\big( p_{\n}, \qbpn, \thn\big)   }}{\partial{\qbpn}^2} = -\thn  \leq 0,~~~~
	\end{aligned}
	\end{equation}
	we have
	\begin{equation}
	\begin{aligned}
	0 &=\left.\frac{\partial{ \UA\big( p_{\n}, \qbpn, \thn^1\big)  }}{\partial{\qbpn}}\right|_{\qbpn=\QBP(\thn^1,\bth_{-\n})} \nonumber\\
	&\leq \left.\frac{\partial{ \UA\big( p_{\n}, \qbpn,  \thn^1\big)  }}{\partial{\qbpn}}\right|_{\qbpn=\QBP(\thn^2,\bth_{-\n})} \nonumber\\
	&< \left.\frac{\partial{\UA\big( p_{\n}, \qbpn,  \thn^2\big)  }}{\partial{\qbpn}}\right|_{\qbpn=\QBP(\thn^2,\bth_{-\n})},
	\end{aligned}
	\end{equation}
	{where the first equality is due to BIC when the agent $\n$'s cost parameter $\thn$ has the value of $\thn^1$, the second equality is due to\eqref{eq:second_order_derivative_qth_typeII_simple}, and the inequality is due to \eqref{eq:second_order_derivative_qq_typeII_simple}.
		
		However, based on the BIC condition, if the agent $\n$'s cost parameter $\thn$ has the value of $\thn^2$, then we should have
		\begin{align}
		\left.\frac{\partial{ \UA\big( p_{\n}, \qbpn, \thn^2\big)  }}{\partial{\qapn}}\right|_{\qbpn=\QBP(\thn^2,\bth_{-\n})}  = 0, \nonumber
		\end{align}
		which holds true for all scalar value of $p_{\n}$.
		Hence, for any $\thn^1 > \thn^2$, $\QBP(\thn^1,\bth_{-\n}) \leq \QBP(\thn^2,\bth_{-\n})$.
	}

	Then we need to prove that (\ref{eq:agent_profit_contract_feasi_typeII}) implies BIC and BIR defined in (\ref{eq:IC_requirement_simple_typeII}) and (\ref{eq:IR_requirement_simpe_typeII}). Notice that we have:
	\begin{equation*}
	\begin{aligned}
	&\EX_{\BThs_{-\n}} \bigg[  \UA\big( \pib(\thrn, \bth_{-\n}), \QBP(\thrn, \bth_{-\n}), \thn\big)   \bigg]  \\
	& = \EX_{\BThs_{-\n}} \bigg[ - \int_{\thn}^{\thu}  \frac{\partial{ \UA\big( \pib(\thrn, \bth_{-\n}), \QBP(\thrn, \bth_{-\n}), z \big)  }}{\partial{z}}\dt{z}    \bigg]   \\
	&= \EX_{\BThs_{-\n}} \bigg[  \frac{1}{2}\int_{\thrn}^{\thu} \big[  \QBP(\eta, \bth_{-\n}) \big]^2 \dt{\eta}  -  \int_{\thn}^{\thrn}  \frac{\partial{ \UA\big( \pib(\thrn, \bth_{-\n}), \QBP(\thrn, \bth_{-\n}), z \big)  }}{\partial{z}}\dt{z}   \bigg] \\
	&= \EX_{\BThs_{-\n}} \bigg[  -\frac{1}{2}\int_{\thu}^{\thn} \big[ \QBP(\eta, \bth_{-\n}) \big]^2\dt{\eta} -  \frac{1}{2}\int_{\thn}^{\thrn} \big[ \QBP(\eta, \bth_{-\n})  \big]^2 \dt{\eta} + \frac{1}{2} \int_{\thn}^{\thrn}  \big[ \QBP(\thrn, \bth_{-\n}) \big]^2\dt{z}   \bigg] \\
	&= \EX_{\BThs_{-\n}} \bigg[ \UA\big( \pib(\thn, \bth_{-\n}), \QBP(\thn, \bth_{-\n}), \thn \big) + \frac{1}{2}\int_{\thn}^{\thr}  \bigg( \big[ \QBP(\thrn, \bth_{-\n}) \big]^2 -  \big[ \QBP(\eta, \bth_{-\n}) \big]^2 \bigg) \dt{\eta} \bigg], \\
	\end{aligned}
	\end{equation*}
	where the second equality and the forth equality is obtained by \eqref{eq:agent_profit_contract_feasi_typeII}.
	
	If $\thrn > \thn$, then the above equation is non-positive (because $\QBP(\eta, \bth_{-\n})$ is non-increasing in $\eta$), hence
	\begin{align}
	&\EX_{\BThs_{-\n}} \big[  \UA( \pib(\thrn, \bth_{-\n}), \QBP(\thrn, \bth_{-\n}), \thn)   \big] < \EX_{\BThs_{-\n}} \big[ \UA( \pib(\thn, \bth_{-\n}), \QBP(\thn, \bth_{-\n}), \thn)\big].\nonumber
	\end{align}
	This inequality also holds for $\thrn < \thn$ by a similar argument. Therefore, the two condition imply BIC.
	
	BIR is verified by putting $\thn$ back to \eqref{eq:agent_profit_contract_feasi_typeII}. $\Halmos$

	Then based on Lemma \ref{lemma:feasibility_typeII}, we have the following lemma.
	\begin{lemma}\label{lemma:optimility_typeII}
		The optimisation problem in (\ref{eq:optimal_mechanism_typeII_proof}) has the following equivalent formulation:
		\begin{align}\label{eq:optimal_contrao_typeII}
		\max_{ \{ \QBP(\bth)\}, \forall \thn\in\BThs } & \EX_{\BThs} \bigg[ -   \frac{ 1 }{ 1/{\var_0} +  \sum_{\n\in\Nset}{ \QBP(\thn, \bth_{-\n})}} - \frac{1}{2}\sum_{\n \in \Nset} \big[ \QBP(\thn, \bth_{-\n}) \big]^2  \thn  - \frac{1}{2}\sum_{\n \in \Nset}  \big[ \QBP(\thn, \bth_{-\n}) \big]^2  \frac{  F(\thn)  }{  f(\thn) }   \bigg],\nonumber\\
		\mathrm{s.t.~} & \QBP(\thn, \bth_{-\n}) \mathrm{~is~nonincreasing~in~}\thn.~~
		\end{align}
	\end{lemma}
	
	\emph{Proof}:
	The proof of Lemma \ref{lemma:optimility_typeII} is as follows.
	The expected payoff of the principal can be written as:
	\begin{equation*}
	\begin{aligned}
	&\EX_{\BThs} \bigg[   -   \frac{ 1 }{ 1/{\var_0} +  \sum_{\n\in\Nset}{ \QBP(\thn, \bth_{-\n})}}  - \frac{1}{2}\sum_{\n \in \Nset} \big[ \QBP(\thn, \bth_{-\n}) \big]^2 \cdot \thn   -\sum_{\n \in \Nset} \UA\big( \pib(\thn, \bth_{-\n}), \QBP(\thn, \bth_{-\n}), \thn\big)   \bigg] \\
	&=\EX_{\BThs} \bigg[   -   \frac{ 1 }{ 1/{\var_0} +  \sum_{\n\in\Nset}{ \QBP(\thn, \bth_{-\n})}} - \frac{1}{2}\sum_{\n \in \Nset} \big[ \QBP(\thn, \bth_{-\n}) \big]^2   - \frac{1}{2}\sum_{\n \in \Nset}   \int_{\thn}^{\thu} \big[ \QBP(x, \bth_{-\n}) \big]^2\dt{x}  \bigg] \\
	\end{aligned}
	\end{equation*}
	Using integration by parts and Lemma \ref{lemma:feasibility_typeII}, we can get the conclusion. $\Halmos$

	Based on Lemma \ref{lemma:optimility_typeII}, the principal's problem thus reduces to choosing the desired effort $ \QBP(\thn, \bth_{-\n})$ for each agent $\n \in \Nset$.
	
	Let $\qbpn = \QBP(\thn, \bth_{-\n})$ and
	\begin{align}
	M\big(\qbpo, \ldots, \qbpN \big) = & -   \frac{ 1 }{ 1/{\var_0} +  \sum_{\n\in\Nset}{ \qbpn}}- \frac{1}{2}\sum_{\n \in \Nset} \big[ \qbpn \big]^2 \cdot \thn - \frac{1}{2}\sum_{\n \in \Nset}  \big[ \qbpn \big]^2  \frac{  F(\thn)  }{  f(\thn) }. \nonumber
	\end{align}
	Let $G = [\partial^2{M}/\partial{\qbpi}\partial{\qbpj}]$ be the matrix of second order derivatives, and it is a symmetric matrix with negative diagonal terms as
	\begin{align}
	&\frac{  \partial^2 M}{ \partial{\qbpi}\partial{\qbpj}} = -\frac{2 }{ \big[ { 1/{\var_0} +  \sum_{\n\in\Nset}{ \qbpn  }} \big]^3 }, j \neq i, \nonumber\\
	&\frac{  \partial^2 M}{ {\partial{\qbpi} }^2} = -\frac{2 }{ \big[ { 1/{\var_0} +  \sum_{\n\in\Nset}{ \qbpn }} \big]^3 } - \th_{i} - \frac{  F(\th_{i})  }{  f(\th_{i}) }. \nonumber
	\end{align}

	As we can verify that, for $k = 1, \ldots, \N$, the $k$th leading principal minors of $G$ alternate in sign, hence $G$ is negative definite and $M$ is strictly concave. Thus, the {\pp}'s desired effort levels from agents $\qbpn = \QBP(\thn, \bth_{-\n})$, $\forall \n \in \N$ are the solution of the below equations:
	\begin{align}
	\label{eq:optmal_quality_typeII}
	&\frac{ 1 }{ \big[ 1/{\var_0} +  \sum_{\n\in\Nset}{ \qbpn } \big]^2} - \QBP(\thn, \bth_{-\n}) \cdot  \thn - \QBP(\thn, \bth_{-\n}) \cdot \frac{F(\thn)}{f(\thn)} = 0, \n = 1,2,\ldots, \N
	\end{align}
	
	Using Cramer's rule and the assumption that the function $\F$ is log concave in $\th$ , we can verify that
	\begin{align}
	\frac{ \partial{\QBP(\thn, \bth_{-\n})}  }{  \partial{\thn} } = - \frac{   \partial^2{M}/\partial{\qbpn}\partial{\thn}   }{  \partial^2{M}/{\partial{\qbpn}}^2  } \leq 0,
	\end{align}
	which shows that $\QBP(\thn, \bth_{-\n})$ derived from \eqref{eq:optmal_quality_typeII} is non-increasing in $\thn$, so that it is the feasible solution of \eqref{eq:optimal_contrao_typeII}.
	
	The solution of \eqref{eq:optmal_quality_typeII} is
	\begin{align}
	\QBP (\thn, \bth_{-\n}) = &\frac{ 1 }{ \thn + \frac{\F(\thn)}{\f(\thn)}  } \cdot  \frac{1}{\big[W(\bth) \big]^2},
	\end{align}
	where the function $W: [\thl, \thu]^{\N} \rightarrow \Rsetp$ is the solution of the below equation:
	\begin{align}\label{eq:aggregate_effort_typeII_proof}
	\big[ W(\bth) \big]^3 - \frac{1}{\vars} \cdot \big[ W(\bth) \big]^2 - \sum_{m \in \Nset} \frac{ 1 }{ \th_m + \frac{\F(\th_m)}{\f(\th_m)}  } =0.
	\end{align}

	The real root of the above cubic equation is as follows.
	\begin{align}\label{eq:aggregate_effort_type_II_proof_xx}
	W(\bthr) = &\frac{ 1 }{ 3 \sigma^2_0}
	+ \sqrt[3]{   \frac{1}{ 27 \sigma^6_0 } + \frac{1}{2}  \bigg[  \sum_{m \in \Nset} \frac{ 1 }{ \thr_{m} + \frac{\F(\thr_{m})}{\f(\thr_{m})} }    \bigg]  + \sqrt{ \lambda(\bthr)  }    } + \sqrt[3]{   \frac{1}{ 27 \sigma^6_0 } + \frac{1}{2}  \bigg[  \sum_{m \in \Nset} \frac{ 1 }{ \thr_{m} + \frac{\F(\thr_{m})}{\f(\thr_{m})} }    \bigg]  - \sqrt{  \lambda(\bthr)  }    },
	\end{align}
	where function $\lambda: [\thl, \thu]^{\N} \rightarrow \Rsetp$ is given as
	\begin{align}
	\lambda(\bthr) = \frac{1}{ 27 \sigma^6_0 } \bigg[  \sum_{m \in \Nset} \frac{ 1 }{ \thr_{m} + \frac{\F(\thr_{m})}{\f(\thr_{m})} }    \bigg] + \frac{1}{4}\bigg[  \sum_{m \in \Nset} \frac{ 1 }{ \thr_{m} + \frac{\F(\thr_{m})}{\f(\thr_{m})} }    \bigg]^2. \nonumber
	\end{align}
	According to \eqref{eq:agent_profit_contract_feasi_typeII}, we have
	\begin{align}\label{eq:payment_typeI_proof}
	&\EX_{\BThs_{-\n}} \big[ \pib(\thn,\bth_{-\n}) - \frac{1}{2} \cdot [ \QBP(\thn, \bth_{-\n}) ]^2 \cdot \thn  \big] =  \frac{1}{2}\EX_{\BThs_{-\n}} \big[  \int_{\thn}^{\thu}   \big[ \QBP(x, \bth_{-\n}) \big]^2  \dt{x}  \big]. \nonumber
	\end{align}
	From the above equation, we can derive the optimal payment function as given in \eqref{eq:reward_fun_special_II_pi_proof}.


	\section{Proof of  Theorem~\ref{thm:optimal_general}: General Setting}
	The proof will proceed in four steps. The first three steps show that our mechanism incentivizes the agents to be truthful, and the fourth step proves optimality of our mechanism. First, we show that irrespective of what an agent reports as his cost parameter, and irrespective of the effort he exerts, the agent is always incentivized to report his true observation. We follow this up and show that irrespective of the effort that an agent exerts, he is always incentivized to report his cost parameter correctly. The third step completes the proof of truthfulness, showing that under truthful reporting of the cost parameter and the observation, in our mechanism, an agent is always incentivized to exert precisely the effort as desired by the principal. Finally, we show that among all mechanisms that ensure truthful reports, our mechanism maximizes the principal's expected utility.
	
	\textbf{Step 1. Truthful reporting of observation under COPE}
	
	We first show that the agent will choose
	\begin{align}
	\yn = \arginf_{\yrn(\yn)} \EX\big[  \la(\x, \yrn) \big].
	\end{align}
	to maximize his expected payoff, given his exerting effort $\qn$ and own observation $\yn$.
	
	As shown in \eqref{eq:reward_fun_general}, $\pi(\thrn, \bth_{-\n})$, $\Sa(\thrn, \bth_{-\n})$ and $\Sb(\thrn, \bth_{-\n})$ are independent of $\yrn$. Moreover, the value of $\Sa(\thrn, \bth_{-\n})$ is always positive. Hence, when the agent $\n$ determines his reporting observation strategy to maximize his expected payoff, i.e.,
	\begin{align}
	&\yrn \in \argmax \EX \big[ \pi(\thrn, \bth_{-\n})   - \Sa(\thrn, \bth_{-\n}) \cdot \la(\x, \yrn)+ \Sb(\thrn, \bth_{-\n}) \big] - \C\big(  \qn, \thn \big), \nonumber
	\end{align}
	where the expectation is taken with respect to $\x$ and cost parameters $\BThs_{-\n} = [\Ths_1, \ldots, \Ths_{\n-1}, \Ths_{\n+1}, \ldots, \Ths_{\N}]^{T}$ of all the agents except agent $\n$, it is equivalent for the agent $\n$ to choose the reporting strategy such that
	\begin{align}
	\yrn \in \argmin \EX_{\x}[ \la(\x, \yrn) ].
	\end{align}

	\rev{According to \eqref{Bayes_risk_assumtion_general} and \eqref{eq:utility_ag_new_general} in Section \ref{sec:assumption_general_cases}}, the value of $\EX_{\x}[  \la(\x, \yrn) ]$ is minimized when $\yrn = \yn$, and the expected value is
	\begin{align}
	\EX_{\x}[  \la(\x, \yrn) ] = \ha \big(  \qn \big). \nonumber
	\end{align}
	
	
	\textbf{Step 2. Truthful reporting of cost parameter under COPE}
	
	We now show that the agent will truthfully reveals its cost type.
	The expected payoff of the agent whose cost type is $\th_{\no}$ but reports $\thr_{\no}$ is:
	\begin{equation}\label{eq:agent_expect_bidding_general}
	\begin{aligned}
	&\EX_{\{\x, \y_{\n},\BThs_{-\n}\}} \big[  \UA( \x, \thr_{\n}, \q_{\n},\y_{\n}, \th_{\n},  \bth_{-\n})  \big] \\
	&=\EX_{\BThs_{-\n}} \bigg[ \pi(\thrn,\bth_{-\n}) - \Sa(  \thrn,\bth_{-\n} ) \cdot \ha\big( \qn \big)  + \Sb(\thrn, \bth_{\n}) - \C( \qn, \thn) \bigg].
	\end{aligned}
	\end{equation}
	
	For notational convenience, we define the function $\UAE: \Rset \times [\thl, \thu] \times \Rsetp \times [\thl, \thu]^{\N} \rightarrow \Rsetp$ as
	\begin{align}\label{eq:agent_expected_payoff_new_general_proof}
	\UAE( \thrn, \qn, \thn, \bth_{-\n}  ) &=
	\big[ \pi(\thrn,\bth_{-\n}) - \Sa(  \thrn,\bth_{-\n} ) \cdot \ha\big( \qn \big)  + \Sb(\thrn, \bth_{\n}) - \C( \qn, \thn) \big],
	\end{align}
	where $\bth_{-\n}$ are the random variables of all agents cost type except that of agent $\n$. By comparing \eqref{eq:agent_expect_bidding_general} to \eqref{eq:agent_expected_payoff_new_general_proof}, the expected payoff of the agent $\n$ is
	\begin{equation*}
	\EX_{\{\x, \y_{\n},\BThs_{-\n}\}} \big[  \UA( \x, \thr_{\n}, \q_{\n},\y_{\n}, \th_{\n},  \bth_{-\n})  \big] = \EX_{\BThs_{-\n}} \big[  \UAE( \thr_{\n}, \q_{\n}, \th_{\n}, \bth_{-\n}  ) \big].
	\end{equation*}
	
	Then, by the mean value theorem, we have:
	\begin{equation*}\label{eq:mean_value_general}
	\begin{aligned}
	&\EX_{\BThs_{-\n}} \bigg[  \UAE( \thn, \thn, \bth_{-\n} )  \bigg]  - \EX_{\BThs_{-\n}} \bigg[  \UAE( \thrn, \thn, \bth_{-\n} )  \bigg] = \EX_{\BThs_{-\n}} \bigg[  \frac{ \partial{ \UAE( \eta, \thn, \bth_{-\n}  )} }{\partial{\eta}}  \bigg] \cdot (  \thn - \thrn )
	\end{aligned}
	\end{equation*}
	We further have:
	\begin{equation}\label{eq:compare_mean_value_general}
	\begin{aligned}
	&\EX_{\BThs_{-\n}} \bigg[  \frac{ \partial{ \UAE( \eta, \thn, \bth_{-\n}  )} }{\partial{\eta}}  \bigg]  \\
	&= \EX \bigg[ \frac{ \ha(\qn)  }{  \dt{\ha(\qpn)}/\dt{\qpn}  } \cdot \bigg( 1 - \frac{ \ha (\qpn)  }{ \ha(\qn)}  \bigg) \cdot \bigg(  \frac{\partial{\c( \qpn, \eta)}}{\partial{\eta}} - \frac{ \c(\qpn, \eta) }{ \dt{\ha(\qpn)}/\dt{\qpn}  } \cdot \frac{\dt^2{\ha(\qpn)}}{\dt{\qpn}^2 } \cdot \frac{\partial{\qpn}}{\partial{\eta}} \bigg) \bigg]
	\end{aligned}
	\end{equation}

	\rev{As the agent is selfish, he will exert effort $\qn$ to maximize his expected payoff. Hence, the agent's exerted effort can be obtained by taking the first order derivative of (\ref{eq:agent_expected_payoff_new_general_proof}) with respect to ${\qn}$ and setting it to zero, which is}
	\begin{equation}\label{eq:h1_func_bid_general}
	\begin{aligned}
	\frac{ \dt{ \ha\big( \qn \big)  } }{ \dt{\qn} } \cdot \c( \qpn, \thrn ) = \frac{ \dt{ \ha\big( \qpn \big)  } }{\dt{\qpn} } \cdot \c( \qn, \thn )
	\end{aligned}
	\end{equation}
	
	As we assume that $\frac{ \dt{\ha(z)} }{ \dt{z} } \leq 0$, then based on \eqref{eq:h1_func_bid_general}, we have (i) if $\thrn > \thn$, $\qpn < \qn$, (ii) if $\thrn < \thn$, $\qpn> \qn$, and (iii) if $\thrn = \thn$, $\qpn = \qn$.

	As we assume that $\dt{\ha(\qpn)}/\dt{\qpn} \leq 0$, ${  \dt^2{ \ha(  \qpn  ) } }/{  \dt{  {\qpn}^2  }   } \geqslant 0$, and later we will prove in Lemma \ref{lemma:optimility_general} that $\QP (\thn, \bth_{-\n})$ is nonincreasing in $\thn$,
	if \eqref{eq:bid_assumption_general} holds, i.e.,
	${  \partial{  \c\left( \QP (\eta, \bth_{-\n}), \eta \right)  }  }/{  \partial{\eta}  } \leqslant 0$,
	then the above equation \eqref{eq:compare_mean_value_general} is negative when $\thrn > \thn$. Based on this, we have
	$\EX_{\BThs_{-\n}} \big[  \UAE( \thn, \thn, \bth_{-\n} )  \big] > \EX_{\BThs_{-\n}} \big[  \UAE( \thrn, \thn, \bth_{-\n} )  \big]$.
	This inequality also holds for $\thrn < \thn$, by a similar argument. Therefore, an agent will truthfully report his own cost parameter.
	
	\textbf{Step 3.  Incentivize agent to exert precisely the effort as desired by the principal under COPE}
	As we have proved in Step 2 that the agent $\n$ would truthfully report his cost type ($\thrn = \thn$).
	Next we will show that the agent $\n$ exerts effort such that $\qn = \qapn$ would maximize his expected payoff as follows.
	\begin{equation}\label{eq:ag_payoff_contract_expect_updated_general}
	\begin{aligned}
	&\EX \big[ \UAE( \x, \qapn, \yn, \thrn,  \bthr_{-\n}) \big]  =  \pi(\thrn,\bthr_{-\n})   - \Sa(  \thrn ) \cdot \ha \big(  \qn \big)  + \Sb(\thrn) - \C(\qn, \thn),
	\end{aligned}
	\end{equation}
	where the expectation is taken with respect to $\BThs_{-\n}$.
	
	It can be verified that \eqref{eq:ag_payoff_contract_expect_updated_general} is concave in $\qn$
	as we assume that \begin{align}
	\frac{\dt^2{\ha \big(  \qn \big)}}{ {\dt{\qn} }^2} \geq 0. \nonumber
	\end{align}
	Hence, by taking the first order derivative of \eqref{eq:ag_payoff_contract_expect_updated_general} with respect to $\qn$, we have
	\begin{align}\label{eq:agent_FOC_general}
	&\frac{\partial}{\partial{\qn}}\EX \big[ \UAE( \x, \qapn, \yn, \thrn  \bthr_{-\n}) \big]  =  \frac{ \c\big( \qpn, \thrn ~\big) }{  \dt{ \ha\big( \qpn \big)  }/\dt{\qpn}  } \cdot \frac{ \dt{ \ha\big( \qn \big)  } }{ \dt{\qn} } - c(\qn, \thn).
	\end{align}
	
	We can verify that the value of \eqref{eq:agent_FOC_general} equals to zero when $\qn = \qpn$. Hence, agent $\n$ will exert the effort as the principal desires to maximize his expected payoff.
	
	
	\textbf{Step 4. Maximize the principal's expected utility under COPE}
	
	Then we look at the expected payoff of the principal. To maximize the expected utility for the prediction, the principal solves
	\begin{align}
	&\max_{\xe}\EX \big[ -  \lp\big(  \x,\xe(  \by, \bqp ) | ( \yrn, \bqp ) \big].
	\end{align}
	
	The principal employs the Bayes estimate $\xe$ as follows:
	\begin{align}
	\xe\big(\byr, \bq^{*}\big) = \arginf_{\xe} \EX \big[  \lp\big(  \x,\xe(  \byr, \bqp)  \big) \big].
	\end{align}
	It follows that the expected utility of the principal is
	\begin{align}
	\hp \big(  \bqp \big) =  \inf_{\xe} \EX \big[  \lp\big(  \x,\xe(  \by, \bqp )  \big) \big].
	\end{align}
	
	We then show that the desired effort level $\QP(\thrn, \bthr_{-\n})$ defined as the solution of  \eqref{eq:optimal_contrac_equivalent_general} and the function $\pi(\thrn, \bthr_{-\n})$ defined in \eqref{eq:reward_fun_general_pi_proof} can maximize the {\pp}'s expected payoff and satisfy BIC and BIR conditions.
	
	Notice that agent $\n$ exerts effort such that $\qn = \qpn$ and reports $ \yn \in \argmin \EX_{\x}[ \la(\x, \yrn) ]$, the expected payment function defined in \eqref{eq:reward_fun_general} is reduced to
	\begin{align}
	\EX[ \P(  \x, \yn, \thn, \bth_{-\n} ) ] =
	\pi(\thn, \bth_{-\n}), \nonumber
	\end{align}
	where the expectation is taken with respect to $\x$ and $\yn$.
	
	Then the expected payoff of agent $\n$ is rewritten as
	\begin{align}\label{eq:payoff_agent_simple_general}
	&\EX_{\BThs_{-\n}} \big[ \UAE\big( \pi(\thrn, \bth_{-\n}), \QP(\thrn, \bth_{-\n}), \thn  \big) \big] = \EX_{\BThs_{-\n}} \big[  \pi(\thrn, \bth_{-\n}) - \C\big( { \QP(\thrn, \bth_{-\n} ) , \thn \big) }  \big],
	\end{align}
	and
	the BIC and BIR conditions, i.e., \eqref{eq:IC_requirement} and \eqref{eq:IR_requirement} can be rewritten as
	\begin{align}
	\EX_{\BThs_{-\n}} \big[ \UAE\big( \pi(\thn, \bth_{-\n}), \QP(\thn, \bth_{-\n}), \thn  \big) \big] &\geq \EX_{\BThs_{-\n}} \big[ \UAE\big( \pi(\thrn, \bth_{-\n}), \QP(\thrn, \bth_{-\n}), \thn  \big) \big]  ~\forall \thrn \neq \thn,\label{eq:IC_requirement_simple_general}\\
	\EX_{\BThs_{-\n}} \big[ \UAE\big( \pi(\thn, \bth_{-\n}), \QP(\thn, \bth_{-\n}), \thn  \big) \big] &\geq 0,\ \forall \thn.\label{eq:IR_requirement_simpe_general}
	\end{align}

	Then the expected payoff of the principal is
	\begin{align}\label{eq:payoff_pp_general}
	&\EX [ \UP(\x, \bqp, \by, \bth) ] =  -   \hp \big(  \bqp \big) -  \sum_{\n\in\Nset} \pi(\thn, \bth_{-\n})\nonumber，
	\end{align}
	where the expectation is taken with respect to $\x$ and $\by$.
	
	Recall that $\qpn = \QP(\thn, \bth_{-\n})$, we then rewrite \eqref{eq:optimal_mechanism} as
	\begin{equation}\label{eq:optimal_mechanism_general_proof}
	\begin{aligned}
	\sup_{\{ \QP(\bth), \pi(\bth) \}, \forall \thn\in\BThs } ~&\EX [ \UP(\x, \bqp, \by, \bthr) ] ,\\
	\mathrm{subject~to:~~} & \mathrm{BIC~and~BIR~in~(\ref{eq:IC_requirement_simple_general})~and~(\ref{eq:IR_requirement_simpe_general}).}
	\end{aligned}
	\end{equation}
	
	For the feasible region defined by BIC and BIR, we can characterise an equivalent formulation in the following lemma:
	\begin{lemma}\label{lemma:feasibility_general}
		The solution of \eqref{eq:optimal_mechanism_general_proof} is feasible if and only if it satisfies the following conditions for all $\thn\in[\thl, \thu]$:
		\begin{itemize}
			\item The expected payoff of agent $\n$ is
			\begin{align}\label{eq:agent_profit_contract_feasi_general}
			\EX_{\BThs_{-\n}} \bigg[ \UAE\big( \pi(\thn, \bth_{-\n}), \QP(\thn, \bth_{-\n}), \thn  \big) \big] =  \EX_{\BThs_{-\n}} \big[  \int_{\thn}^{\thu}  \frac{ \partial{\C\big( \QP(z, \bth_{-\n}), \eta \big) }}{ \partial{\eta}  }  \dt{z}  \bigg],
			\end{align}
			\item $\QP(\thn, \bth_{-\n} )$ is non-increasing in $\thn$.
		\end{itemize}
	\end{lemma}
	
	\emph{Proof}:
	We first show that BIC and BIR imply the condition \ref{eq:agent_profit_contract_feasi_general}.
	
	The first derivative of \eqref{eq:payoff_agent_simple_general} is
	\begin{align}\label{eq:illustr_tmp_general}
	&\frac{\partial{\EX_{\bth_{-\n}} \big[ \UAE( \pi(\thr, \bth_{-\n}), \QP(\thr, \bth_{-\n}), \eta)  \big]}}{\partial{\eta}}  =  \EX_{\bth_{-\n}} \bigg[ - \frac{ \partial{  \C(\QP(\thr, \bth_{-\n}), \eta )     }   } { \partial{\eta}  } \bigg] \leq 0.
	\end{align}
	
	Then, for any $\thn^1 > \thn^2$, we have
	\begin{equation}\label{eq:illstr_tmp2_general}
	\begin{aligned}
	\EX_{\BThs_{-\n}}  \big[  \UAE( \pi(\thn^1, \bth_{-\n}), \QP(\thn^1, \bth_{-\n}), \thn^1)  \big] &\leq
	\EX_{\BThs_{-\n}} \big[  \UAE( \pi(\thn^1, \bth_{-\n}), \QP(\thn^1, \bth_{-\n}), \thn^2) \big]\\
	& \leq \EX_{\BThs_{-\n}} \big[  \UAE( \pi(\thn^2, \bth_{-\n}), \QP(\thn^2, \bth_{-\n}), \thn^2)  \big],
	\end{aligned}
	\end{equation}
	where the first inequality is due to \eqref{eq:illustr_tmp_general} and the second is from the BIC condition defined in \eqref{eq:IC_requirement_simple_general}.
	
	Recall that the BIR condition is
	\begin{align}\label{eq:illustrate_IR_sufficient_appendix_general_xx}
	\EX_{\BThs_{-\n}} \big[ \UAE\big( \pia(\thn, \bth_{-\n}), \QAP(\thn, \bth_{-\n}), \thn  \big) \big] \geq 0,\ \forall \thn \in [\thl, \thu],
	\end{align}
	which implies that, for the agent $\n\in\Nset$ with any value $\thn \in [\thl, \thu]$, his expected payoff should be nonnegative.
	Then the expected payoff of the agent $\n$ with cost parameter $\thu$ must be binding at zero. Otherwise, the principal can reduce the $\pia(\thu, \bth_{-\n})$ by a small value of $\delta >0$, which does not violate the constraint of \eqref{eq:illustrate_IR_sufficient_appendix_general_xx} but raises the principal's expected payoff. Hence, we have
	\begin{equation}\label{eq:IR_proof_xxx_general}
	\begin{aligned}
	\EX_{\BThs_{-\n}} \big[  \UAE( \pi(\thu, \bth_{-\n}), \QP(\thu, \bth_{-\n}), \thu)  \big] &= 0.
	\end{aligned}
	\end{equation}


	Let $\UAE( \thn, \bth_{-\n} )  =  \UAE\big( \pi(\thn, \bth_{-\n}), \QP(\thn, \bth_{-\n}), \thn \big)$. From BIC condition, we have
	\begin{align}
	&\EX_{\BThs_{-\n}} \big[  \UAE( \thn, \bth_{-\n} )  \big] = \max_{\thrn}  \EX_{\BThs_{-\n}} \big[   \UAE\big( \pi(\thrn, \bth_{-\n}), \QP(\thrn, \bth_{-\n}), \thn\big)  \big].\nonumber
	\end{align}
	By using the envelope theorem, we have:
	\begin{equation*}
	\begin{aligned}
	\frac{\partial{\EX_{\BThs_{-\n}} \big[  \UAE( \thn, \bth_{-\n} )  \big] } }{ \partial{\eta}} &= \left.\frac{\partial{\EX_{\BThs_{-\n}} \big[ \UAE( \pi(\thrn, \bth_{-\n}), \QP(\thrn, \bth_{-\n}), \thn)  \big]}}{\partial{\thn}}\right|_{\thrn = \thn} \\
	& =  \EX_{\BThs_{-\n}} \bigg[ - \frac{ \partial{  \C(\QP(\thn, \bth_{-\n}), \thn )     }   }{ \partial{\thn}  }\bigg],
	\end{aligned}
	\end{equation*}
	where $\thn$ is a parameter. By integrating both sides from $\thn$ to $\thu$ and using \eqref{eq:IR_proof_xxx_general} and the assumption that the random variable $\thn$ of the agent $\n$ is independent for every $\n\in \Nset$, we get
	\begin{equation}
	\begin{aligned}
	&\EX_{\BThs_{-\n}} \big[ \UAE\big( \pi(\thn, \bth_{-\n}), \QP(\thn, \bth_{-\n}), \thn  \big) \big] =  \EX_{\BThs_{-\n}} \bigg[ \int_{\thn}^{\thu}   \frac{ \partial{  \C(\QP(z,\bth_{-\n}), \eta )     }   }{ \partial{\eta}  }  \dt{z}\bigg].
	\end{aligned}
	\end{equation}
	
	{We prove that $\QP(\thn, \bth_{\n})$ is nonincreasing in $\thn$ by contradiction. Let $p_{\n}$ as the shorthand notation for $\pi( \thn, \bth_{-\n} )$. Suppose for any $\thn^1 > \thn^2$, we have $\QP(\thn^1,\bth_{-\n}) > \QP(\thn^2,\bth_{-\n})$.  }
	Because
	\begin{equation*}\label{eq:second_order_derivative_qth_general_simple}
	\begin{aligned}
	\frac{\partial^2{  \UAE\big( p_{\n}, \qpn, \thn\big)   }}{\partial{\qpn}\partial{\thn}} =  -  \frac{\partial{\c( \q, \thn )}}{\partial{\thn}} & < 0 , \mathrm{~~and~~}
	\end{aligned}
	\end{equation*}
	\begin{equation*}\label{eq:second_order_derivative_qq_general_simple}
	\begin{aligned}
	\frac{\partial^2{ \UAE\big( p_{\n}, \qpn, \thn\big)   }}{\partial{\qpn}^2}  = -\frac{\partial{\c( \q, \thn )}}{\partial{\q}} < 0,
	\end{aligned}
	\end{equation*}
	we then have
	\begin{equation}
	\begin{aligned}
	0 &=\left.\frac{\partial{ \UAE\big( p_{\n}, \qpn, \thn^1\big)  }}{\partial{\qpn}}\right|_{\qpn=\QP(\thn^1,\bth_{-\n})} \nonumber\\
	&< \left.\frac{\partial{ \UAE\big( p_{\n}, \qpn,  \thn^1\big)  }}{\partial{\qpn}}\right|_{\qpn=\QP(\thn^2,\bth_{-\n})} \nonumber\\
	&< \left.\frac{\partial{\UAE\big( p_{\n}, \qpn,  \thn^2\big)  }}{\partial{\qpn}}\right|_{\qpn=\QP(\thn^2,\bth_{-\n})},
	\end{aligned}
	\end{equation}
	where the equality is due to BIC when the agent $\n$'s cost parameter $\thn$ has the value of $\thn^1$, the second equality is due to \eqref{eq:second_order_derivative_qq_general_simple}, and the inequality is due to \eqref{eq:second_order_derivative_qth_general_simple}.
	
	However, based on the BIC condition, if the agent $\n$'s cost parameter $\thn$ has the value of $\thn^2$, then we should have
	\begin{align}
	\left.\frac{\partial{ \UAE\big( p_{\n}, \qapn, \thn^2\big)  }}{\partial{\qpn}}\right|_{\qpn=\QP(\thn^2,\bth_{-\n})}  = 0, \nonumber
	\end{align}
	which holds true for all scalar values of $p_{\n}$.
	Hence, for any $\thn^1 > \thn^2$, $\QP(\thn^1,\bth_{-\n}) \leq \QP(\thn^2,\bth_{-\n})$.
	
	
	Then we need to prove that (\eqref{eq:agent_profit_contract_feasi_general}) implies BIC and BIR defined in (\eqref{eq:IC_requirement_simple_general}) and (\eqref{eq:IR_requirement_simpe_general}).
	
	BIR is verified by putting $\thn$ back to \eqref{eq:agent_profit_contract_feasi_general}. Besides, by putting $\thn = \thu$ back to \eqref{eq:agent_profit_contract_feasi_general}, we have
	\begin{align}
	&\EX_{\BThs_{-\n}} \bigg[ \UAE\big( \pia(\thu, \bth_{-\n}), \QAP(\thu, \bth_{-\n}), \thu  \big) \bigg] =  0. \nonumber
	\end{align}

	Then we prove that (\eqref{eq:agent_profit_contract_feasi_general}) implies BIC.
	Notice that we have:
	\begin{equation*}
	\begin{aligned}
	&\EX_{\BThs_{-\n}} \bigg[  \UAE\big( \pia(\thrn, \bth_{-\n}), \QAP(\thrn, \bth_{-\n}), \thn\big)   \bigg]  \\
	&\overset{1}{=} \EX_{\BThs_{-\n}} \bigg[ - \int_{\thn}^{\thu}  \frac{\partial{ \UAE\big( \pia(\thrn, \bth_{-\n}), \QAP(\thrn, \bth_{-\n}), z \big)  }}{\partial{z}}\dt{z}    \bigg]   \\
	&\overset{2}{=}  \EX_{\BThs_{-\n}} \bigg[ \UAE\big( \pia(\thrn, \bth_{-\n}), \QAP(\thrn, \bth_{-\n}), \thrn\big)  -  \int_{\thn}^{\thrn}  \frac{\partial{ \UAE\big( \pia(\thrn, \bth_{-\n}), \QAP(\thrn, \bth_{-\n}), z \big)  }}{\partial{z}}\dt{z}   \bigg] \\
	&\overset{3}{=}  \EX_{\BThs_{-\n}} \bigg[  \int_{\thrn}^{\thu} \frac{ \partial{  \C(\QP(\eta,\bth_{-\n}), z )     }   }{ \partial{z}  } \dt{\eta}  -  \int_{\thn}^{\thrn}  \frac{\partial{ \UAE\big( \pia(\thrn, \bth_{-\n}), \QAP(\thrn, \bth_{-\n}), z \big)  }}{\partial{z}}\dt{z}   \bigg] \\
	&\overset{4}{=}  \EX_{\BThs_{-\n}} \bigg[  -\int_{\thu}^{\thn} \frac{ \partial{  \C(\QP(\eta,\bth_{-\n}), z )     }   }{ \partial{z}  } \dt{\eta} -  \int_{\thn}^{\thrn} \frac{ \partial{  \C(\QP(\eta,\bth_{-\n}), z )     }   }{ \partial{z}  } \dt{\eta} +  \int_{\thn}^{\thrn}  \frac{ \partial{  \C(\QP(\thrn,\bth_{-\n}), z )     }   }{ \partial{z}  } \dt{z}   \bigg] \\
	&\overset{5}{=}  \EX_{\BThs_{-\n}} \bigg[ \UAE\big( \pia(\thn, \bth_{-\n}), \QAP(\thn, \bth_{-\n}), \thn \big) + \int_{\thn}^{\thr}  \bigg( \frac{ \partial{  \C(\QP(\thrn,\bth_{-\n}), z )     }   }{ \partial{z}  } -  \frac{ \partial{  \C(\QP(\eta,\bth_{-\n}), z )     }   }{ \partial{z}  } \bigg) \dt{\eta} \bigg], \\
	\end{aligned}
	\end{equation*}
	where the third equality and the fifth equality is obtained by \eqref{eq:agent_profit_contract_feasi_general}.
	
	If $\thrn > \thn$, then the above equation is non-positive. This is because $\QP(\eta, \bth_{-\n})$ is non-increasing in $\eta$ and ${  \partial{ \C(  \q, \th  ) } }/{  \partial{  \th  }   } > 0$. Hence,
	\begin{align}
	&\EX_{\BThs_{-\n}} \bigg[  \UAE( \pi(\thrn, \bth_{-\n}), \QP(\thrn, \bth_{-\n}), \thn)   \bigg] < \EX_{\BThs_{-\n}} \bigg[ \UAE( \pi(\thn, \bth_{-\n}), \QP(\thn, \bth_{-\n}), \thn)\bigg].\nonumber
	\end{align}
	This inequality also holds for $\thrn < \thn$ by a similar argument. Therefore, the two condition imply BIC. $\Halmos$
	

	Then based on Lemma \ref{lemma:feasibility_general} and let $\qpn = \QP(\thn, \bth_{-\n})$, we have the following Lemma.
	\begin{lemma}\label{lemma:optimility_general}
		The optimisation problem in (\ref{eq:optimal_mechanism_general_proof}) has the following equivalent formulation:
		\begin{align}\label{eq:optimal_contrac_equivalent_general}
		\max_{ \bqp} & \EX_{\BThs} \bigg[  -    \hp \big(  \bqp \big) -  \sum_{\n \in \Nset} \C\big(  \qpn , \thn \big) - \sum_{\n \in \Nset} \frac{  \partial{  \C\big(  \qpn, \thn  \big)}  }{  \partial{\thn}   } \cdot \frac{  F(\thn)  }{  f(\thn) }   \bigg] \nonumber\\
		\mathrm{s.t.~} & \qpn \mathrm{~is~nonincreasing~in~}\thn, \forall \n \in \Nset.~
		\end{align}
		where the expectation is taken with respect to $\BThs$.
	\end{lemma}
	
	
	\emph{proof}:	
	The proof of Lemma \ref{lemma:optimility} is as follows.
	The expected payoff of the principal can be written as:
	\begin{equation}\label{eq:optimiliayt_linear_proof_xx}
	\begin{aligned}
	&\EX_{\BThs} \bigg[    -   \hp \big(  \bqp \big)  -  \sum_{\n \in \Nset} \C\big(\qpn, \thn \big)   -\sum_{\n \in \Nset} \UA\big( \pia(\thn, \bth_{-\n}), \qpn, \thn\big)   \bigg] \\
	&=\EX_{\BThs} \bigg[    -  \hp \big(  \bqp \big)  - \sum_{\n \in \Nset}  \C\big(\qpn, \thn \big)   - \sum_{\n \in \Nset}   \int_{\thn}^{\thu} \frac{ \partial{\C\big( \qpn, \eta \big) }}{ \partial{\eta}  } \dt{x}  \bigg], \\
	\end{aligned}
	\end{equation}
	where the expectation is taken with respect to $\bth$.
	Notice that
	\begin{equation*}
	\begin{aligned}
	\EX_{\thn} \bigg[   \int_{\thn}^{\thu} \frac{ \partial{\C\big( \qpn, \eta \big) }}{ \partial{\eta}  } \dt{x} \bigg] &=
	\int_{\thl}^{\thu}  \int_{z}^{\thu} \frac{ \partial{\C\big( \qpn, \eta \big) }}{ \partial{\eta}  } \dt{x} \cdot f(z) \dt{z}
	= \int_{\thl}^{\thu} F(z) \frac{ \partial{\C\big( \qpn, \eta \big) }}{ \partial{\eta}  } \dt{z} \\
	&= \int_{\thl}^{\thu} \frac{F(z)}{f(z)} \frac{ \partial{\C\big( \qpn, \eta \big) }}{ \partial{\eta}  } f(z)\dt{z}  = \EX_{\thn} \bigg[   \frac{F(\thn)}{f(\thn)} \frac{ \partial{\C\big( \qpn, \eta \big) }}{ \partial{\eta}  } \bigg],
	\end{aligned}
	\end{equation*}
	where the first equation is obtained by using integration by parts. Then by applying the above equation to \eqref{eq:optimiliayt_linear_proof_xx} and the fact that $\{\thn\}_{\n \in \Nset}$ are assumed to be random, independently and identically distributed on support $[\thl, \thu]$, we can get the conclusion. $\Box$
	
	Based on Lemma \ref{lemma:optimility_general}, the principal's problem thus reduces to choosing the desired effort $\qpn= \QP(\thn, \bth_{-\n})$ for each agent $\n \in \Nset$.
	We first consider the problem in \eqref{eq:optimal_contrac_equivalent_general} without the constraint. If the optimal solution to this unconstrained problem is increasing, then it is also an optimal solution to the constrained problem.
	
	Let $\qpn = \QP(\thn, \bth_{-\n})$ and
	\begin{align}
	M\big(\qp_1, \ldots, \qp_{\N} \big) = &   -  \hp \big(  \bqp \big)  -  \sum_{\n \in \Nset} \C\big(\qpn, \thn \big)  - \sum_{\n \in \Nset} \frac{  \partial{  \C\big(  \qpn, \thn  \big)}  }{  \partial{\thn}   } \cdot \frac{  F(\thn)  }{  f(\thn) }. \nonumber
	\end{align}
	
	\rev{
		As we assume that ${  \partial^2{ \hp(  \bqp ) } }/{  \partial{  {\qp_i} }\partial{\qp_j} } \geqslant 0, \forall j \neq i$,
		${ \partial{\C(\qpn, \thn)} }/{ \partial{\qpn}  } >0$ and ${ \partial^2{\C(\qpn, \thn)}  }/{ \partial{\qpn}\partial{\thn}} \geq 0$, we can check that
		\begin{align}\label{eq:second_order_matrix_general_proof}
		\frac{  \partial^2 M}{ \partial{\qp_i}\partial{\qp_j} } &= - \frac{ \partial^2{\hp \big(  \bqp \big)} }{ \partial{\qp_i}\partial{\qp_j} } \leq 0, j \neq i, \nonumber\\
		\frac{  \partial^2 M}{ {\partial{\qp_i} }^2} &= - \frac{ \partial^2{\hp \big(  \bqp \big)} }{ {\partial{\qp_i} }^2 } - \c(\qp_i, \th_i)    - \frac{ \partial{\c(\qp_i, \th_i)}  }{ \partial{\th_i}} \cdot \frac{F(\th_i)}{f(\th_i)} \leq 0.
		\end{align}
		Let $G = [\partial^2{M}/\partial{\qp_{i}}\partial{\qp_{j}}]$, $i, j = 1,\ldots, \N$, be the matrix of second order derivatives. Matrix $G$ is symmetric with negative diagonal terms as shown in \eqref{eq:second_order_matrix_general_proof}.}
	
	As we can verify that, for $k =1, \ldots, \N$, the $k$th leading principal minors of $G$ alternate in sign, so that $G$ is negative definite and $W$ is strictly concave. The computational complexity of finding the optimal solution of  \eqref{eq:optimal_contrac_equivalent_general} will depend on the specific structure of the functions.
	
	Using Cramer's rule, the assumption that the c.d.f. function $\F$ is log concave in $\th$,  and the assumption that ${ \partial{\C(\qp_i, \th_i)} }/{ \partial{\qp_i}  } >0$ and ${ \partial^2{\c(\qp_i, \th_i)}  }/{ \partial{\qp_i}\partial{\th_i}} \geq 0$, we can verify that
	\begin{align}
	\frac{ \partial{\QBP(\thn, \bth_{-\n})}  }{  \partial{\thn} } = - \frac{   \partial^2{M}/\partial{\qbpn}\partial{\thn}   }{  \partial^2{M}/{\partial{\qbpn}}^2  } \leq 0, \forall \n \in \Nset,
	\end{align}
	which shows that $\QP(\thn, \bth_{-\n})$ derived by solving \eqref{eq:optimal_contrac_equivalent_general} is nonincreasing in $\thn$, so that it is a feasible solution of \eqref{eq:optimal_contrac_equivalent_general}.

	
	
	According to \eqref{eq:agent_profit_contract_feasi_general}, we have
	\begin{align}\label{eq:payment_typeI_proof}
	&\EX_{\BThs_{-\n}} \bigg[ \pi(\thn,\bth_{-\n}) - \C\big( \QP(\thn, \bth_{-\n}) , \thn \big)  \bigg] =  \EX_{\BThs_{-\n}} \bigg[  \int_{\thn}^{\thu}  \frac{ \partial{\C\big( \QP(z, \bth_{-\n}), \eta \big) }}{ \partial{\eta}  }  \dt{z}  \bigg] . \nonumber
	\end{align}
	From the above equation, we can derive the optimal payment function as given in \eqref{eq:reward_fun_general_pi_proof}.
	

\end{APPENDIX}

\theendnotes


\end{document}